\documentclass{article}

\usepackage{arxiv}

\usepackage[utf8]{inputenc} 
\usepackage[T1]{fontenc}    
\usepackage{hyperref}       
\usepackage{url}            
\usepackage{booktabs}       
\usepackage{amsfonts}       
\usepackage{nicefrac}       
\usepackage{microtype}      
\usepackage{lipsum}
\usepackage{graphicx}
\graphicspath{ {./images/} }
\usepackage{longtable}
\usepackage{geometry}
\usepackage{fancyhdr}
\usepackage{afterpage}
\usepackage{graphicx,pdflscape}
\usepackage{amsmath,amssymb,amsbsy}
\usepackage{algorithmicx}
\usepackage{algorithm}
\usepackage{algpseudocode}
\usepackage{dcolumn,array}
\usepackage{tocloft}
\usepackage{subcaption}
\title{Bayesian Learning in a Multiscale Nonlinear State-Space Model}

\author{
 Nayely V\'{e}lez-Cruz \\
  School of Complex Systems\\
Arizona State University\\
  Tempe, AZ 85282 \\
  \texttt{nvelezcr@asu.edu} \\
   \And
 Manfred D. Laubichler \\
  School of Complex Systems\\
  Arizona State University\\
  Tempe, AZ 85282 \\
  \texttt{Manfred.Laubichler@asu.edu} \\
}

\begin{document}
\maketitle
\begin{abstract}
The ubiquity of multiscale interactions in complex systems is well-recognized, with development and heredity serving as a prime example of how processes at different temporal scales influence one another. This work introduces a novel multiscale state-space model to explore the dynamic interplay between systems interacting across different time scales. We propose a Bayesian learning framework to estimate unknown states by learning unknown process noise covariances within this multiscale model. We develop a Particle Gibbs with Ancestor Sampling (PGAS) algorithm for inference. We demonstrate through simulations the efficacy of our approach.
\end{abstract}


\section{Introduction}
In many biological systems, the developmental processes of individuals play a crucial role in shaping the traits, characteristics, and growth patterns of subsequent generations. Throughout various stages of growth and maturation, organisms undergo significant changes that impact their overall fitness and reproductive success. These developmental stages, ranging from early cellular differentiation to reproductive maturity, each contribute uniquely to the organism's ability to survive and transmit biological information to offspring. Conversely, hereditary processes also influence the developmental stages of subsequent generations, creating a feedback loop where the heritable traits and adaptations of individuals as well as their health statuses such as disease resistance, metabolic efficiency, or physiological robustness can impact the developmental trajectories of future generations. This feedback loop between developmental processes and heredity continually shapes evolutionary trajectories, driving adaptation and resilience in populations over time. 

The interplay between development and heredity in driving evolutionary change is inherently a multiscale phenomenon. At one time scale, developmental processes operate within the lifespan of individual organisms, directly affecting their traits and fitness. At the intergenerational time scale, these developmental outcomes are passed down across generations through heredity. This necessitates the development of multiscale models to capture the interactions between processes occurring at different temporal scales, which can enable us to better understand how these interactions shape evolutionary trajectories. As well, statistical inference methods which can capture inherent uncertainties in both developmental and hereditary transmission can aid in the construction of predictive models of evolutionary dynamics. To this extent, Bayesian learning (\cite{gelman1995bayesian}) offers a powerful framework for developing models that can both accommodate the multiscale nature of evolutionary processes while also providing a systematic approach to inferring unknown states and parameters within these complex systems. 

Bayesian inference in state-space models has been widely used in a range of biological applications, from gene regulatory network inference (\cite{wu2011state}, \cite{noor2012inferring}) to ecology (\cite{pedersen2011estimation}, \cite{wang2007latent}). However, to the author's knowledge, there is no existing approach that integrates development and heredity into a unified modeling framework with Bayesian inference to learn unknown states and parameters at both time scales. In this work, we introduce a novel multiscale state-space model designed to capture the interaction between developmental and hereditary processes across different time scales with feedback between the scales. The model integrates fine-scale states that represent individual developmental stages and coarse-scale states that reflect hereditary traits across generations. We develop a Bayesian learning approach to estimate the unknown states by learning the process noise covariances. More specifically, we develop a Particle Gibbs with Ancestor Sampling (PGAS) algorithm, which combines particle filtering with ancestor sampling and Gibbs sampling for effective state and parameter estimation.

\section{Problem Formulation}

\subsection{Multiscale State-Space Model}
We develop a novel multiscale model characterized by latent fine-scale states \( \mathbf{x}^t_{d,k} \) and coarse-scale states \( \tilde{\mathbf{X}}_{d,t} \). Let $\textbf{x}^{t}_{d,k}$ denote the fine time scale state for individual $d \in \{1,...,D\}$ at fine time point $k$ in coarse time point $t$, where $\textbf{x}^{t}_{k,d} \in \mathbb{R}^{N_{\textbf{x}}}$ and $ 
\textbf{x}^{t}_{k,d} = \{x^{t}_{1,k,d},...,x^{t}_{N_{\textbf{x}},k,d} \}$. We denote the entire fine time scale trajectory by $\textbf{X}^{t}_{d,(K)} = \{\textbf{x}^{t}_{1,d}, \textbf{x}^{t}_{2,d},..,\textbf{x}^{t}_{K,d}\}$.  Let $\tilde{\textbf{X}}_{d,t} \in \mathbb{R}^{M_{\tilde{\textbf{X}}}}$ denote the coarse time scale state in generation $t$. The objective is to estimate the unknown states at both time scales by learning the unknown process noise covariances for both time scales. The model is described by the following transition and measurement equations:
\begin{align}
\label{mssmchap3}
\mathbf{x}^t_{d,k} &= f(\mathbf{x}^{t}_{d,k-1}, \tilde{\mathbf{X}}_{d,t-1}) + \mathbf{w}^t_{d,k-1}, \\
\tilde{\mathbf{X}}_{d,t} &= g(\tilde{\mathbf{X}}_{d,t-1},  \mathbf{X}^t_{d,(K)}) + \mathbf{W}_{d,t-1}, \\
\mathbf{y}^t_{d,k} &= \mathbf{x}^t_{d,k} + \mathbf{v}^t_{d,k}, \\
\tilde{\mathbf{Y}}_{d,t} &= \tilde{\mathbf{X}}_{d,t} + \mathbf{V}_{d,t},
\end{align}
where \( \mathbf{w}^t_{d,k} \sim \mathcal{N}(\mathbf{0}, \Sigma_f) \) and \( \mathbf{W}_{d,t} \sim \mathcal{N}(\mathbf{0}, \Sigma_{c,d}) \) are fine time scale process and coarse time scale process Gaussian noise terms, respectively. Note that \( \mathbf{v}^t_{d,k} \sim \mathcal{N}(0, \Sigma_v) \) and \( \mathbf{V}_{d,t} \sim \mathcal{N}(0, \Sigma_V) \) are Gaussian distributed noise terms for the fine and coarse time scales, respectively. For our problem, we assume that the fine time scale process noise covariance  $\Sigma_{f}$ is unknown and shared across all individuals $d$ for all coarse time steps $t$. For the coarse time scale, we assume that each individual $d$ has an associated unknown process noise covariance $\Sigma_{c,d}$. We assume that the measurement noise covariances for the fine and coarse time scales, $\Sigma_{v}$ and $\Sigma_{V}$ are known. 

To model the interactions between fine-scale and coarse-scale states, we define the transition functions as follows:
\begin{align}
f(\mathbf{x}^{t}_{d,k-1}, \tilde{\mathbf{X}}_{d,t-1}) &= \cos\left(\mathbf{A} \mathbf{x}^{t}_{d,k-1} + \tilde{\mathbf{X}}_{d,t-1}\right), \\
g(\tilde{\mathbf{X}}_{d,t-1},  \mathbf{X}^t_{d,(K)}) &= \sin\left(\mathbf{B} \left[\tilde{\mathbf{X}}_{d,t-1}\right]_{d=1}^{D} + \frac{1}{\sum_{k=1}^{K} w_k} \sum_{k=1}^{K} w_k \mathbf{x}^t_{d,k}\right),
\end{align}

\noindent
where \( \mathbf{A} \) is an \( n \times n \) adjacency matrix describing interactions between the dimensions of \( \mathbf{x}_{k,d}^{t} \) and  \( \mathbf{B} \) is a \( D \times D \) adjacency matrix describing interactions between the different individuals $d$ in the coarse time scale. In this context, the weights $w_{k}$ represent the weighted contributions of each fine-scale developmental time point to the coarse-scale state. These weights encapsulate the influence of various stages of development on the overall fitness and hereditary characteristics of the organism. 

\subsection{Bayesian Hierarchical Model}
The model is summarized by the following hierarchy in equations 7-12. 
\begin{align}
\textbf{x}^{t}_{d,k} &\vert \textbf{x}^{t}_{d,k-1}, \tilde{\textbf{X}}_{d,t-1}, \Sigma_{f} \sim \mathcal{F}(\textbf{x}^{t}_{d,k} \vert, \textbf{x}^{t}_{d,k-1}, \tilde{\textbf{X}}_{d,t-1}, \Sigma_{f})\\
\tilde{\textbf{X}}_{d,t} &\vert \tilde{\textbf{X}}_{d,t-1} , \textbf{X}_{d,(K)}^{t}, \Sigma_{c,d} \sim \mathcal{G}(\tilde{\textbf{X}}_{d,t} \vert \tilde{\textbf{X}}_{d,t-1}  , \textbf{X}_{d,(K)}^{t}, \Sigma_{c,d})\\
\textbf{y}^{t}_{d,k} &\vert \textbf{x}^{t}_{d,k}, \Sigma_{f}  \sim \mathcal{N}(\textbf{y}^{t}_{d,k} \vert \textbf{x}^{t}_{d,k} )\\
\textbf{Y}_{d,t} &\vert \tilde{\textbf{X}}_{d,t}, \Sigma_{\textbf{V},d}  \sim \mathcal{N}(\textbf{Y}_{d,t} \vert \tilde{\textbf{X}}_{d,t})\\
\Sigma_{f} & \sim \mbox{IW}(\Sigma_{f} \vert \boldsymbol{\Psi}_{f}, \nu_{f})\\
\Sigma_{c,d} & \sim \mbox{IW}(\Sigma_{c,d} \vert \boldsymbol{\Psi}_{c,d}, \nu_{c,d}),  d =1,\dots,D
\end{align}
where $\text{IW}$ denotes the inverse-Wishart distribution with hyperparameters $\boldsymbol{\Psi}_{\cdot}$ and $\nu_{\cdot}$, where $\boldsymbol{\Psi}_{\cdot}$ is the prior scale matrix and $\nu_{\cdot}$ is a scalar denoting the prior degrees of freedom. Note that the state distributions $\mathcal{F}$ and $\mathcal{G}$, and emission distributions are assumed to be Gaussian. The posterior density needed to estimate the unknown states $\textbf{x}_{d,k}^{t}$ and $\tilde{\textbf{X}}_{d,t}$ and the unknown process noise covariances $\Sigma_{f}$ and $\Sigma_{c,d}$ is given by
\begin{equation}
\begin{aligned}
&p(\{\textbf{x}_{d,k}^{t} \}_{d=1, k=1, t=1}^{D,K,T}, \{\tilde{\textbf{X}}_{d,t}\}_{d=1,t=1}^{D,T}, \Sigma_{f}, \{\Sigma_{c,d}\}_{d=1}^{D} \vert \{\textbf{y}_{d,k}^{t} \}_{d=1, k=1, t=1}^{D,K,T}, \{\textbf{Y}_{d,t} \}_{d=1, t=1}^{D,T}, \\
&\{\boldsymbol{\Psi}_{c,d} \}_{d=1}^{D}, \boldsymbol{\Psi}_{f}, \{\nu_{c,d}\}_{d=1}^{D}, \nu_{f})\\
& \propto \prod_{d=1}^{D} \prod_{t=1}^{T} p(\textbf{Y}_{d,t} \vert \tilde{\textbf{X}}_{d,t})p(\tilde{\textbf{X}}_{d,t} \vert \tilde{\textbf{X}}_{d,t-1}, \textbf{X}_{d,(K)}^{t}, \Sigma_{c,d})p(\Sigma_{c,d} \vert \boldsymbol{\Psi}_{c,d}, \nu_{c,d}) \\
&\prod_{k=1}^{K} p(\textbf{y}_{d,k}^{t} \vert \textbf{x}_{d,k}^{t}) p(\textbf{x}_{d,k}^{t} \vert \textbf{x}_{d,k-1}^{t}, \tilde{\textbf{X}}_{d,t-1},\Sigma_{f}) p(\Sigma_{f} \vert \boldsymbol{\Psi}_{f}, \nu_{f}) 
\end{aligned}
\end{equation}

\begin{longtable}{|c|p{0.6\textwidth}|}
\caption{Notation} \label{tab:notationchap4} \\

\hline
\textbf{Notation} & \textbf{Description} \\ \hline
\endfirsthead

\hline
\textbf{Notation} & \textbf{Description} \\ \hline
\endhead

\hline \multicolumn{2}{|r|}{{Continued on next page}} \\ \hline
\endfoot

\hline
\endlastfoot

$D$ & Number of individuals \\ \hline
$k$ & Index for the fine time scale \\ \hline
$t$ & Index for the coarse time scale \\ \hline
$N_{\mathbf{x}}$ & Dimensionality of the fine time scale state \\ \hline
$M_{\tilde{\mathbf{X}}}$ & Dimensionality of the coarse time scale state \\ \hline
$\mathbf{x}^t_{d,k}$ & Fine time scale state for individual $d$ at fine time point $k$ 
within coarse time point $t$ \\ \hline
$\mathbf{x}^t_{d,k} = \{x^t_{1,k,d},...,x^t_{N_{\mathbf{x}},k,d}\}$ & 
Vector of fine time scale state components \\ \hline
$\mathbf{X}^t_{d,(K)} = \{\mathbf{x}^t_{1,d}, \mathbf{x}^t_{2,d},..., \mathbf{x}^t_{K,d}\}$ & 
Entire fine time scale trajectory for individual $d$ \\ \hline
$\tilde{\mathbf{X}}_{d,t} \in \mathbb{R}^{M_{\tilde{\mathbf{X}}}}$ & 
Coarse time scale state for individual $d$ at generation $t$ \\ \hline
$\mathbf{y}^t_{d,k}$ & Fine time scale measurement \\ \hline
$\tilde{\mathbf{Y}}_{d,t}$ & Coarse time scale measurement \\ \hline
$f(\mathbf{x}^{t}_{d,k-1}, \tilde{\mathbf{X}}_{d,t-1})$ & Transition function for fine 
time scale states \\ \hline
$g(\tilde{\mathbf{X}}_{d,t-1}, \mathbf{X}^t_{d,(K)})$ & Transition function for coarse 
time scale states \\ \hline
$\mathbf{w}^t_{d,k}$ & Process noise for the fine time scale state \\ \hline
$\mathbf{W}_{d,t}$ & Process noise for the coarse time scale state \\ \hline
$\Sigma_{f}$ & Covariance matrix for fine time scale process noise \\ \hline
$\Sigma_{c,d}$ & Covariance matrix for coarse time scale process noise for individual $d$ \\ \hline
$\Sigma_{v}$ & Covariance matrix for fine time scale measurement noise \\ \hline
$\Sigma_{V}$ & Covariance matrix for coarse time scale measurement noise \\ \hline
$\mathbf{A}$ & Adjacency matrix describing interactions in the fine time scale state \\ \hline
$\mathbf{B}$ & Adjacency matrix describing interactions among individuals in the coarse 
time scale \\ \hline
$w_k$ & Weights representing contributions of fine-scale time points to the coarse-scale state \\ \hline
$\boldsymbol{\Psi}_{f}, \nu_{f}$ & IW hyperparameters for the prior of $\Sigma_{f}$ \\ \hline
$\boldsymbol{\Psi}_{c,d}, \nu_{c,d}$ & IW hyperparameters for the prior of $\Sigma_{c,d}$ \\ \hline

\end{longtable}
\section{Inference via Particle Gibbs with Ancestor Sampling (PGAS)}
To estimate the states at both time scales by learning the unknown process noise covariances, we employ Particle Gibbs with Ancestor Sampling (PGAS) (\cite{lindsten2014particle}). This approach utilizes Gibbs sampling and particle filtering with ancestor sampling to iteratively estimate the joint posterior distribution \( p(\theta, \mathbf{x}_{1:T} \mid \mathbf{y}_{1:T}) \), where \( \theta \) represents the unknown parameters and \( \mathbf{x}_{1:T} \) denotes the state trajectories. The algorithm alternates between sampling the states given the parameters and measurements, \( p(\mathbf{x}_{1:T} \mid \mathbf{y}_{1:T}, \theta) \), and sampling the parameters given the states and measurements, \( p(\theta \mid \mathbf{x}_{1:T}, \mathbf{y}_{1:T}) \). Within the particle filtering step, ancestor sampling is applied to explore different paths that might have led to the current state, maintaining particle diversity and reducing the risk of collapsing into a few high-weight paths. In the context of our model, the PGAS algorithm relies on the following decomposition (for brevity we omit the conditioning on the inverse-Wishart hyperparameters):
\begin{equation}
\begin{aligned}
    &p(\{\mathbf{x}_{d,k}^{t} \}_{d=1, k=1, t=1}^{D,K,T}, \{\tilde{\mathbf{X}}_{d,t}\}_{d=1,t=1}^{D,T}, \Sigma_{f}, \{\Sigma_{c,d}\}_{d=1}^{D} \mid \{\mathbf{y}_{d,k}^{t} \}_{d=1, k=1, t=1}^{D,K,T}, \{\mathbf{Y}_{d,t} \}_{d=1, t=1}^{D,T}) \nonumber\\
    &= p(\Sigma_{f}, \{\Sigma_{c,d}\}_{d=1}^{D} \mid \{\mathbf{x}_{d,k}^{t} \}_{d=1, k=1, t=1}^{D,K,T}, \{\tilde{\mathbf{X}}_{d,t}\}_{d=1,t=1}^{D,T}, \{\mathbf{y}_{d,k}^{t} \}_{d=1, k=1, t=1}^{D,K,T}, \{\mathbf{Y}_{d,t} \}_{d=1, t=1}^{D,T})\\
    &p_{\boldsymbol{\theta}}(\{\tilde{\mathbf{X}}_{d,t}\}_{d=1,t=1}^{D,T}, \{\mathbf{x}_{d,k}^{t} \}_{d=1, k=1, t=1}^{D,K,T} \mid \{\mathbf{y}_{d,k}^{t} \}_{d=1, k=1, t=1}^{D,K,T}, \{\mathbf{Y}_{d,t} \}_{d=1, t=1}^{D,T})\\
    &= p(\Sigma_{f}, \{\Sigma_{c,d}\}_{d=1}^{D} \mid \{\mathbf{x}_{d,k}^{t}\}_{d=1, k=1, t=1}^{D,K,T},  \{\tilde{\mathbf{X}}_{d,t}\}_{d=1,t=1}^{D,T}, \{\mathbf{y}_{d,k}^{t}\}_{d=1, k=1, t=1}^{D,K,T}, \{\mathbf{Y}_{d,t}\}_{d=1, t=1}^{D,T})\\
    &p_{\boldsymbol{\theta}}(\{\mathbf{x}_{d,k}^{t} \}_{d=1, k=1, t=1}^{D,K,T}, \{\tilde{\mathbf{X}}_{d,t}\}_{d=1,t=1}^{D,T} \mid \{\mathbf{y}_{d,k}^{t} \}_{d=1, k=1, t=1}^{D,K,T}, \{\mathbf{Y}_{d,t} \}_{d=1, t=1}^{D,T})\\
    &= p(\Sigma_{f} \mid \{\mathbf{x}_{d,k}^{t} \}_{d=1, k=1, t=1}^{D,K,T}, \{\mathbf{y}_{d,k}^{t} \}_{d=1, k=1, t=1}^{D,K,T})p(\{\Sigma_{c,d}\}_{d=1}^{D} \mid \{\tilde{\mathbf{X}}_{d,t}\}_{d=1,t=1}^{D,T}, \{\mathbf{Y}_{d,t} \}_{d=1, t=1}^{D,T})\\
    &p_{\Sigma_{f}}(\{\mathbf{x}_{d,k}^{t} \}_{d=1, k=1, t=1}^{D,K,T} \mid \{\mathbf{y}_{d,k}^{t} \}_{d=1, k=1, t=1}^{D,K,T}, \{\tilde{\mathbf{X}}_{d,t}\}_{d=1,t=1}^{D,T-1})\\
    &p_{\Sigma_{c,(D)}}(\{\tilde{\mathbf{X}}_{d,t}\}_{d=1,t=1}^{D,T} \mid \{\mathbf{x}_{d,k}^{t} \}_{d=1, k=1, t=1}^{D,K,T}, \{\mathbf{Y}_{d,t} \}_{d=1, t=1}^{D,T})
\end{aligned}
\end{equation}
Thus, the algorithm iteratively samples the state trajectories for the fine time scale given the measurements and coarse scale states up to time $T-1$,
\begin{equation}
p_{\Sigma_{f}}(\{\mathbf{x}_{d,k}^{t} \}_{d=1, k=1, t=1}^{D,K,T} \mid \{\mathbf{y}_{d,k}^{t} \}_{d=1, k=1, t=1}^{D,K,T}, \{\tilde{\mathbf{X}}_{d,t}\}_{d=1,t=1}^{D,T-1})
\end{equation}
as per the model specifications, the coarse time scale trajectories given the measurements and fine time scale trajectories, 
\begin{equation}
p_{\Sigma_{c,(D)}}(\{\tilde{\mathbf{X}}_{d,t}\}_{d=1,t=1}^{D,T} \mid \{\mathbf{x}_{d,k}^{t} \}_{d=1, k=1, t=1}^{D,K,T}, \{\mathbf{Y}_{d,t} \}_{d=1, t=1}^{D,T}),
\end{equation}
the fine time scale process covariance given the fine time scale states and measurements 
\begin{equation}
p(\Sigma_{f} \mid \{\mathbf{x}_{d,k}^{t} \}_{d=1, k=1, t=1}^{D,K,T}, \{\mathbf{y}_{d,k}^{t} \}_{d=1, k=1, t=1}^{D,K,T})
\end{equation} and finally the coarse time scale process covariance given the coarse time scale states and measurements 
\begin{equation}
p(\{\Sigma_{c,d}\}_{d=1}^{D} \mid \{\tilde{\mathbf{X}}_{d,t}\}_{d=1,t=1}^{D,T}, \{\mathbf{Y}_{d,t} \}_{d=1, t=1}^{D,T}),
\end{equation}
where $(D)$ denotes all $d=1,...,D$. Both the fine and coarse time scale trajectory sampling steps are performed within the particle filter with ancestor sampling (Algorithm 2). These steps are summarized in Algorithm 1. 

{
\renewcommand{\baselinestretch}{0.9} 
\small 
\begin{algorithm}
\caption{PGAS for Bayesian Learning of MsSSMs}
\begin{algorithmic}[1]
\State Initialize parameters $\Sigma_{f}(0)$, $\{\Sigma_{c,d}(0)\}_{d=1}^{D}$, and states $\{\mathbf{x}_{d,k}^{t}(0)\}_{d=1, k=1, t=1}^{D,K,T}$ and $\{\tilde{\mathbf{X}}_{d,t}(0)\}_{d=1,t=1}^{D,T}$.
\For{iteration $r = 0$ to $R$}
    \State Using Algorithm \ref{alg:pgaskernel}, draw 
    \begin{align*}
        \{\mathbf{x}_{d,k}^{t}(r+1)\}_{d=1, k=1, t=1}^{D,K,T} &\sim p_{\Sigma_{f}(r)}\left(\{\mathbf{x}_{d,k}^{t}(r)\}_{d=1, k=1, t=1}^{D,K,T} \mid \{\mathbf{y}_{d,k}^{t}\}_{d=1, k=1, t=1}^{D,K,T}, \{\tilde{\mathbf{X}}_{d,t}(r)\}_{d=1,t=1}^{D,T-1}\right) 
    \end{align*}
    and 
    \begin{align*}
        \{\tilde{\mathbf{X}}_{d,t}(r+1)\}_{d=1,t=1}^{D,T} &\sim p_{\Sigma_{c,d}(r)}\left(\{\tilde{\mathbf{X}}_{d,t}(r)\}_{d=1,t=1}^{D,T} \mid \{\mathbf{Y}_{d,t}\}_{d=1,t=1}^{D,T}, \{\mathbf{X}_{d,(K)}^{t}(r)\}_{d=1, t=1}^{D,T}\right)
    \end{align*}
    \State Draw 
    \begin{align*}
        \Sigma_{f}(r+1) &\sim p\left(\Sigma_{f} \mid \{\mathbf{x}_{d,k}^{t}(r+1)\}_{d=1, k=1, t=1}^{D,K,T}, \{\mathbf{y}_{d,k}^{t}\}_{d=1, k=1, t=1}^{D,K,T}, \right)
    \end{align*}
    \State Draw 
    \begin{align*}
        \{\Sigma_{c,d}(r+1)\}_{d=1}^{D} &\sim p\left(\{\Sigma_{c,d}\}_{d=1}^{D} \mid \{\tilde{\mathbf{X}}_{d,t}(r+1)\}_{d=1,t=1}^{D,T}, \{\mathbf{Y}_{d,t}\}_{d=1,t=1}^{D,T}\right)
    \end{align*}
\EndFor
\State \textbf{Output:} Estimates $\{\mathbf{x}_{d,k}^{t}\}_{d=1, k=1, t=1}^{D,K,T}$, $\{\tilde{\mathbf{X}}_{d,t}\}_{d=1,t=1}^{D,T}$.
\end{algorithmic}
\end{algorithm}
}
Note that since the inverse-Wishart is a conjugate prior to the Gaussian, the resulting posterior covariance will also be inverse-Wishart distributed and can thus be computed in closed form as follows. For the fine time scale we draw $\Sigma_{f} \sim \text{IW}(\mathbf{\Psi}_{f,K}, \nu_{f,K})$ where the hyperparameters are computed as follows:
\begin{equation}
S_{f} = \sum_{k=1}^{K} (\mathbf{x}_{d,k}^{t} - f(\mathbf{x}_{d,k-1}^{t}, \tilde{\textbf{X}}_{d,t-1})) (\mathbf{x}_{d,k}^{t} - f(\mathbf{x}_{d,k-1}^{t}, \tilde{\textbf{X}}_{d,t-1}))^{T},
\end{equation}
\begin{equation}
\mathbf{\Psi}_{f,K} = \mathbf{\Psi}_{f,0} + S_{f},
\end{equation}
\begin{equation}
\nu_{f,K} = \nu_{f,0} + K.
\end{equation}

Similarly, for the coarse scale, we draw $\Sigma_{c,d} \sim \text{IW}(\mathbf{\Psi}_{c,d,T}, \nu_{c,d,T})$
\begin{equation}
S_{c,d} = \sum_{t=1}^{T} (\tilde{\mathbf{X}}_{d,t} - g(\tilde{\mathbf{X}}_{d,t-1}, \textbf{X}_{d,(K)}^{t})) (\tilde{\mathbf{X}}_{d,t} - g(\tilde{\mathbf{X}}_{d,t-1},  \textbf{X}_{d,(K)}^{t}))^T,
\end{equation}
\begin{equation}
\mathbf{\Psi}_{c,d,T} = \mathbf{\Psi}_{c,d,0} + S_{c,d},
\end{equation}
\begin{equation}
\nu_{c,d,T} = \nu_{c,d,0} + T.
\end{equation}
where \(S_{f}\) and \(S_{c,d}\) are the sum of squares matrices for the fine and coarse time scales, respectively, \(\bar{\mathbf{x}}_{d}^{t}\) and \(\bar{\tilde{\mathbf{X}}}_{d}\) are the means of the fine and coarse scale estimates, \(\mathbf{\Psi}_{\text{fine}}\) and \(\mathbf{\Psi}_{\text{coarse}}\) are the updated scale matrices for the inverse-Wishart distribution for the fine and coarse time scales, and \(\nu_{\text{fine}}\) and \(\nu_{\text{coarse}}\) are the updated degrees of freedom for the inverse-Wishart distribution for the fine and coarse time scales.

\subsubsection{Particle Filter with Ancestor Sampling}
In PGAS, the particle filter is conditioned on a reference trajectory, which helps the algorithm explore the state space more effectively, leading to more accurate state and parameter estimates. At each time step, the reference trajectory is connected with one of the \( N-1 \) particles from the previous time step (the ancestors) by sampling an ancestor index based on the respective importance weights of the particles (\cite{berntorp2017particle}). This maintains a dependence on the reference trajectory while allowing for connections with particles between successive time steps and ensures that the algorithm can generate samples from the smoothing distribution $p_{\theta}(\textbf{x}_{1:T} \vert \textbf{y}_{1:T})$. 

For each coarse time step \( t \) and each individual \( d \), the algorithm processes each fine time step \( k \) as follows: Initially, the algorithm generates \( N-1 \) particles \( \{\hat{\mathbf{x}}_{t,d,1:k-1}^{(i)}\}_{i=1}^{N-1} \), each corresponding to a trajectory up to \( k-1 \), by sampling with replacement from the set of previous particles \( \{\mathbf{x}_{t,d,1:k-1}^{(i)}\}_{i=1}^{N} \) based on their normalized importance weights \( \{w_{t,d,k-1}^{(i)}\} \). Next, in the ancestor sampling step, the algorithm draws an index \( J \) with probability proportional to the product of its weight \( w_{t,d,k-1}^{(i)} \) and the transition density \( f_{\theta}(\mathbf{x}_{t,d,k}^{*} \mid \mathbf{x}_{t,d,k-1}^{(i)}, \tilde{\mathbf{X}}_{t-1,d}^{(i)}, \Sigma_{f}) \). The ancestor trajectory for the reference particle (the \( N \)-th particle) is then set to the trajectory of the particle at index \( J \), which is \( \mathbf{x}_{t,d,1:k-1}^{(J)} \). In the particle propagation step, \( N-1 \) new states \( \mathbf{x}_{t,d,k}^{(i)} \) are drawn using Equation (1) 
for $N-1$ particles. The \( N \)-th particle is set to the reference particle \( \mathbf{x}_{t,d,k}^{*} \). The complete trajectory for each particle is then updated by concatenating the ancestor trajectory up to \( k-1 \) with the newly sampled state \( \mathbf{x}_{t,d,k}^{(i)} \). Since we choose the proposal density to be the prior, the weights for the current fine time step are computed using the likelihood as \( w_{t,d,k}^{(i)} \propto p(\mathbf{y}_{t,d,k} \mid \mathbf{x}_{t,d,k}^{(i)}) \). The same process occurs for the coarse time scale, with the final step involving resampling the particles based on their final weights to set the new reference trajectory. These steps are summarized in Algorithm \ref{alg:pgaskernel}.

{
\renewcommand{\baselinestretch}{0.2} 
\small 

\begin{algorithm}
\caption{PGAS Kernel for msSSM}
\label{alg:pgaskernel}
\begin{algorithmic}[1]
     \item \textbf{Input:} Reference trajectory $\mathbf{x}_{d,t,1:K}^*$ and $\tilde{\mathbf{X}}_{d,1:T}^*$
        \State Draw $\textbf{x}_{t,d,1}^{(i)} \sim p(\textbf{x}_{t,d,1}^{(i)})$ for $i = 1, \ldots, N-1$, for $d = 1,...,D$, and for $t=1,...,T$.
        \State Draw $\tilde{\mathbf{X}}_{d,1}^{(i)} \sim p(\tilde{\textbf{X}}_{d,1})$ for $i = 1, \ldots, N-1$ and for for $d = 1,...,D$.
        \State Set $\textbf{x}_{t,d,1}^{(N)} = \textbf{x}_{t,d,1}^*$ and $\tilde{\mathbf{X}}_{d,1}^{(N)} = \tilde{\mathbf{X}}_{d,1}^{*}$ 
        \State Set fine time scale weights $w_{t,d,1}^{(i)} \propto p(\textbf{y}_{t,d,1} \mid \textbf{x}_{t,d,1}^{(i)})$ for $i = 1, \ldots, N$ and for $t=1,...,T$
        \State Set coarse time scale weights $w_{1,d}^{(i)} \propto p(\textbf{Y}_{1,d} \mid \tilde{\textbf{X}}_{1,d}^{(i)})$ for $i = 1, \ldots, N$
    \For{each coarse time step $t = 2$ to $T$}
        \For{each individual $d = 1$ to $D$}
            \For{each fine time step $k = 2$ to $K$}
                \State Generate $\{\hat{\mathbf{x}}_{t,d,1:k-1}^{(i)}\}_{i=1}^{N-1}$ by sampling $N-1$ times with replacement from $\{\mathbf{x}_{t,d,1:k-1}^{(i)}\}_{i=1}^{N}$ with probabilities proportional to the importance weights $\{w_{t,d,k-1}^{(j)}\}_{i=1}^{N}$
                \State Draw $J$ with $\Pr(J = i) \propto w_{t,d,k-1}^{i} f_{\theta}(\mathbf{x}_{t,d,k}^{*} | \mathbf{x}_{t,d,k-1}^{(i)}, \tilde{\mathbf{X}}_{t-1,d}^{(i)}, \Sigma_{f})$
                \State Set $\hat{\mathbf{x}}_{t,d,1:k-1}^{(N)} \leftarrow \mathbf{x}_{t,d,1:k-1}^{(J)}$
                \State Sample $\mathbf{x}_{t,d,k}^{(i)} \sim p(\mathbf{x}_{t,d,k} \mid \hat{\mathbf{x}}_{t,d,k-1}^{(i)}, \tilde{\mathbf{X}}_{t-1,d}^{(i)}, \Sigma_{f})$ for $i=1 \dots N-1$
                \State Set $\mathbf{x}_{t,d,k}^{(N)} \leftarrow \mathbf{x}_{t,d,k}^*$
                \State Set $\mathbf{x}_{t,d,1:k}^{(i)} = (\hat{\mathbf{x}}_{t,d,1:k-1}^{(i)}, \mathbf{x}_{t,d,k}^{(i)})$ for $i=1 \dots N$
                \State Compute weights $w_{t,d,k}^{(i)} \propto p(\mathbf{y}_{t,d,k} | \mathbf{x}_{t,d,k}^{(i)})$
            \EndFor
            \State Generate $\{\hat{\mathbf{X}}_{d,1:t-1}^{(i)}\}_{i=1}^{N-1}$ by sampling $N-1$ times with replacement from $\{\mathbf{X}_{d,1:t-1}^{(i)}\}_{i=1}^{N}$ with probabilities proportional to the importance weights $\{\tilde{w}_{t-1,d}^{(j)}\}_{j=1}^{N}$
            \State Draw $J_c$ with $\Pr(J_c = i) \propto \tilde{w}_{t-1,d}^{(i)} g_{\theta}(\tilde{\mathbf{X}}_{t,d}^{*} | \mathbf{X}_{t-1,d}^{i}, \{\textbf{x}_{t,d, 1:K}^{(i)} \}, \Sigma_{c,d})$
            \State Set $\hat{\tilde{\mathbf{X}}}_{1:t-1,d}^{(N)} \leftarrow \tilde{\mathbf{X}}_{1:t-1,d}^{(J_c)}$
            \State Sample $\tilde{\mathbf{X}}_{t,d}^{(i)} \sim p(\tilde{\mathbf{X}}_{t,d} | \hat{\tilde{\mathbf{X}}}_{t-1,d}^{(i)}, \{\textbf{x}_{t,d, 1:K}^{(i)}\}, \Sigma_{c,d})$ for $i=1 \dots N-1$
            \State Set $\tilde{\mathbf{X}}_{t,d}^{(N)} \leftarrow \tilde{\mathbf{X}}_{t,d}^*$
            \State Set $\tilde{\mathbf{X}}_{1:t,d}^{(i)} = (\hat{\tilde{\mathbf{X}}}_{1:t-1,d}^{(i)}, \tilde{\mathbf{X}}_{t,d}^{(i)})$ for $i=1 \dots N$
            \State Compute weights $\tilde{w}_{t,d}^{(i)} \propto p(\tilde{\mathbf{Y}}_{t,d} | \tilde{\mathbf{X}}_{t,d}^{(i)})$
        \EndFor
    \EndFor
    \For{each individual $d = 1$ to $D$}
        \For{each coarse time step $t = 1$ to $T$}
            \State Draw $J_f$ with $\Pr(J_f = i) \propto w_{t,d,K}^{(i)}$
            \State \Return $\mathbf{x}_{d,t,1:K}^{*} = \mathbf{x}_{d,t,1:K}^{(J_f)}$
        \EndFor
        \State Draw $J_c$ with $\Pr(J_c = i) \propto w_{T,d}^{(i)}$
        \State \Return $\tilde{\mathbf{X}}_{1:T,d}^{*} = \tilde{\mathbf{X}}_{1:T,d}^{(J_c)}$
    \EndFor
\end{algorithmic}
\end{algorithm}
}

\subsection{Results and Discussion}

\subsubsection{Simulation Settings}
In this section, we detail the simulation settings employed for the multiscale state-space model. For the fine time scale, the true process noise covariance matrix \( \Sigma_{f}^{\text{true}} \) was set to \( 0.2 \times \mathbf{I}_{N_{\textbf{x}}} \), where \( N_{\textbf{x}} = 3 \) denotes the dimensionality of the fine-scale states. The measurement noise covariance for the fine scale, \( \Sigma_v \), was similarly defined as \( 0.0003 \times \mathbf{I}_{N_{\textbf{x}}} \). For the coarse time scale, the true process noise covariance matrices \( \Sigma_{c,d}^{\text{true}} \) are defined separately for each individual \( d \) as follows:
\[
\Sigma_{c,1}^{\text{true}} = 0.3 \times \mathbf{I}_{M_{\tilde{\textbf{X}}}}, \quad \Sigma_{c,2}^{\text{true}} = 0.5 \times \mathbf{I}_{M_{\tilde{\textbf{X}}}}, \quad \Sigma_{c,3}^{\text{true}} = 0.7 \times \mathbf{I}_{M_{\tilde{\textbf{X}}}}, \quad \Sigma_{c,4}^{\text{true}} = 0.2 \times \mathbf{I}_{M_{\tilde{\textbf{X}}}}
\]
where \( M_{\tilde{\textbf{X}}} = 3 \) represents the dimensionality of the coarse-scale states. Similarly, the measurement noise covariance for each individual \( d \) in the coarse scale, \( \Sigma_{V,d} \), are defined as follows:
\[
\Sigma_{V,1} = 0.0003 \times \mathbf{I}_{M_{\tilde{\textbf{X}}}}, \quad \Sigma_{V,2} = 0.0005 \times \mathbf{I}_{M_{\tilde{\textbf{X}}}}, \quad \Sigma_{V,3} = 0.0007 \times \mathbf{I}_{M_{\tilde{\textbf{X}}}}, \quad \Sigma_{V,4} = 0.0009 \times \mathbf{I}_{M_{\tilde{\textbf{X}}}}
\] 
The prior distributions for the process noise covariances were chosen to be inverse-Wishart distributions, as they are conjugate priors to the Gaussian. The initial fine scale process noise covariance is \( \Sigma_{f}(0) \sim \text{IW}(\mathbf{V}_{f,0}, \nu_{f,0}) \), where the scale matrix \( \mathbf{V}_{0,f} \) was set to \( 0.1 \times \mathbf{I}_{N_{\textbf{x}}} \) and the degrees of freedom \( \nu_{0,f} \) were set to \( N_{\textbf{x}} + 1 \). Similarly, for the coarse scale for each $d$, the initial process noise covariance matrices are \( \Sigma_{c,d}(0) \sim  \text{IW}(\mathbf{V}_{c,d,0}, \nu_{c,d,0}) \), with the scale matrix \( \mathbf{V}_{0,c,d} \) set to \( 0.1 \times \mathbf{I}_{M_{\tilde{\textbf{X}}}} \) and the degrees of freedom \( \nu_{0,c,d} \) set to \( M_{\tilde{\textbf{X}}} + D + 1 \). The initial states of the particles for both the fine and coarse scales are Gaussian as:
\[
\{\mathbf{x}^{(i),1}_{d,1}\}_{i=1}^{N_{\text{part}}} \sim \mathcal{N}(\mathbf{x}^{1}_{d,1}, \Sigma_{f}^{1}),
\]
where \(\{\mathbf{x}^{(i),1}_{d,1}\}_{i=1}^{N_{\text{part}}}\) denotes the set of particles \( i = 1,...,N_{\text{part}} \) for individual \( d \) at fine time point $k=1$ and coarse time point $t=1$. Similarly, the initial coarse-scale states for each individual \( d \) were sampled as follows:
\[
\{\tilde{\mathbf{X}}^{(i)}_{d,1}\}_{i=1}^{N_{\text{part}}} \sim \mathcal{N}(\tilde{\mathbf{X}}_{d,1}, \Sigma_{c,d}^{1}).
\]
Lastly, the adjacency matrices \( \mathbf{A} \) and \( \mathbf{B} \), which govern the interactions within the fine and coarse scales, respectively, were initialized randomly. To perform the inference using the PGAS algorithm, a total of 800 particles were employed for the particle filtering step. The PGAS algorithm was executed for $R = 10,000$ iterations, with the first 10\% being discarded as burn-in.

\subsubsection{Results and Discussion}
\begin{figure}[h!]
    \centering
    \begin{subfigure}[b]{0.45\textwidth}
        \centering
        \includegraphics[width=\textwidth]{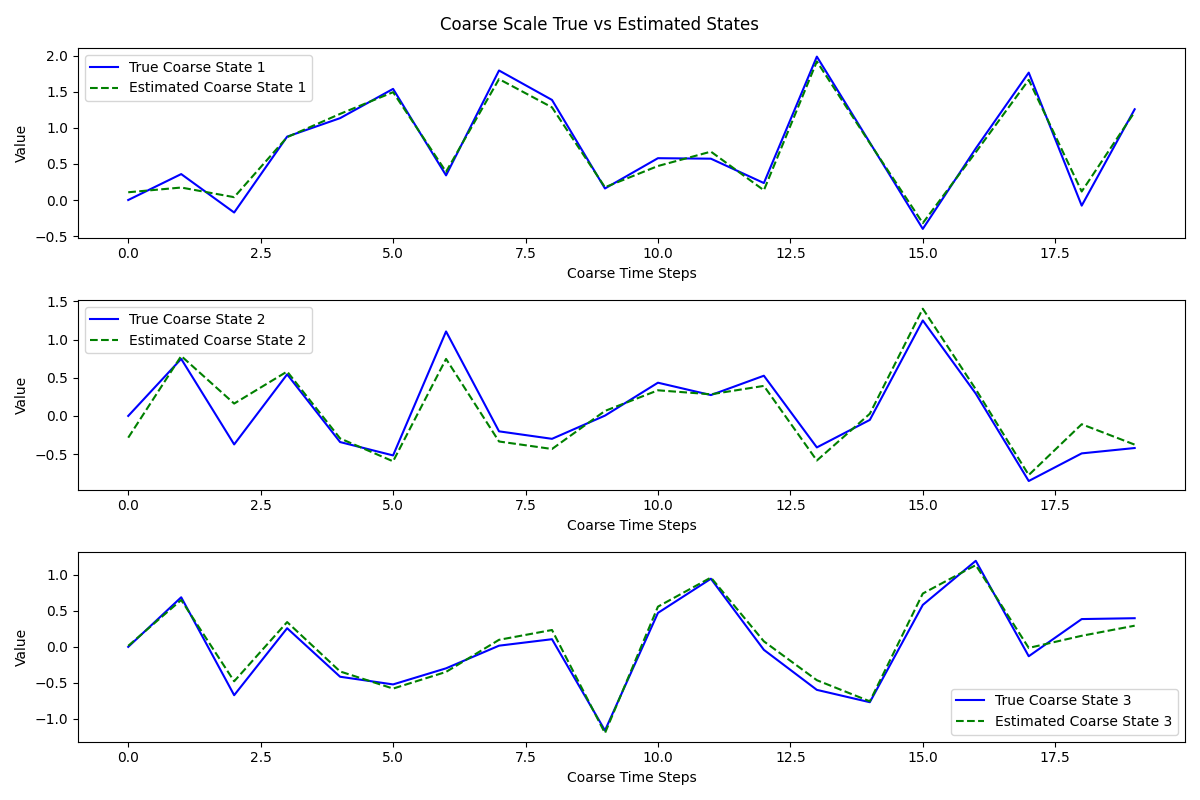}
        \caption{Individual $d=0$}
        \label{fig:d0coarse}
    \end{subfigure}
    \hfill
    \begin{subfigure}[b]{0.45\textwidth}
        \centering
        \includegraphics[width=\textwidth]{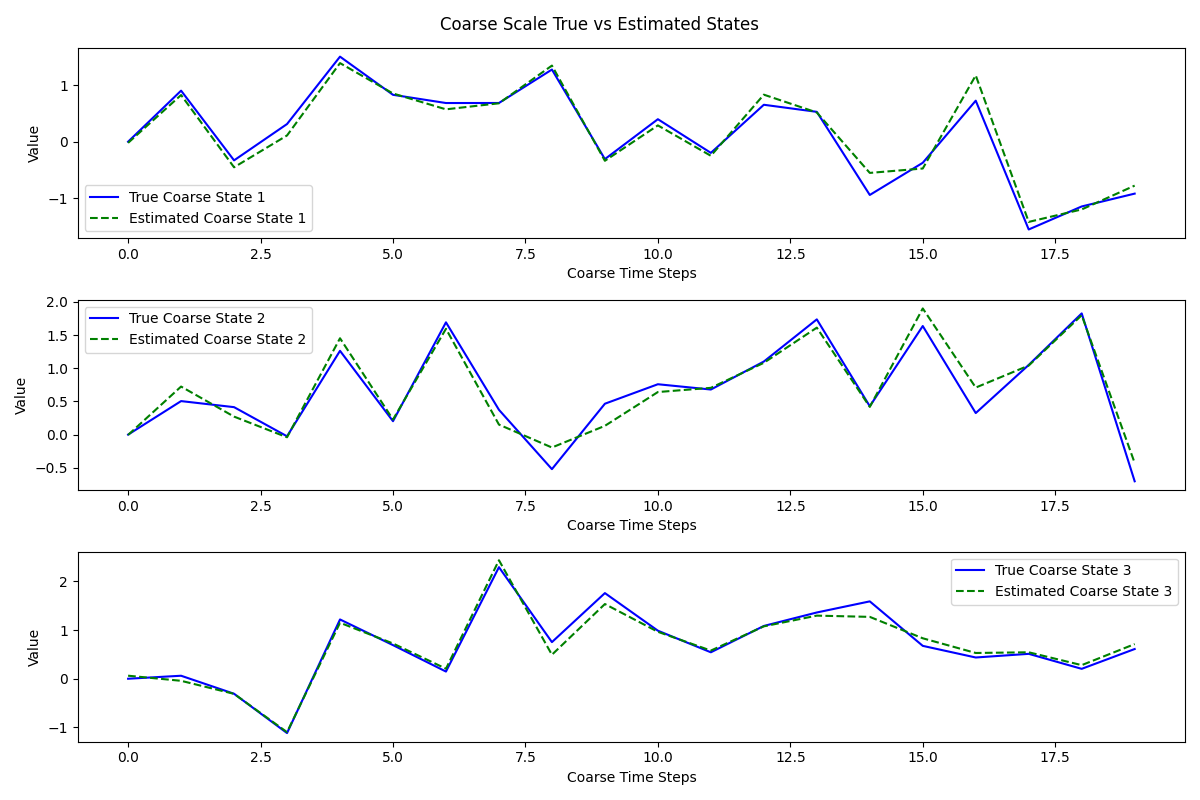}
        \caption{Individual $d=1$}
        \label{fig:d1coarse}
    \end{subfigure}

    \vspace{0.3cm} 

    \begin{subfigure}[b]{0.45\textwidth}
        \centering
        \includegraphics[width=\textwidth]{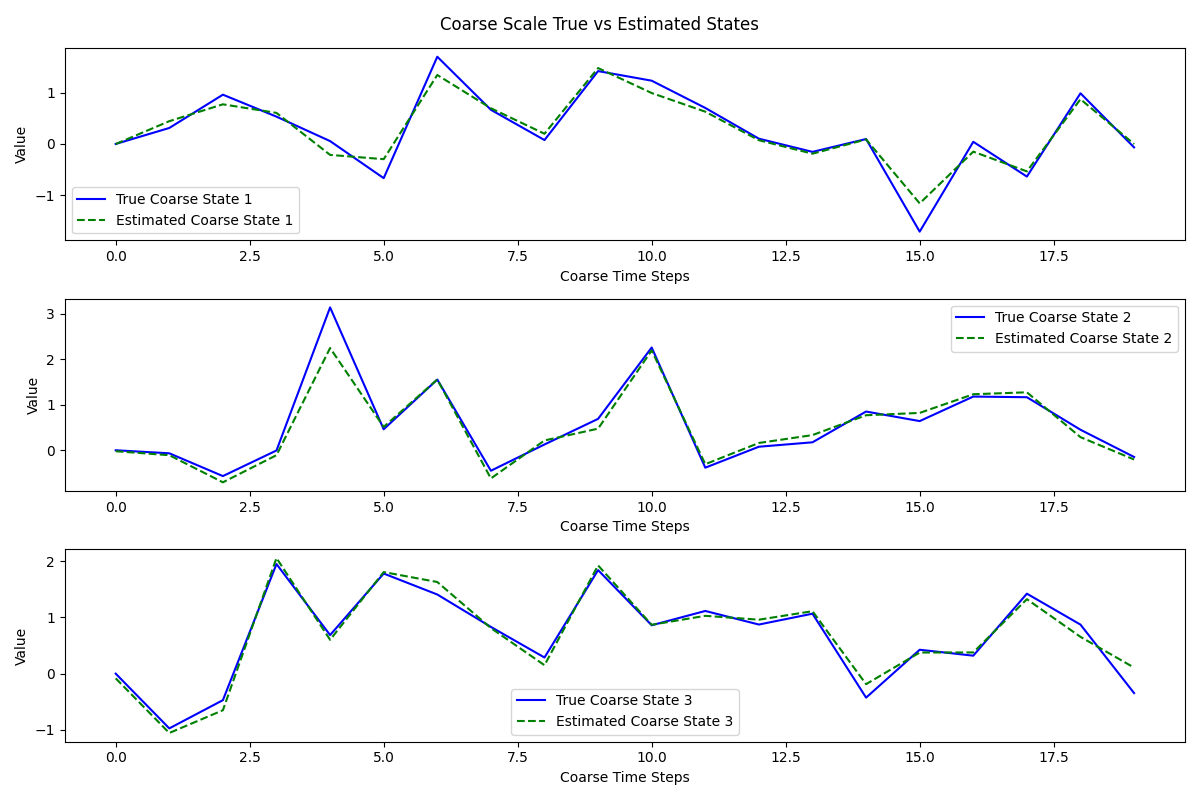}
        \caption{Individual $d=2$}
        \label{fig:d2coarse}
    \end{subfigure}
    \hfill
    \begin{subfigure}[b]{0.45\textwidth}
        \centering
        \includegraphics[width=\textwidth]{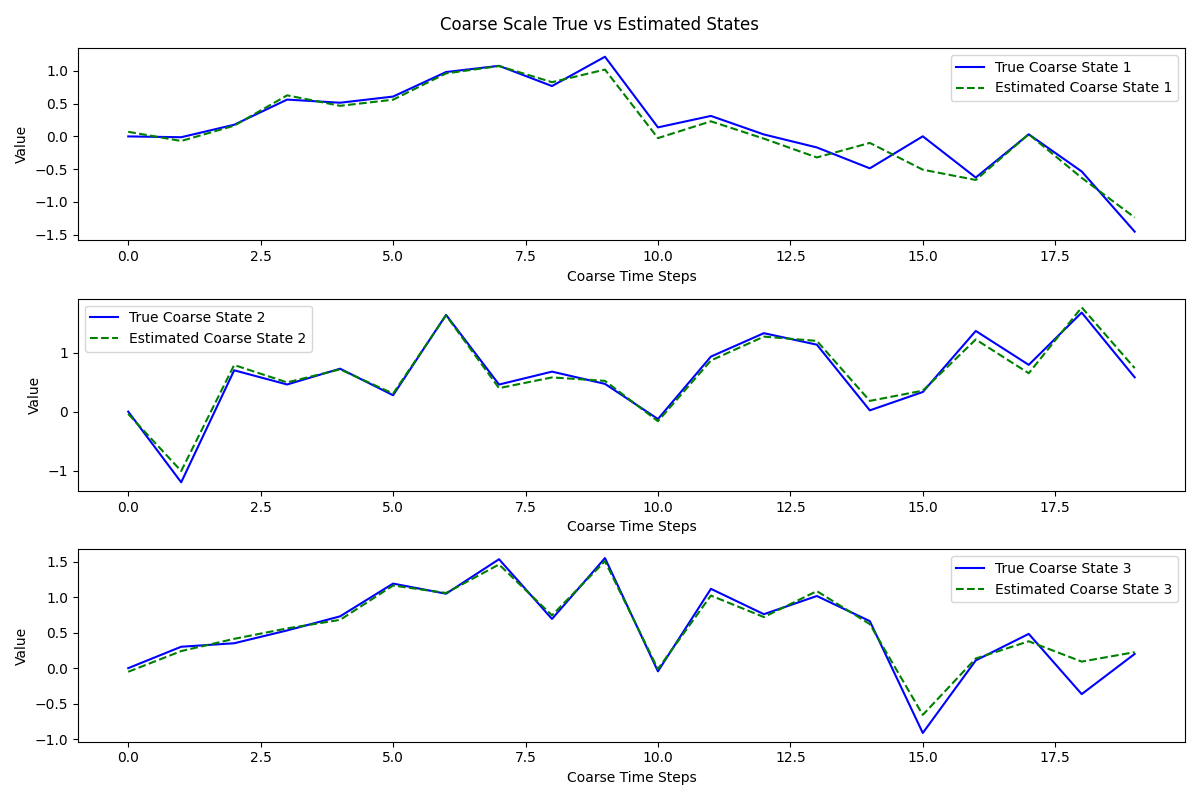}
        \caption{Individual $d=3$}
        \label{fig:d3coarse}
    \end{subfigure}

    \caption{True vs. estimated coarse time scale trajectories for individuals $d=0, d=1, d=2$, and $d=3$.}
    \label{fig:coarse_trajectories}
\end{figure}

\begin{figure}[h!]
    \centering
    \includegraphics[width=\textwidth]{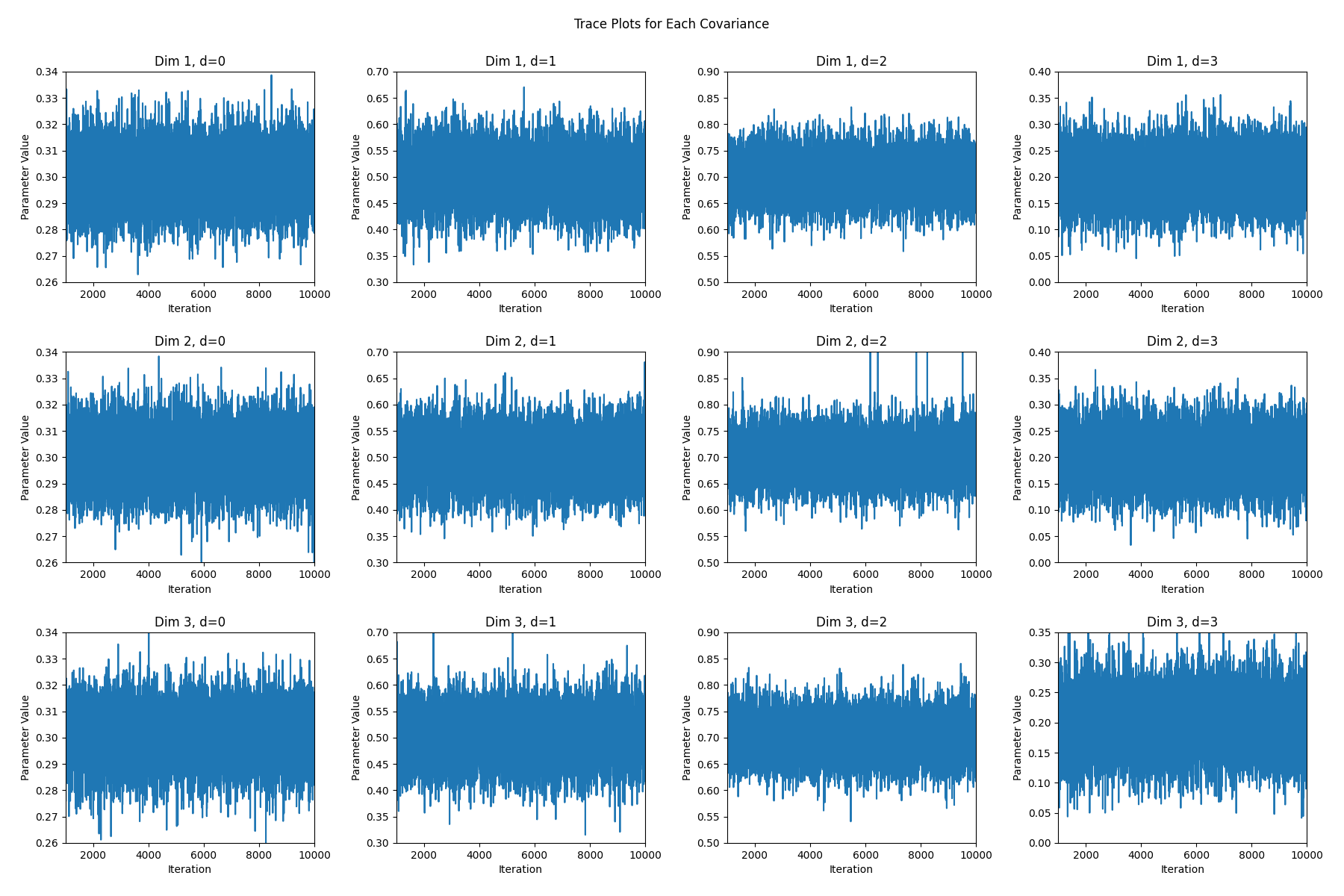}
  \caption{Trace plots for each coarse scale dimension and each individual $d$. The true process noise covariance matrices are $\Sigma_{c,1}^{\text{true}} = 0.3 \times \mathbf{I}_{M_{\tilde{\textbf{X}}}}, \; \Sigma_{c,2}^{\text{true}} = 0.5 \times \mathbf{I}_{M_{\tilde{\textbf{X}}}}, \; \Sigma_{c,3}^{\text{true}} = 0.7 \times \mathbf{I}_{M_{\tilde{\textbf{X}}}}, \; \Sigma_{c,4}^{\text{true}} = 0.2 \times \mathbf{I}_{M_{\tilde{\textbf{X}}}}$.}
    \label{fig:coarsetrace}
\end{figure}

\begin{figure}[h!]
    \centering
    \includegraphics[width=0.6\textwidth]{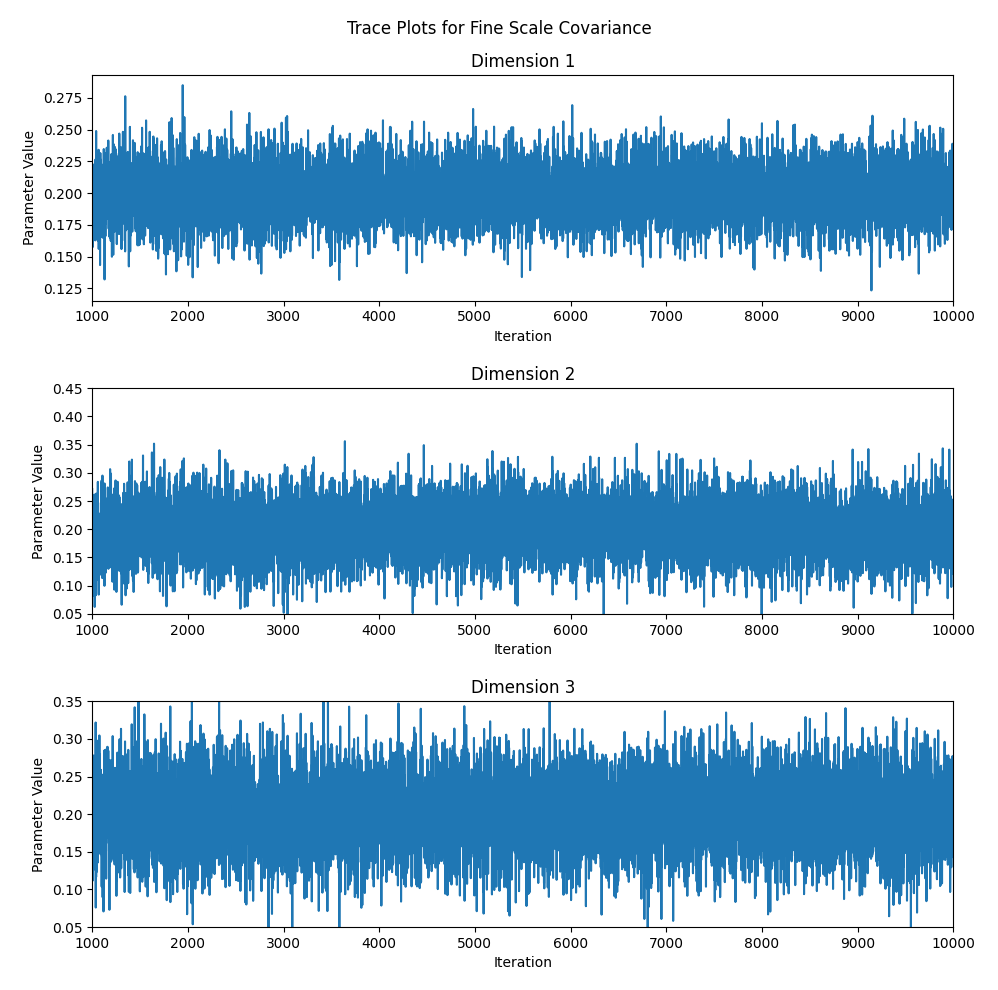}
   \caption{Trace plots for each fine scale dimension. The true covariance is $0.2 \cdot I_{N_{\textbf{x}}}$.}
    \label{fig:tracefine}
\end{figure}

The results demonstrate the efficacy of our algorithm in estimating both the fine and coarse scale trajectories through learning the unknown process noise covariances at each scale. Figures \ref{fig:d0coarse}-\ref{fig:d3coarse} show the true versus estimated states for each individual \(d\) in the coarse time scale. The root mean square error (RMSE) averaged across all coarse time scale points \(t\) for each individual and for each dimension \(m \in M_{\tilde{\textbf{X}}}\) is shown in Table \ref{tab:coarsecaleRMSE}. Similarly, the RMSE averaged across all fine time scale points \(k\) within each coarse time step \(t\) for each \(d\) is shown in Tables \ref{tab:finescaleRMSEd0}-\ref{tab:finescaleRMSE_d3}. The true versus estimated trajectories are shown in Figures \ref{fig:d0fine}-\ref{fig:d3fine} for coarse time point \(t=11\). The trace plots for each dimension and each $d$ for the coarse scale are in Figure \ref{fig:coarsetrace}, and for the fine time scale are in Figure \ref{fig:tracefine}.

Overall, the trace plots for each coarse scale dimension and each individual $d$ demonstrate good convergence, but with higher variability in some cases (e.g. dimension 2, $d=2$). Similarly, the trace plots for each fine scale dimension also demonstrate good convergence, but with higher variability for dimensions 2 and 3. The results suggest that the model is successfully learning the process noise covariances. The low RMSE values across most individuals and dimensions indicate that the algorithm is effective in capturing the latent states across both scales of the system while learning the noise with a high degree of precision. However, slight variations in RMSE across different dimensions and individuals suggest that the model's performance could be further optimized, particularly for specific cases where higher errors were observed. In general, the results demonstrate that our approach is effective in modeling and estimating multiscale complex systems with feedback between each scale. Future work could focus on refining the algorithm and further reducing the process noise covariances.

\begin{figure}[h!]
    \centering
    \begin{subfigure}[b]{0.45\textwidth}
        \centering
        \includegraphics[width=\textwidth]{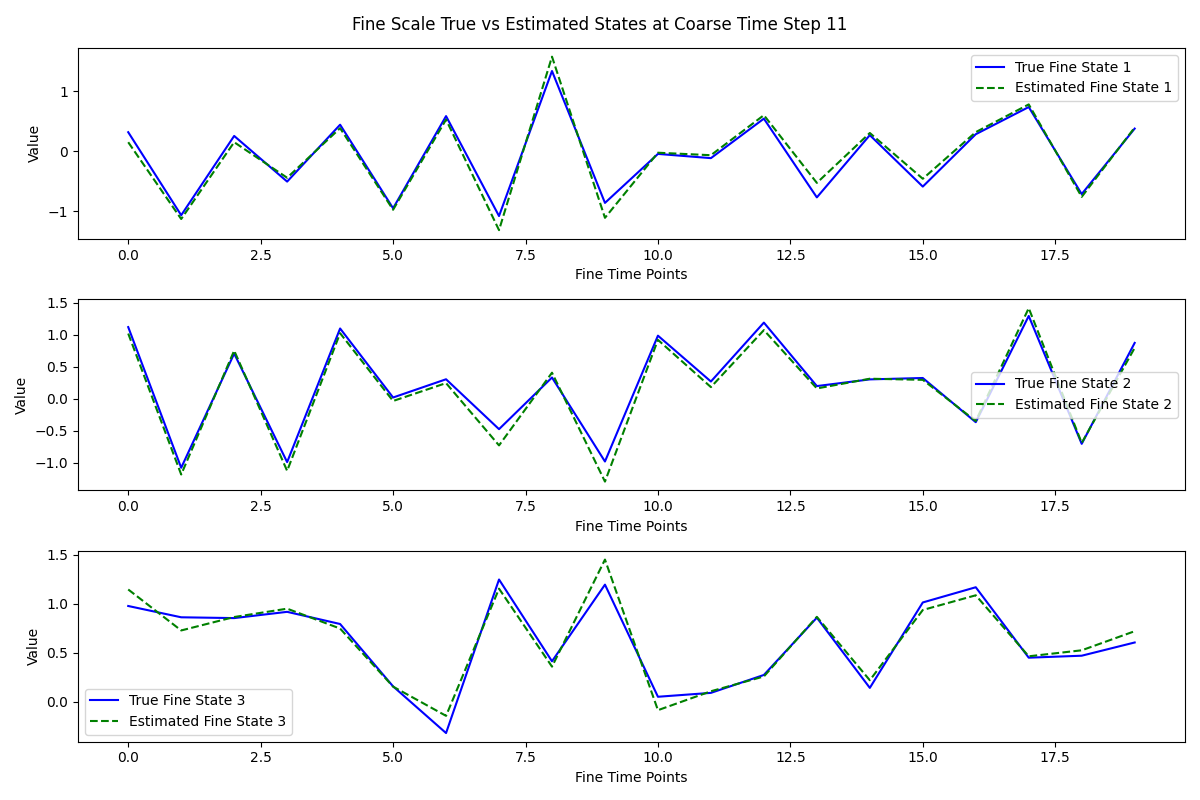}
        \caption{Individual $d=0$}
        \label{fig:d0fine}
    \end{subfigure}
    \hfill
    \begin{subfigure}[b]{0.45\textwidth}
        \centering
        \includegraphics[width=\textwidth]{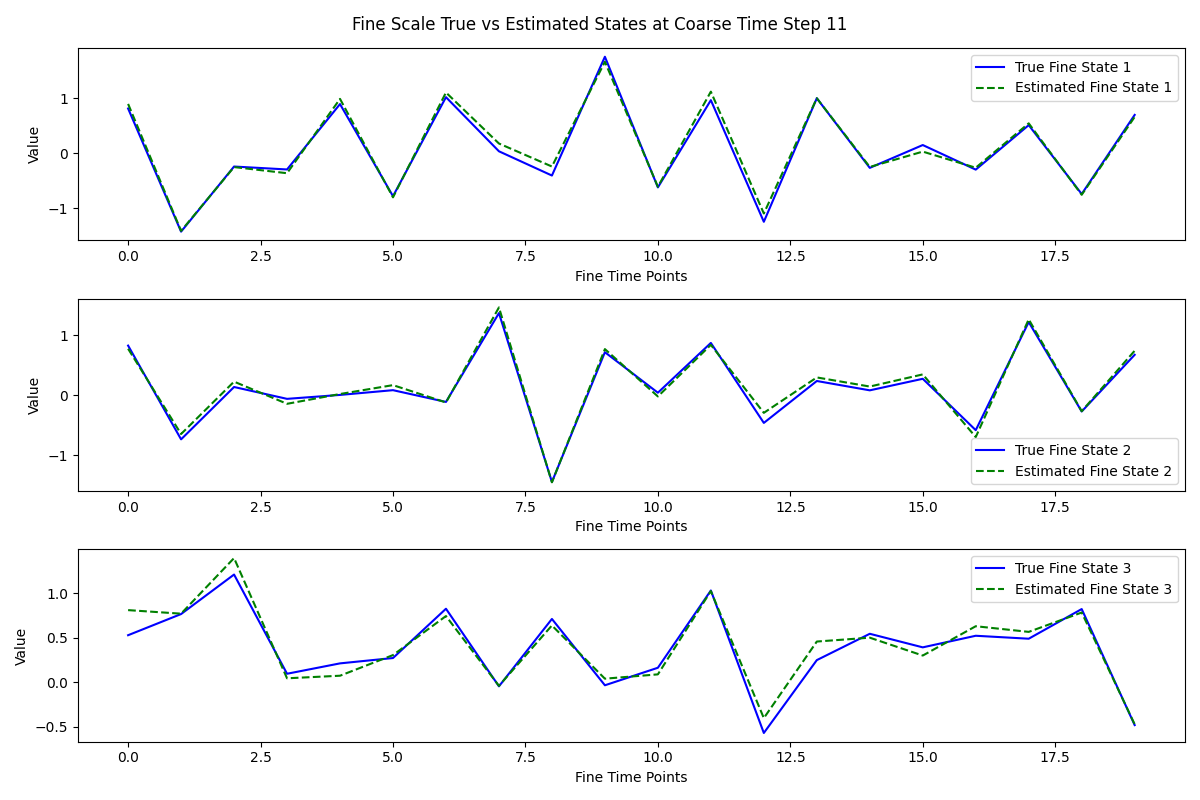}
        \caption{Individual $d=1$}
        \label{fig:d1fine}
    \end{subfigure}

    \vspace{0.3cm} 

    \begin{subfigure}[b]{0.45\textwidth}
        \centering
        \includegraphics[width=\textwidth]{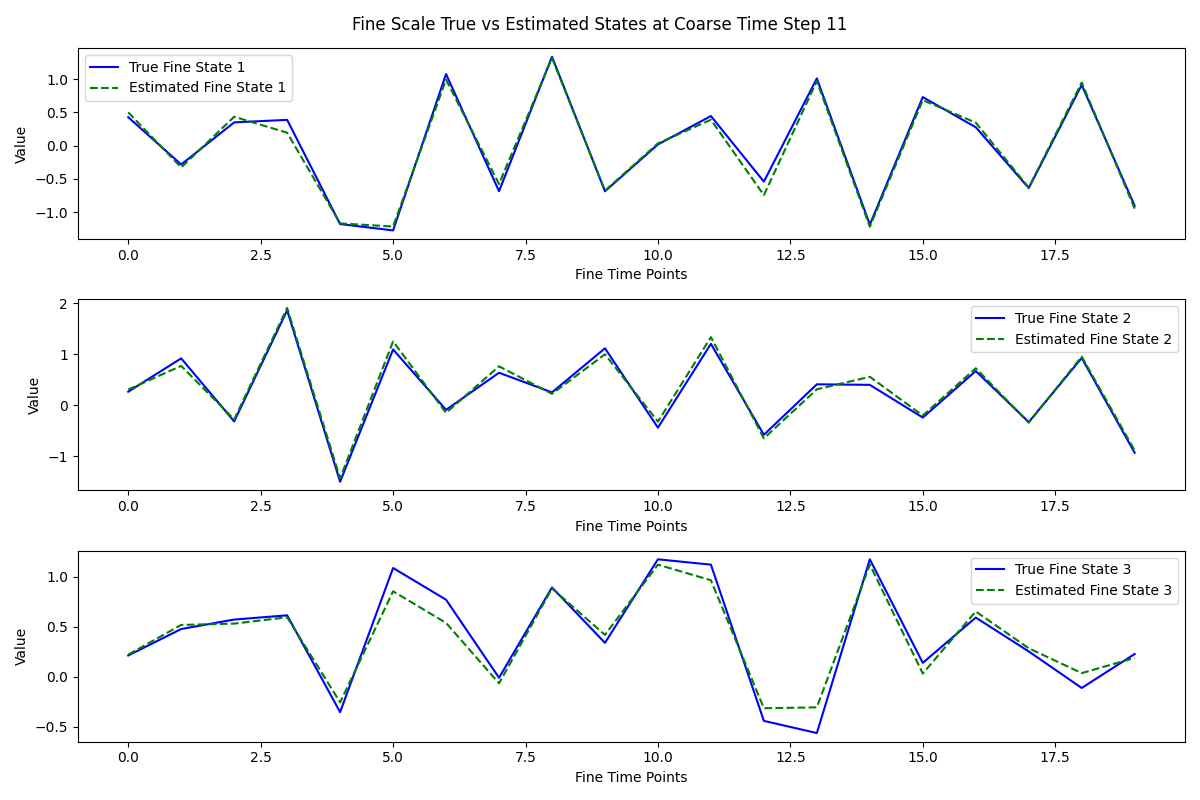}
        \caption{Individual $d=2$}
        \label{fig:d2fine}
    \end{subfigure}
    \hfill
    \begin{subfigure}[b]{0.45\textwidth}
        \centering
        \includegraphics[width=\textwidth]{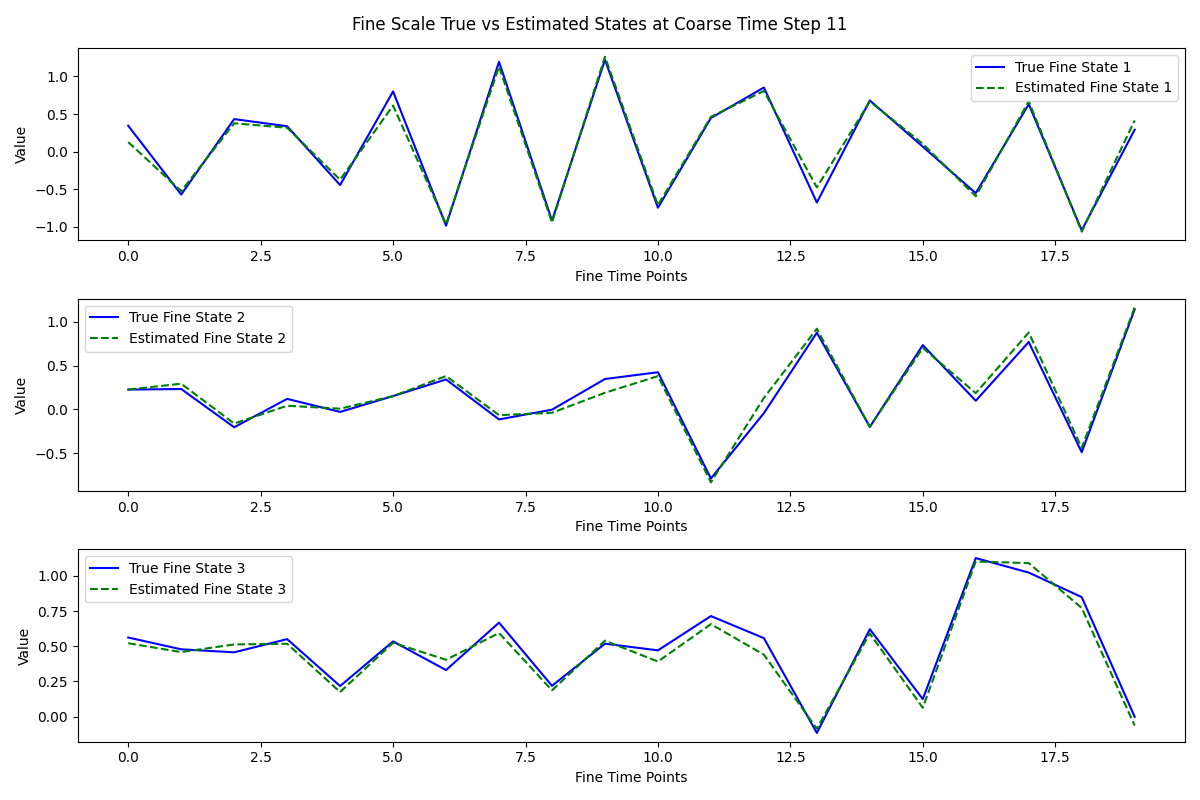}
        \caption{Individual $d=3$}
        \label{fig:d3fine}
    \end{subfigure}

    \caption{True vs. estimated fine time scale trajectories at coarse time step $t=11$ for individuals $d=0, d=1, d=2$, and $d=3$.}
    \label{fig:fine_trajectories}
\end{figure}

\begin{table}[h!]
\centering
\begin{tabular}{|c|c|c|c|}
\hline
Series & dim\_1 & dim\_2 & dim\_3 \\ \hline
0 & 0.107 & 0.123 & 0.102 \\ \hline
1 & 0.090 & 0.174 & 0.194 \\ \hline
2 & 0.095 & 0.107 & 0.126 \\ \hline
3 & 0.103 & 0.098 & 0.106 \\ \hline
\end{tabular}
\caption{Root mean square error averaged across all coarse time points $t$ for each series and each dimension $m \in M_{\tilde{\textbf{X}}}$.}
\label{tab:coarsecaleRMSE}
\end{table}

\begin{table}[h!]
\centering
\begin{subtable}[b]{0.45\textwidth}
    \centering
    \begin{tabular}{|c|c|c|c|}
    \hline
    Time Step & dim\_1 & dim\_2 & dim\_3 \\ \hline
    0  & 0.170 & 0.130 & 0.154 \\ \hline
    1  & 0.069 & 0.080 & 0.088 \\ \hline
    2  & 0.092 & 0.073 & 0.062 \\ \hline
    3  & 0.120 & 0.114 & 0.051 \\ \hline
    4  & 0.712 & 0.376 & 0.660 \\ \hline
    5  & 0.097 & 0.119 & 0.112 \\ \hline
    6  & 0.060 & 0.061 & 0.078 \\ \hline
    7  & 0.078 & 0.095 & 0.061 \\ \hline
    8  & 0.073 & 0.074 & 0.073 \\ \hline
    9  & 0.078 & 0.082 & 0.068 \\ \hline
    10 & 0.058 & 0.066 & 0.078 \\ \hline
    11 & 0.062 & 0.087 & 0.104 \\ \hline
    12 & 0.122 & 0.112 & 0.069 \\ \hline
    13 & 0.081 & 0.071 & 0.089 \\ \hline
    14 & 0.055 & 0.072 & 0.072 \\ \hline
    15 & 0.094 & 0.097 & 0.088 \\ \hline
    16 & 0.081 & 0.075 & 0.066 \\ \hline
    17 & 0.132 & 0.317 & 0.123 \\ \hline
    18 & 0.315 & 0.240 & 0.151 \\ \hline
    19 & 0.127 & 0.128 & 0.090 \\ \hline
    \end{tabular}
    \caption{Individual $d=0$}
    \label{tab:finescaleRMSEd0}
\end{subtable}
\hfill
\begin{subtable}[b]{0.45\textwidth}
    \centering
    \begin{tabular}{|c|c|c|c|}
    \hline
     Time Step & dim\_1 & dim\_2 & dim\_3 \\ \hline
    0  & 0.107 & 0.120 & 0.104 \\ \hline
    1  & 0.093 & 0.092 & 0.078 \\ \hline
    2  & 0.077 & 0.056 & 0.053 \\ \hline
    3  & 0.087 & 0.096 & 0.119 \\ \hline
    4  & 0.176 & 0.179 & 0.140 \\ \hline
    5  & 0.062 & 0.133 & 0.061 \\ \hline
    6  & 0.094 & 0.071 & 0.089 \\ \hline
    7  & 0.079 & 0.095 & 0.087 \\ \hline
    8  & 0.065 & 0.113 & 0.082 \\ \hline
    9  & 0.094 & 0.063 & 0.057 \\ \hline
    10 & 0.051 & 0.076 & 0.062 \\ \hline
    11 & 0.127 & 0.057 & 0.052 \\ \hline
    12 & 0.081 & 0.086 & 0.069 \\ \hline
    13 & 0.441 & 0.200 & 0.353 \\ \hline
    14 & 0.071 & 0.066 & 0.077 \\ \hline
    15 & 0.154 & 0.066 & 0.091 \\ \hline
    16 & 0.070 & 0.127 & 0.097 \\ \hline
    17 & 0.087 & 0.079 & 0.056 \\ \hline
    18 & 0.055 & 0.046 & 0.043 \\ \hline
    19 & 0.077 & 0.084 & 0.084 \\ \hline
    \end{tabular}
    \caption{Individual $d=1$}
    \label{tab:finescaleRMSE_d1}
\end{subtable}

\vspace{0.5cm} 

\begin{subtable}[b]{0.45\textwidth}
    \centering
    \begin{tabular}{|c|c|c|c|}
    \hline
    Time Step & dim\_1 & dim\_2 & dim\_3 \\ \hline
    0  & 0.112 & 0.097 & 0.075 \\ \hline
    1  & 0.119 & 0.062 & 0.077 \\ \hline
    2  & 0.066 & 0.066 & 0.090 \\ \hline
    3  & 0.077 & 0.077 & 0.060 \\ \hline
    4  & 0.786 & 0.598 & 0.355 \\ \hline
    5  & 0.089 & 0.061 & 0.064 \\ \hline
    6  & 0.064 & 0.105 & 0.075 \\ \hline
    7  & 0.083 & 0.104 & 0.089 \\ \hline
    8  & 0.054 & 0.072 & 0.066 \\ \hline
    9  & 0.076 & 0.084 & 0.089 \\ \hline
    10 & 0.105 & 0.044 & 0.060 \\ \hline
    11 & 0.053 & 0.058 & 0.062 \\ \hline
    12 & 0.062 & 0.062 & 0.052 \\ \hline
    13 & 0.056 & 0.056 & 0.076 \\ \hline
    14 & 0.068 & 0.075 & 0.070 \\ \hline
    15 & 0.127 & 0.123 & 0.114 \\ \hline
    16 & 0.121 & 0.081 & 0.093 \\ \hline
    17 & 0.161 & 0.067 & 0.116 \\ \hline
    18 & 0.066 & 0.089 & 0.069 \\ \hline
    19 & 0.085 & 0.086 & 0.126 \\ \hline
    \end{tabular}
    \caption{Individual $d=2$}
    \label{tab:finescaleRMSE_d2}
\end{subtable}
\hfill
\begin{subtable}[b]{0.45\textwidth}
    \centering
    \begin{tabular}{|c|c|c|c|}
    \hline
    Time Step & dim\_1 & dim\_2 & dim\_3 \\ \hline
    0  & 0.116 & 0.149 & 0.095 \\ \hline
    1  & 0.115 & 0.067 & 0.107 \\ \hline
    2  & 0.100 & 0.086 & 0.060 \\ \hline
    3  & 0.070 & 0.083 & 0.075 \\ \hline
    4  & 0.592 & 0.623 & 0.969 \\ \hline
    5  & 0.067 & 0.059 & 0.080 \\ \hline
    6  & 0.044 & 0.065 & 0.074 \\ \hline
    7  & 0.139 & 0.074 & 0.060 \\ \hline
    8  & 0.088 & 0.080 & 0.091 \\ \hline
    9  & 0.077 & 0.074 & 0.079 \\ \hline
    10 & 0.090 & 0.075 & 0.080 \\ \hline
    11 & 0.062 & 0.045 & 0.060 \\ \hline
    12 & 0.046 & 0.059 & 0.083 \\ \hline
    13 & 0.095 & 0.074 & 0.065 \\ \hline
    14 & 0.069 & 0.053 & 0.065 \\ \hline
    15 & 0.065 & 0.097 & 0.066 \\ \hline
    16 & 0.060 & 0.051 & 0.054 \\ \hline
    17 & 0.082 & 0.140 & 0.083 \\ \hline
    18 & 0.140 & 0.092 & 0.090 \\ \hline
    19 & 0.069 & 0.078 & 0.065 \\ \hline
    \end{tabular}
    \caption{Individual $d=3$}
    \label{tab:finescaleRMSE_d3}
\end{subtable}

\caption{Root mean square error (RMSE) averaged across all fine time scale points $k$, within each coarse time step $t$ for individuals $d=0$, $d=1$, $d=2$, and $d=3$ across each dimension $n \in N_{\textbf{x}}$.}
\label{tab:finescaleRMSE_all}
\end{table}

\bibliographystyle{unsrt}

\section{Additional Figures}

\begin{figure}[h!]
    \centering
    \begin{subfigure}[b]{0.45\textwidth}
        \centering
        \includegraphics[width=\textwidth]{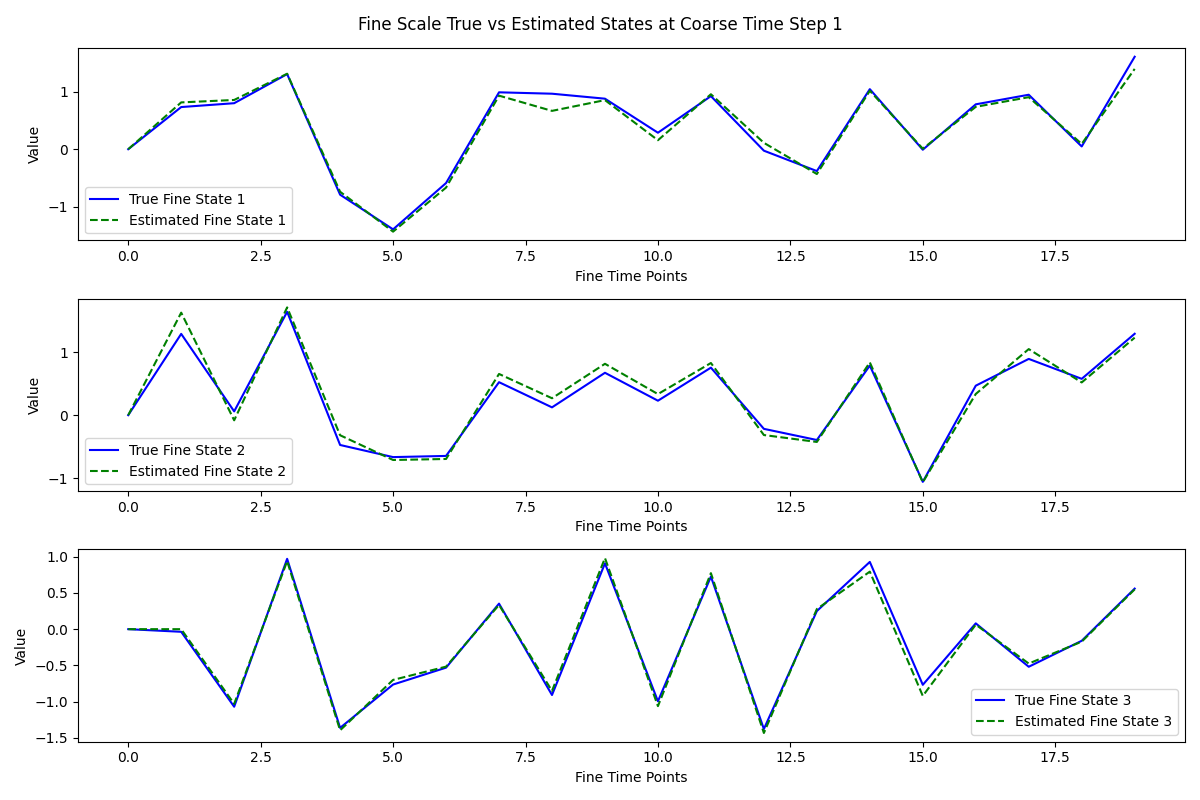}
        \caption{Individual $d=0$}
        \label{fig:d0fine}
    \end{subfigure}
    \hfill
    \begin{subfigure}[b]{0.45\textwidth}
        \centering
        \includegraphics[width=\textwidth]{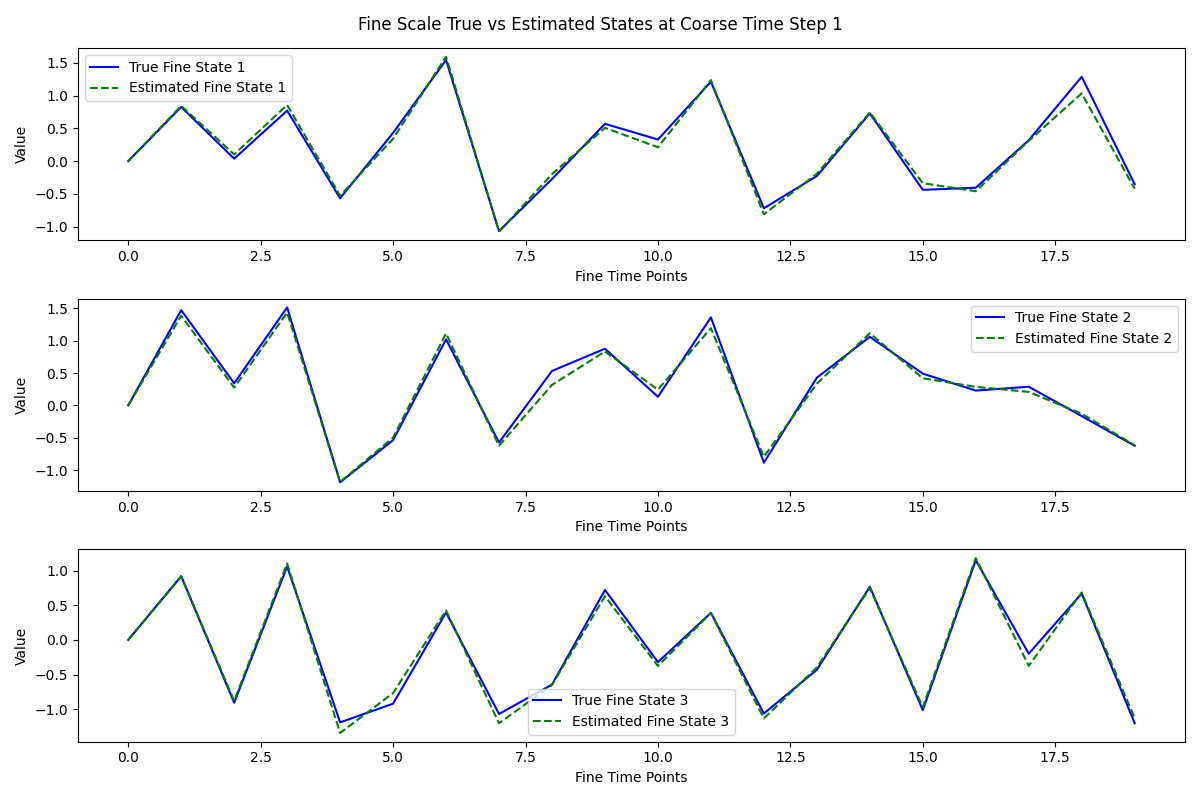}
        \caption{Individual $d=1$}
        \label{fig:d1fine}
    \end{subfigure}

    \vspace{0.3cm} 

    \begin{subfigure}[b]{0.45\textwidth}
        \centering
        \includegraphics[width=\textwidth]{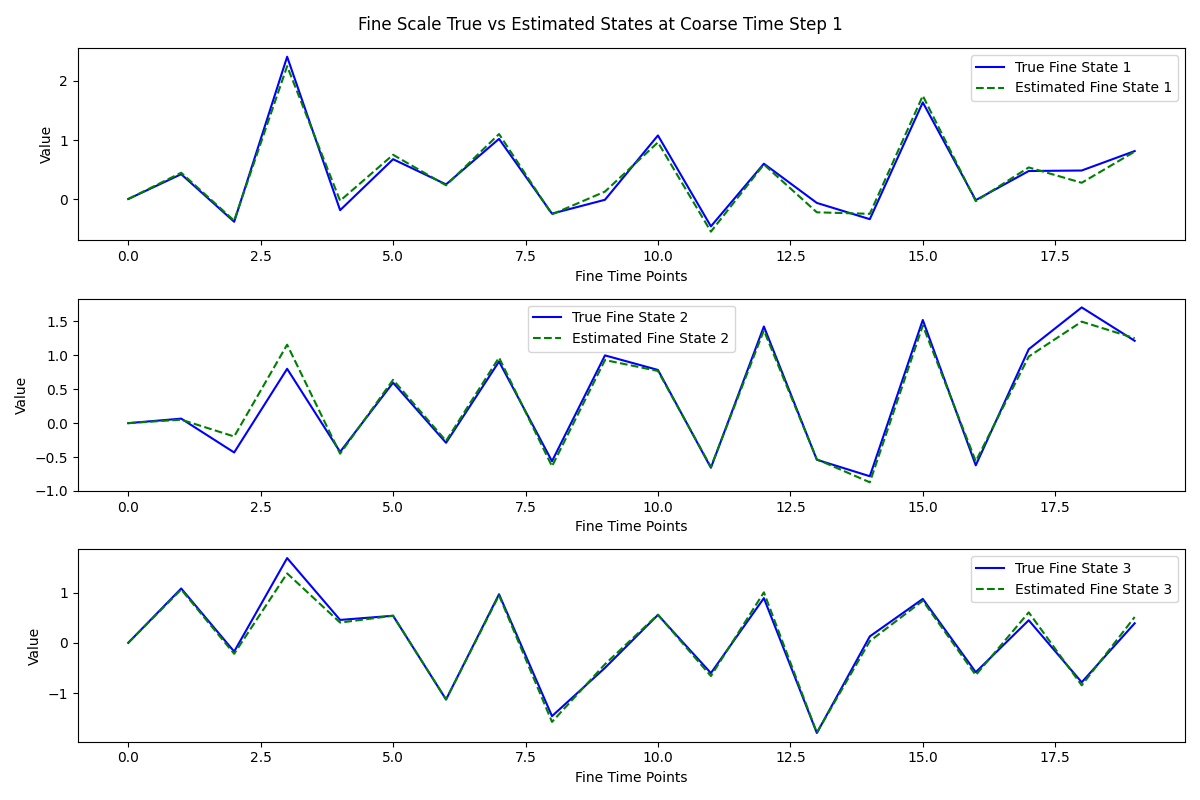}
        \caption{Individual $d=2$}
        \label{fig:d2fine}
    \end{subfigure}
    \hfill
    \begin{subfigure}[b]{0.45\textwidth}
        \centering
        \includegraphics[width=\textwidth]{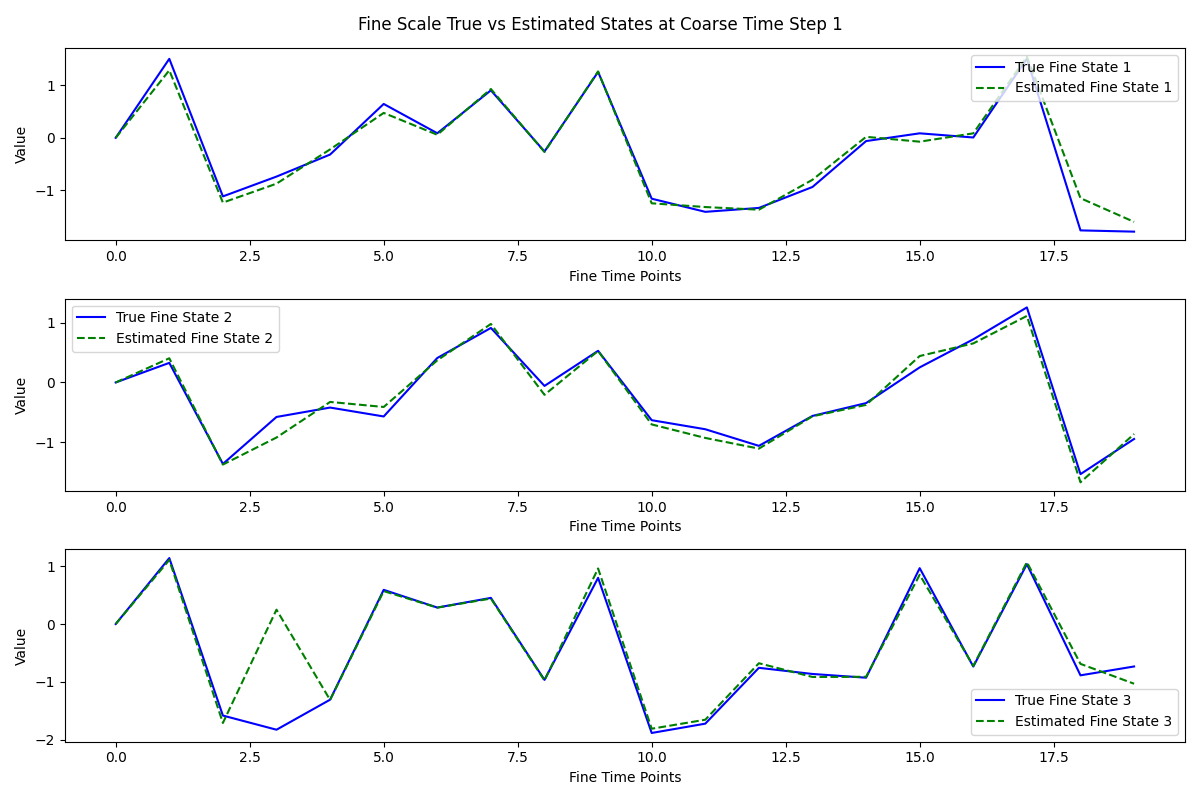}
        \caption{Individual $d=3$}
        \label{fig:d3fine}
    \end{subfigure}

    \caption{True vs. estimated fine time scale trajectories at coarse time step $t=1$ for individuals $d=0, d=1, d=2$, and $d=3$.}
    \label{fig:fine_trajectories}
\end{figure}

\begin{figure}[h!]
    \centering
    \begin{subfigure}[b]{0.45\textwidth}
        \centering
        \includegraphics[width=\textwidth]{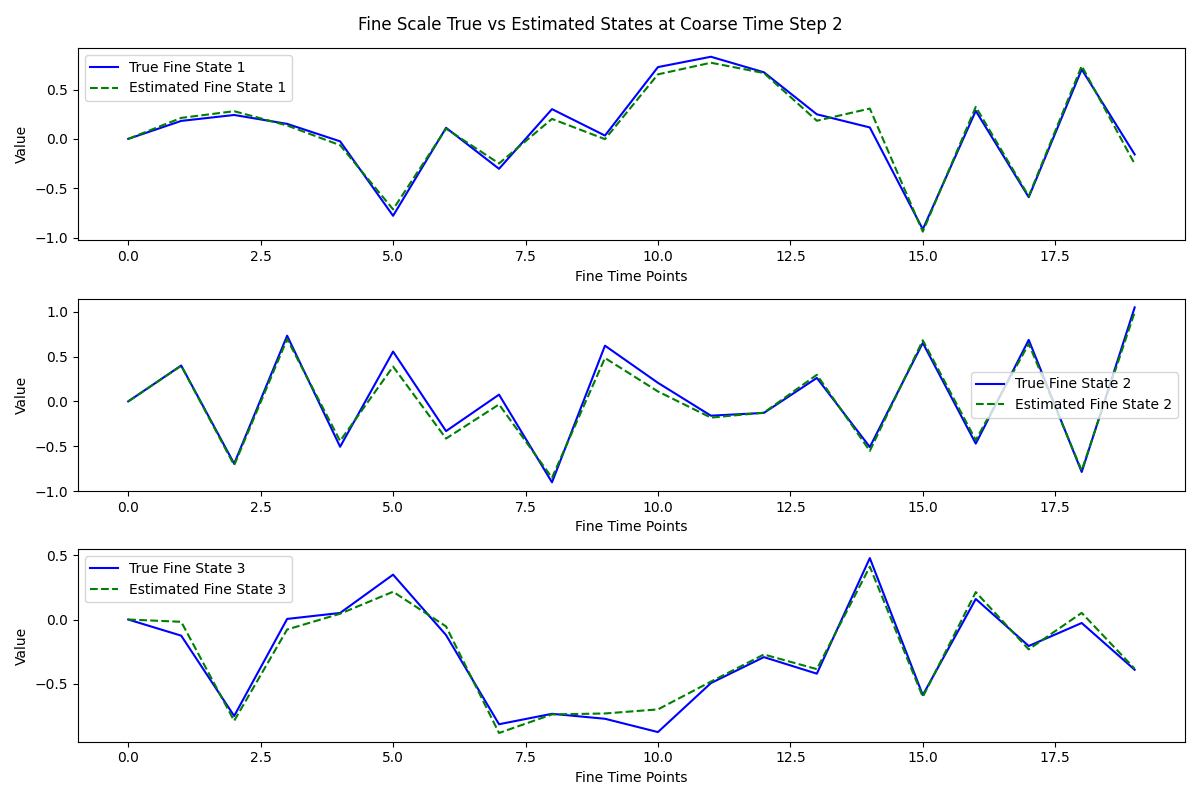}
        \caption{Individual $d=0$}
        \label{fig:d0fine}
    \end{subfigure}
    \hfill
    \begin{subfigure}[b]{0.45\textwidth}
        \centering
        \includegraphics[width=\textwidth]{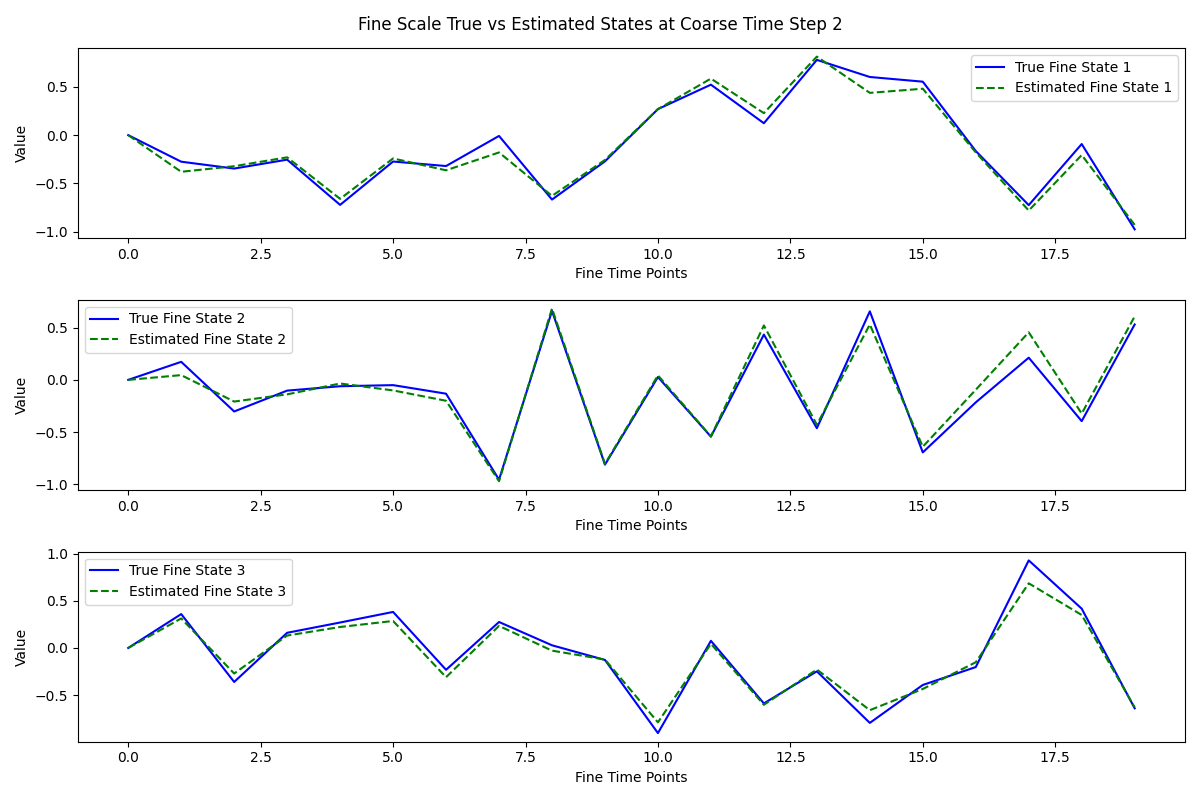}
        \caption{Individual $d=1$}
        \label{fig:d1fine}
    \end{subfigure}

    \vspace{0.3cm} 

    \begin{subfigure}[b]{0.45\textwidth}
        \centering
        \includegraphics[width=\textwidth]{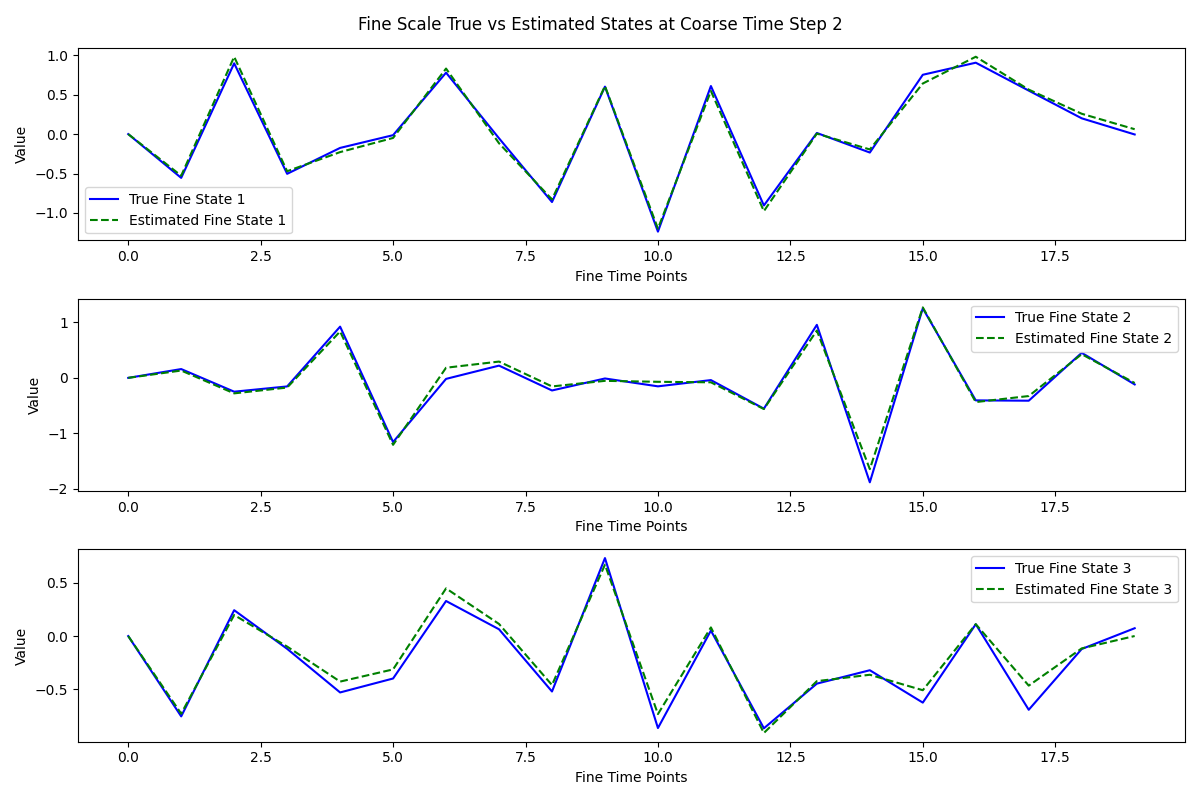}
        \caption{Individual $d=2$}
        \label{fig:d2fine}
    \end{subfigure}
    \hfill
    \begin{subfigure}[b]{0.45\textwidth}
        \centering
        \includegraphics[width=\textwidth]{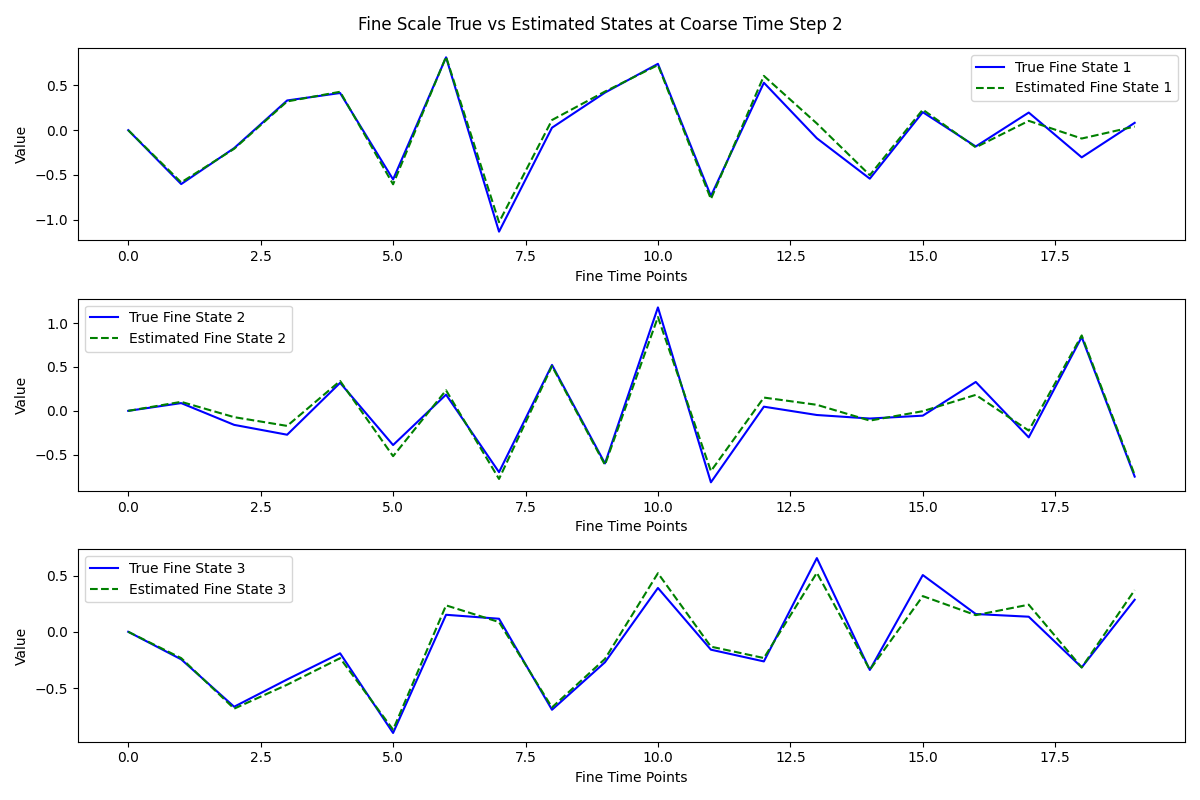}
        \caption{Individual $d=3$}
        \label{fig:d3fine}
    \end{subfigure}

    \caption{True vs. estimated fine time scale trajectories at coarse time step $t=2$ for individuals $d=0, d=1, d=2$, and $d=3$.}
    \label{fig:fine_trajectories}
\end{figure}
\begin{figure}[h!]
    \centering
    \begin{subfigure}[b]{0.45\textwidth}
        \centering
        \includegraphics[width=\textwidth]{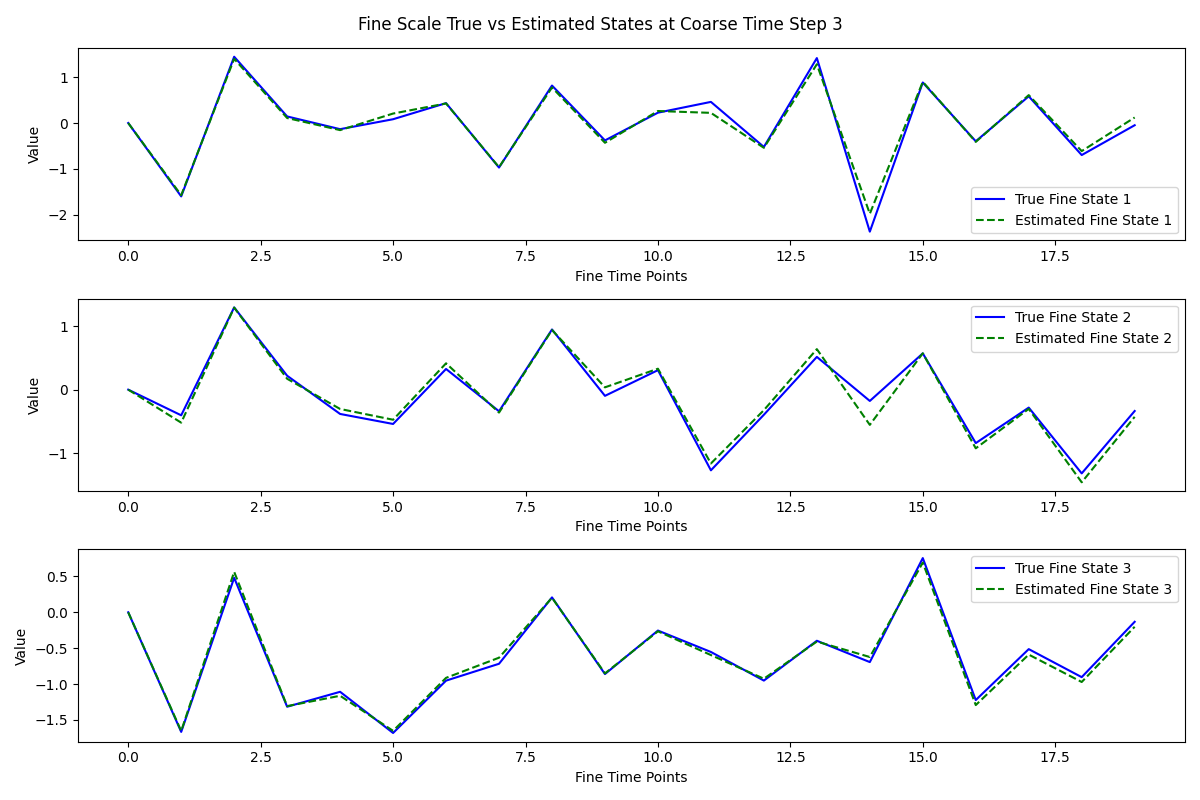}
        \caption{Individual $d=0$}
        \label{fig:d0fine}
    \end{subfigure}
    \hfill
    \begin{subfigure}[b]{0.45\textwidth}
        \centering
        \includegraphics[width=\textwidth]{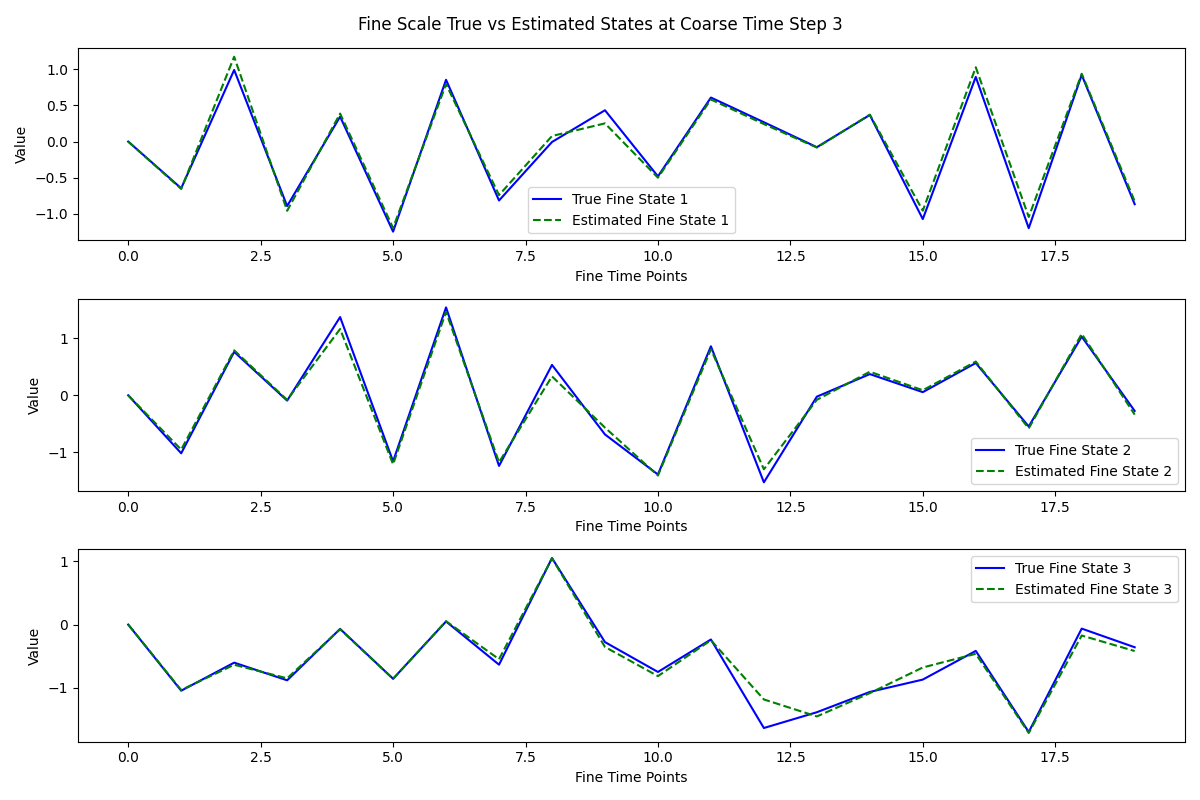}
        \caption{Individual $d=1$}
        \label{fig:d1fine}
    \end{subfigure}

    \vspace{0.3cm} 

    \begin{subfigure}[b]{0.45\textwidth}
        \centering
        \includegraphics[width=\textwidth]{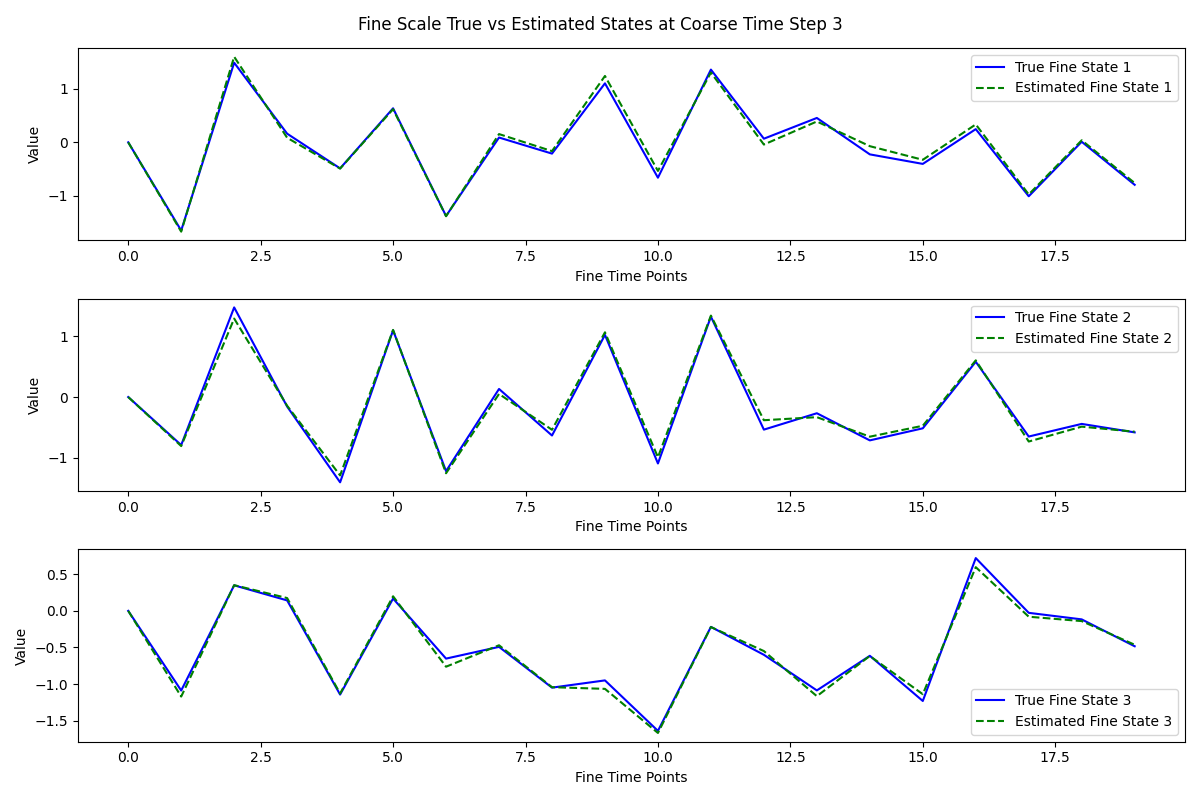}
        \caption{Individual $d=2$}
        \label{fig:d2fine}
    \end{subfigure}
    \hfill
    \begin{subfigure}[b]{0.45\textwidth}
        \centering
        \includegraphics[width=\textwidth]{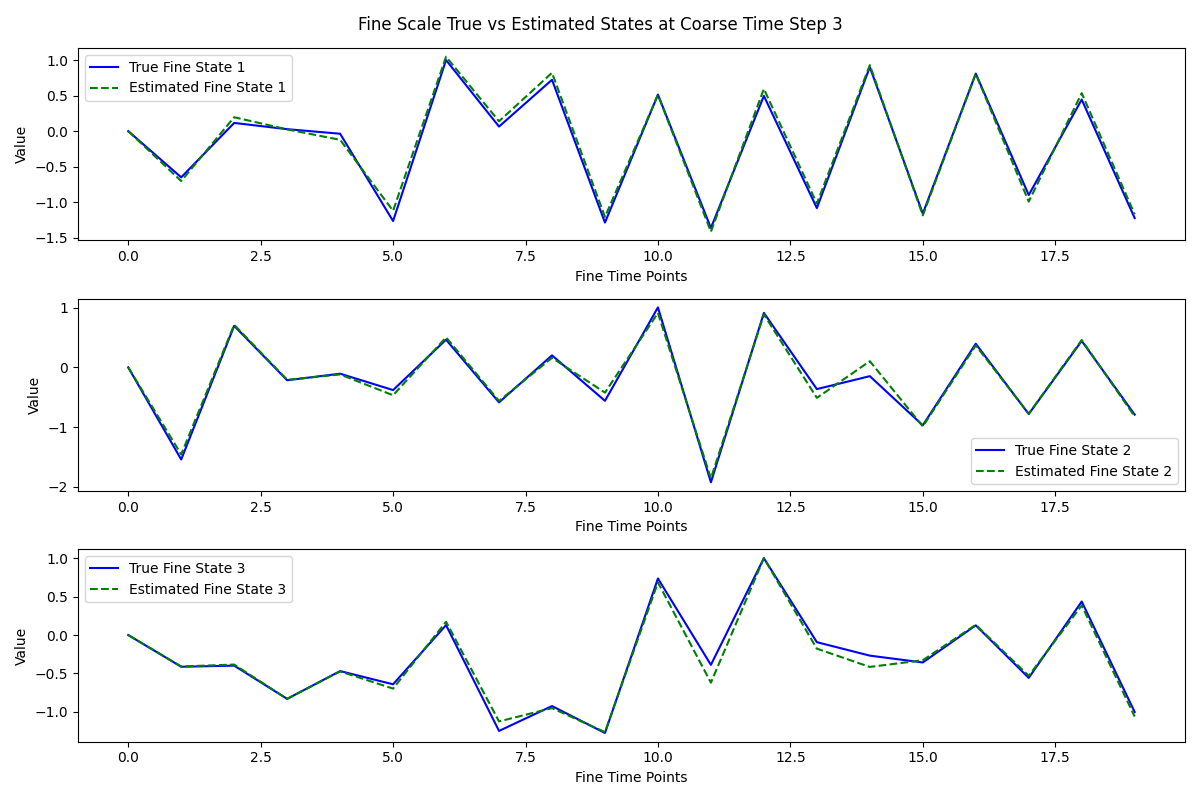}
        \caption{Individual $d=3$}
        \label{fig:d3fine}
    \end{subfigure}

    \caption{True vs. estimated fine time scale trajectories at coarse time step $t=3$ for individuals $d=0, d=1, d=2$, and $d=3$.}
    \label{fig:fine_trajectories}
\end{figure}
\begin{figure}[h!]
    \centering
    \begin{subfigure}[b]{0.45\textwidth}
        \centering
        \includegraphics[width=\textwidth]{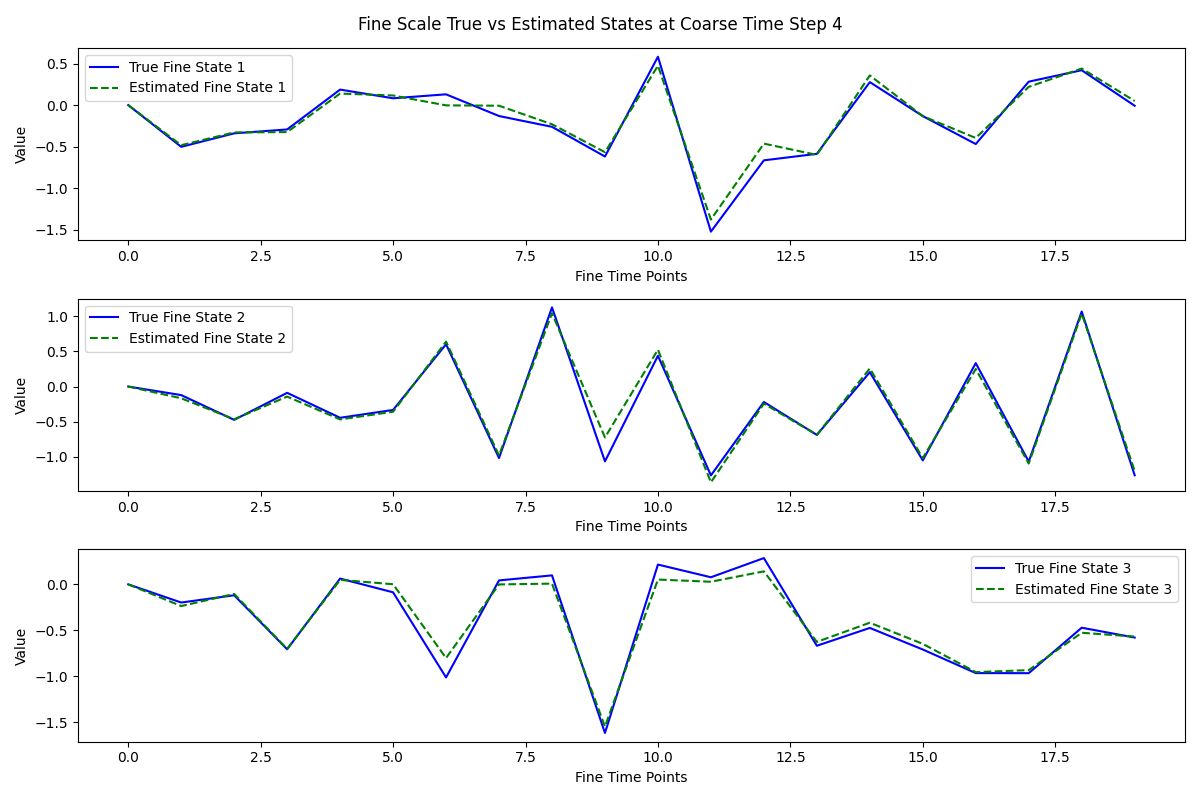}
        \caption{Individual $d=0$}
        \label{fig:d0fine}
    \end{subfigure}
    \hfill
    \begin{subfigure}[b]{0.45\textwidth}
        \centering
        \includegraphics[width=\textwidth]{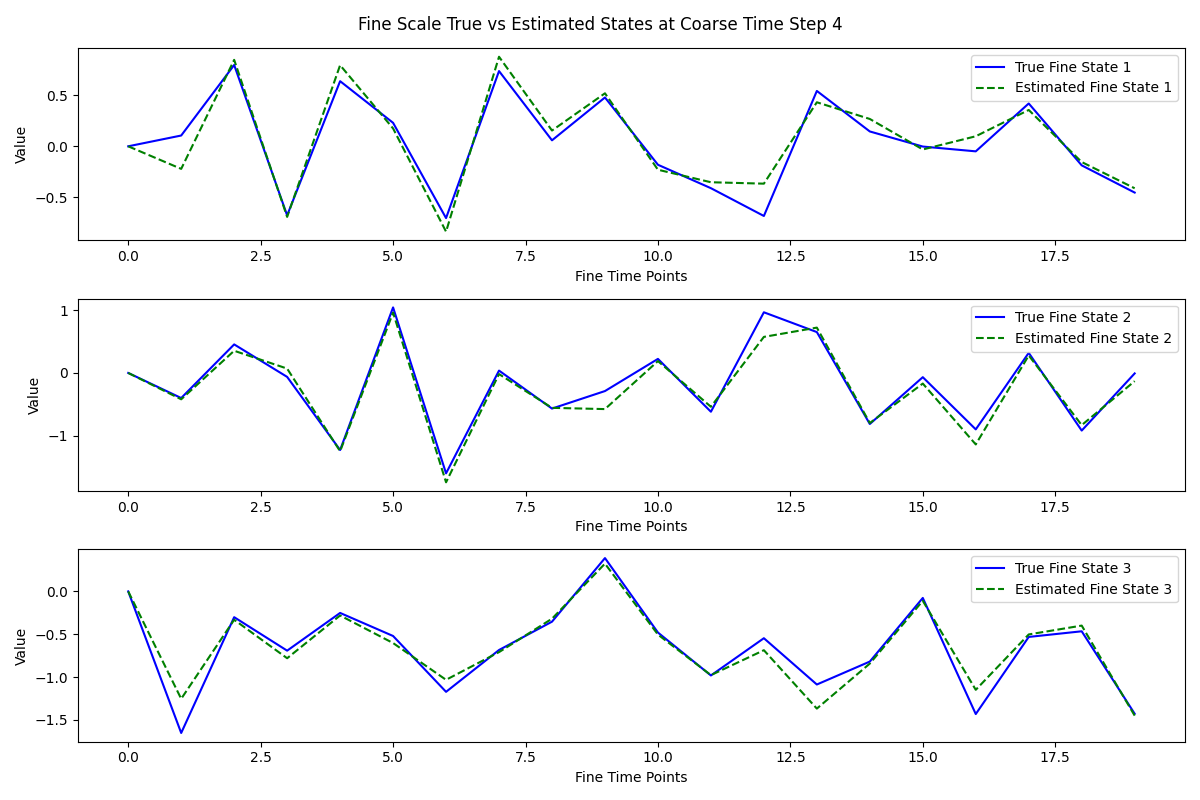}
        \caption{Individual $d=1$}
        \label{fig:d1fine}
    \end{subfigure}

    \vspace{0.3cm} 

    \begin{subfigure}[b]{0.45\textwidth}
        \centering
        \includegraphics[width=\textwidth]{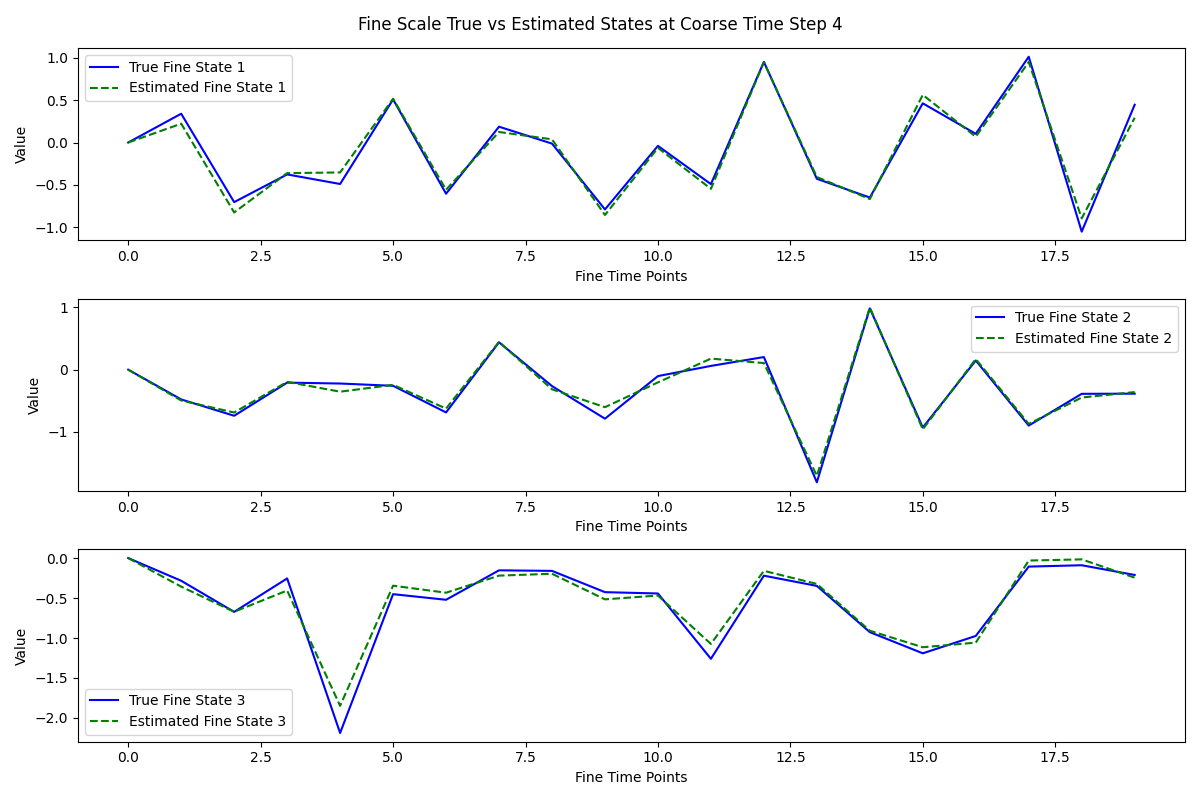}
        \caption{Individual $d=2$}
        \label{fig:d2fine}
    \end{subfigure}
    \hfill
    \begin{subfigure}[b]{0.45\textwidth}
        \centering
        \includegraphics[width=\textwidth]{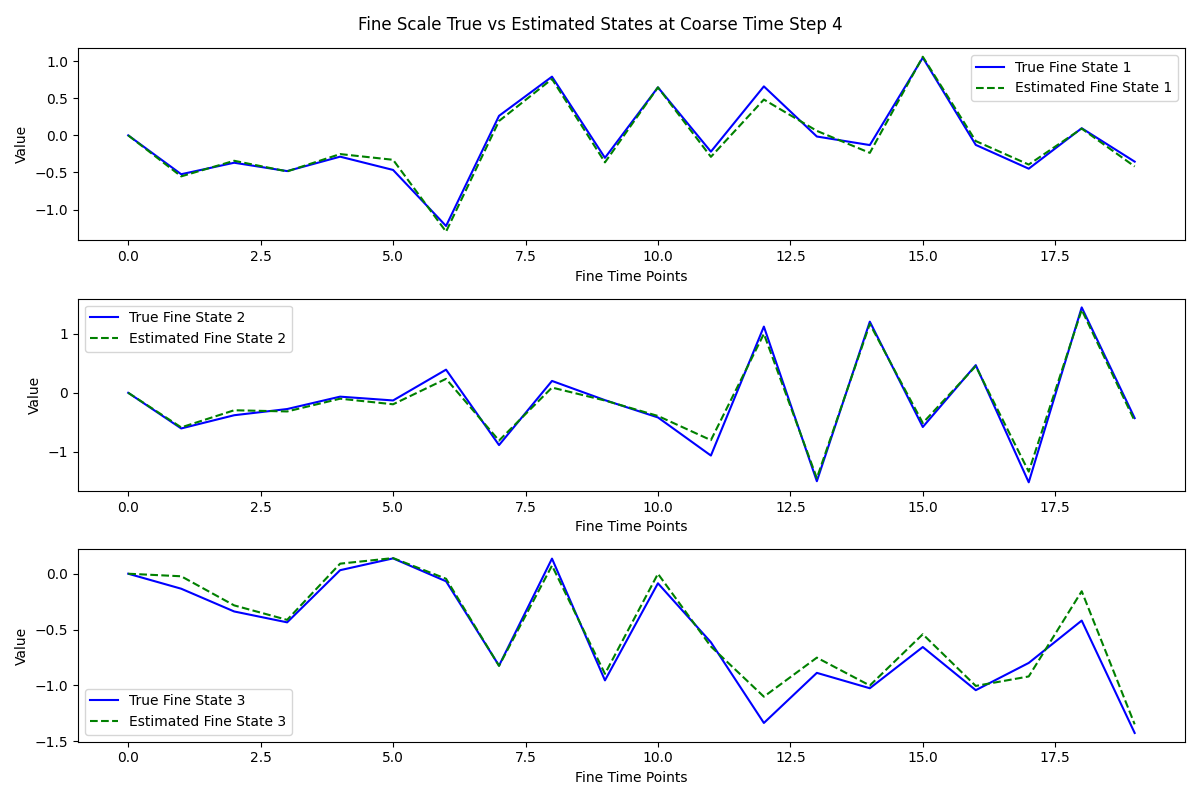}
        \caption{Individual $d=3$}
        \label{fig:d3fine}
    \end{subfigure}

    \caption{True vs. estimated fine time scale trajectories at coarse time step $t=4$ for individuals $d=0, d=1, d=2$, and $d=3$.}
    \label{fig:fine_trajectories}
\end{figure}
\begin{figure}[h!]
    \centering
    \begin{subfigure}[b]{0.45\textwidth}
        \centering
        \includegraphics[width=\textwidth]{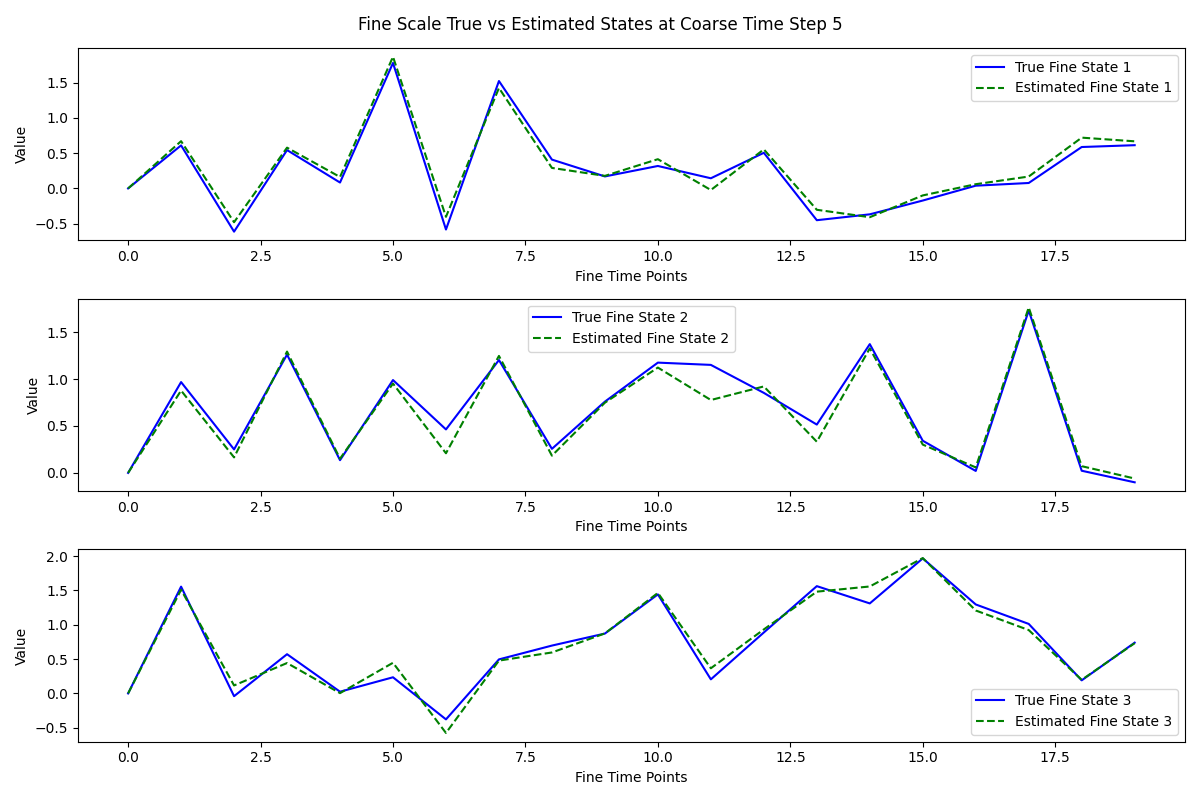}
        \caption{Individual $d=0$}
        \label{fig:d0fine}
    \end{subfigure}
    \hfill
    \begin{subfigure}[b]{0.45\textwidth}
        \centering
        \includegraphics[width=\textwidth]{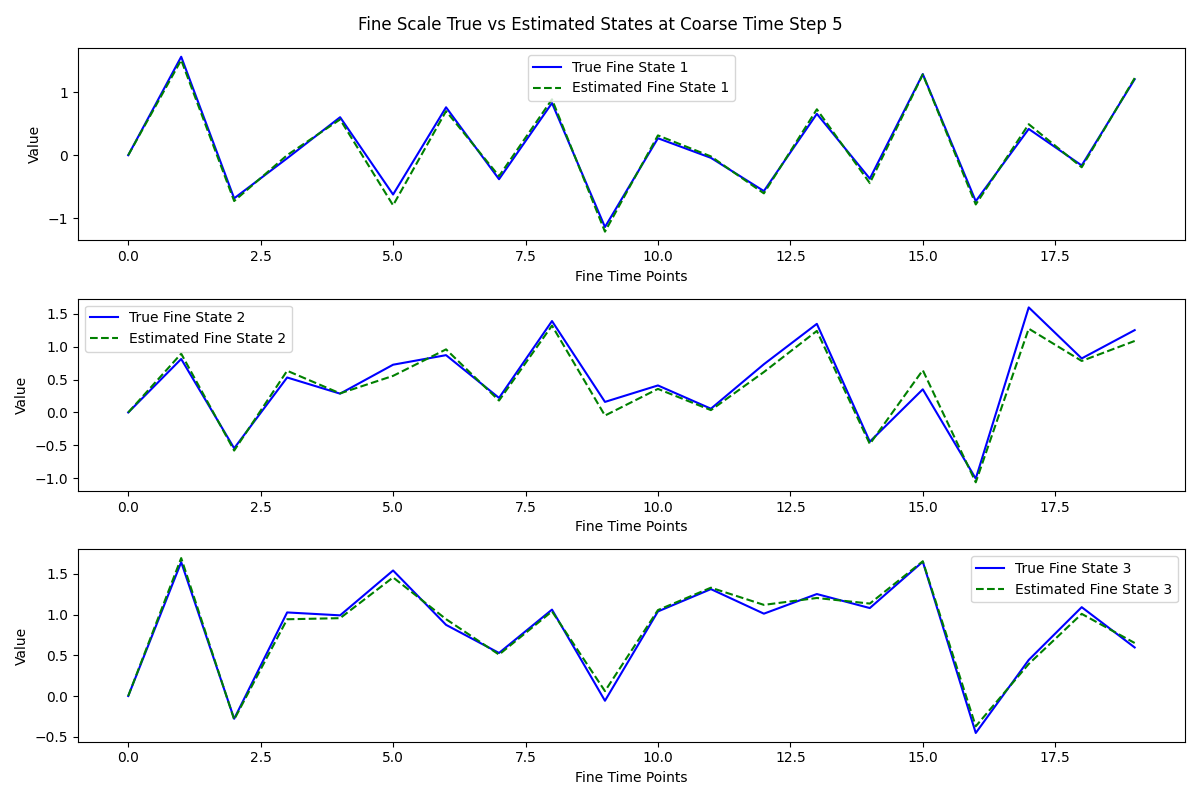}
        \caption{Individual $d=1$}
        \label{fig:d1fine}
    \end{subfigure}

    \vspace{0.3cm} 

    \begin{subfigure}[b]{0.45\textwidth}
        \centering
        \includegraphics[width=\textwidth]{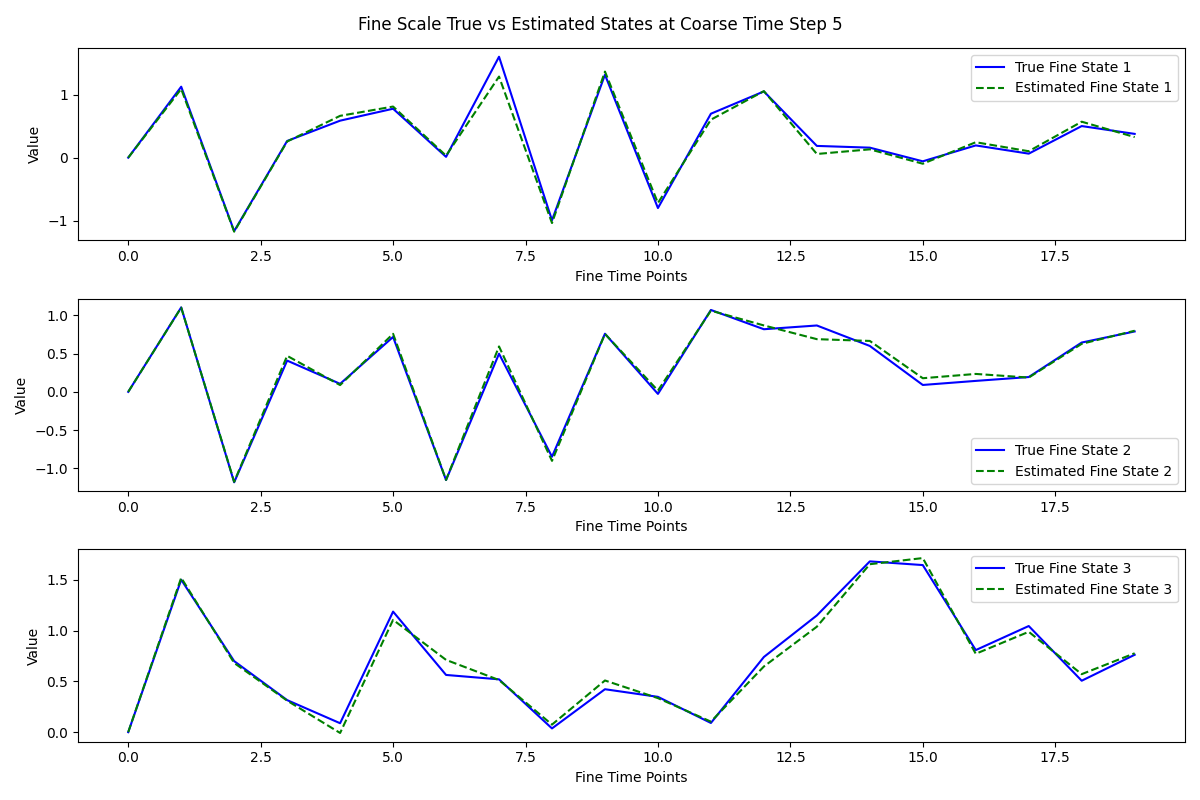}
        \caption{Individual $d=2$}
        \label{fig:d2fine}
    \end{subfigure}
    \hfill
    \begin{subfigure}[b]{0.45\textwidth}
        \centering
        \includegraphics[width=\textwidth]{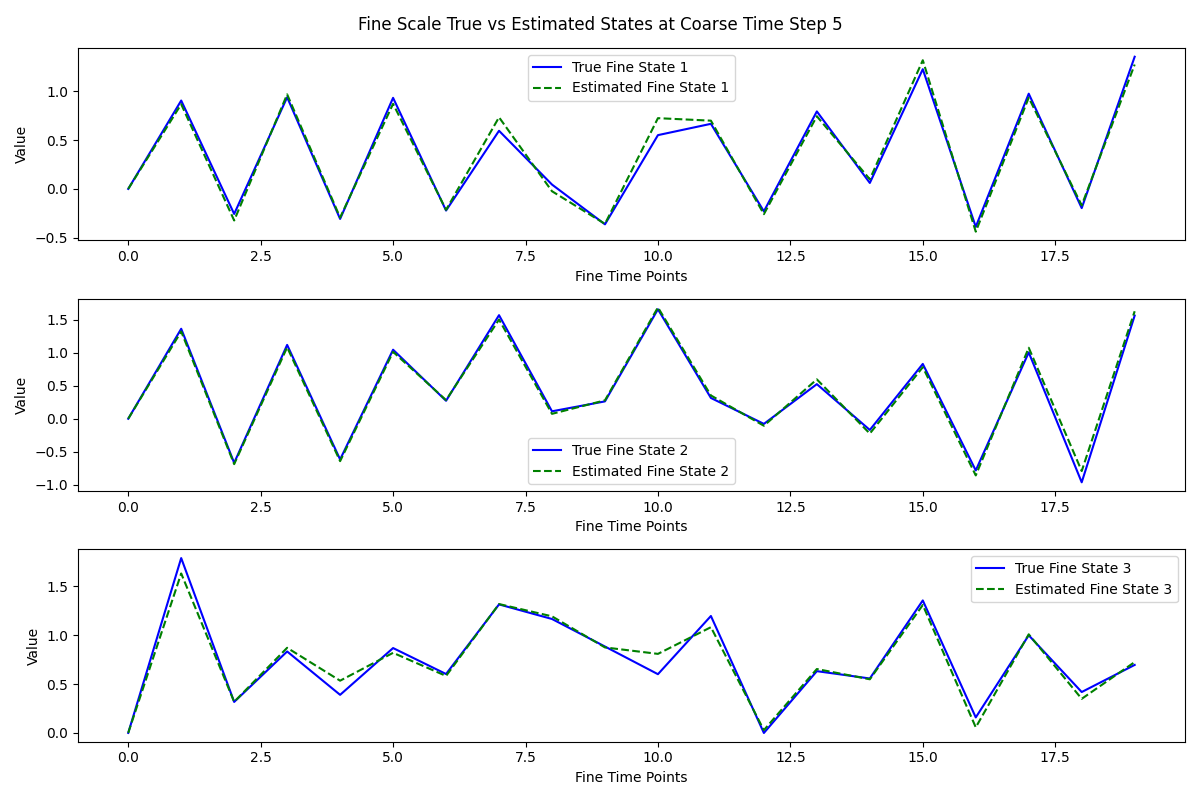}
        \caption{Individual $d=3$}
        \label{fig:d3fine}
    \end{subfigure}

    \caption{True vs. estimated fine time scale trajectories at coarse time step $t=5$ for individuals $d=0, d=1, d=2$, and $d=3$.}
    \label{fig:fine_trajectories}
\end{figure}
\begin{figure}[h!]
    \centering
    \begin{subfigure}[b]{0.45\textwidth}
        \centering
        \includegraphics[width=\textwidth]{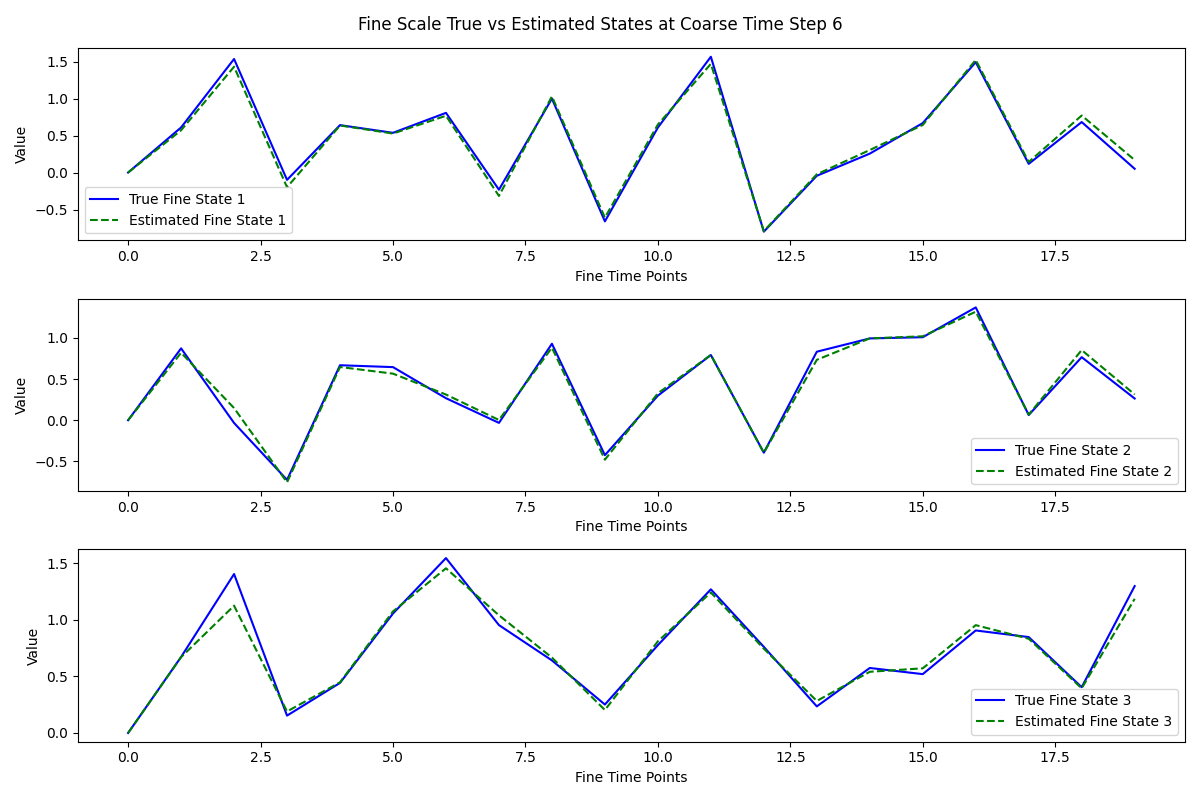}
        \caption{Individual $d=0$}
        \label{fig:d0fine}
    \end{subfigure}
    \hfill
    \begin{subfigure}[b]{0.45\textwidth}
        \centering
        \includegraphics[width=\textwidth]{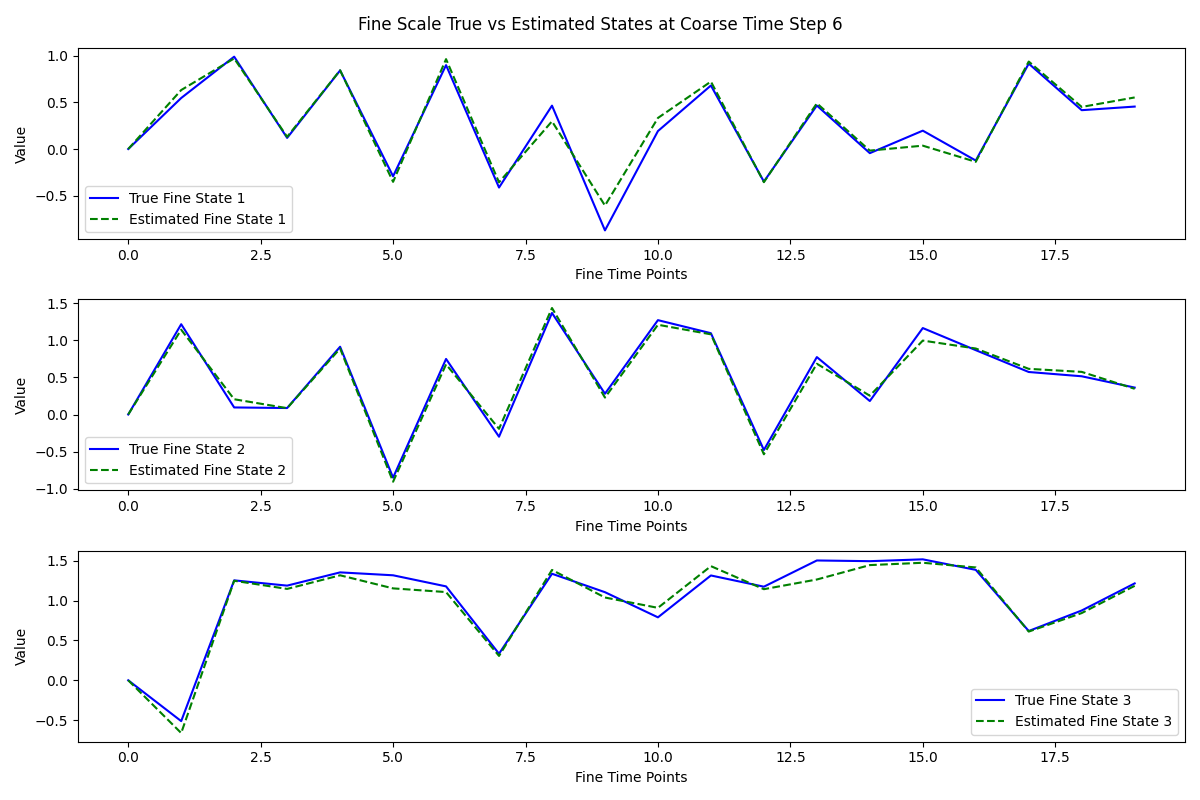}
        \caption{Individual $d=1$}
        \label{fig:d1fine}
    \end{subfigure}

    \vspace{0.3cm} 

    \begin{subfigure}[b]{0.45\textwidth}
        \centering
        \includegraphics[width=\textwidth]{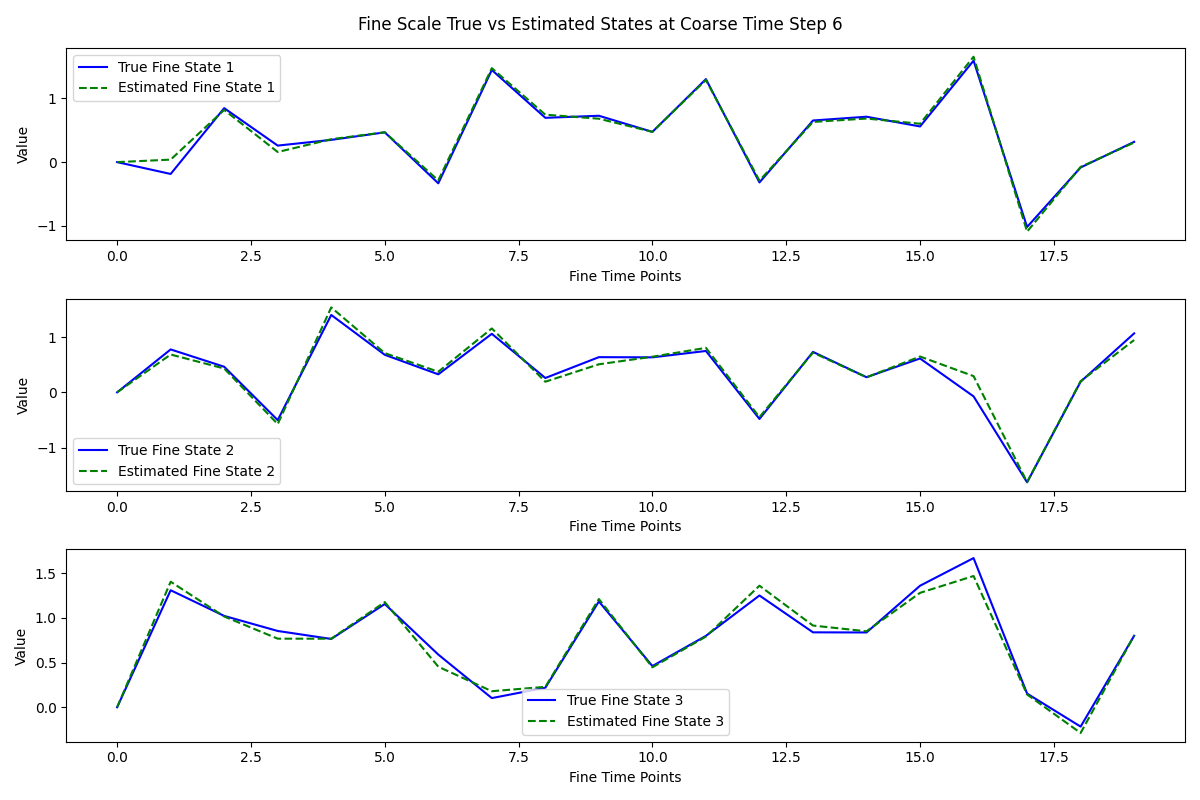}
        \caption{Individual $d=2$}
        \label{fig:d2fine}
    \end{subfigure}
    \hfill
    \begin{subfigure}[b]{0.45\textwidth}
        \centering
        \includegraphics[width=\textwidth]{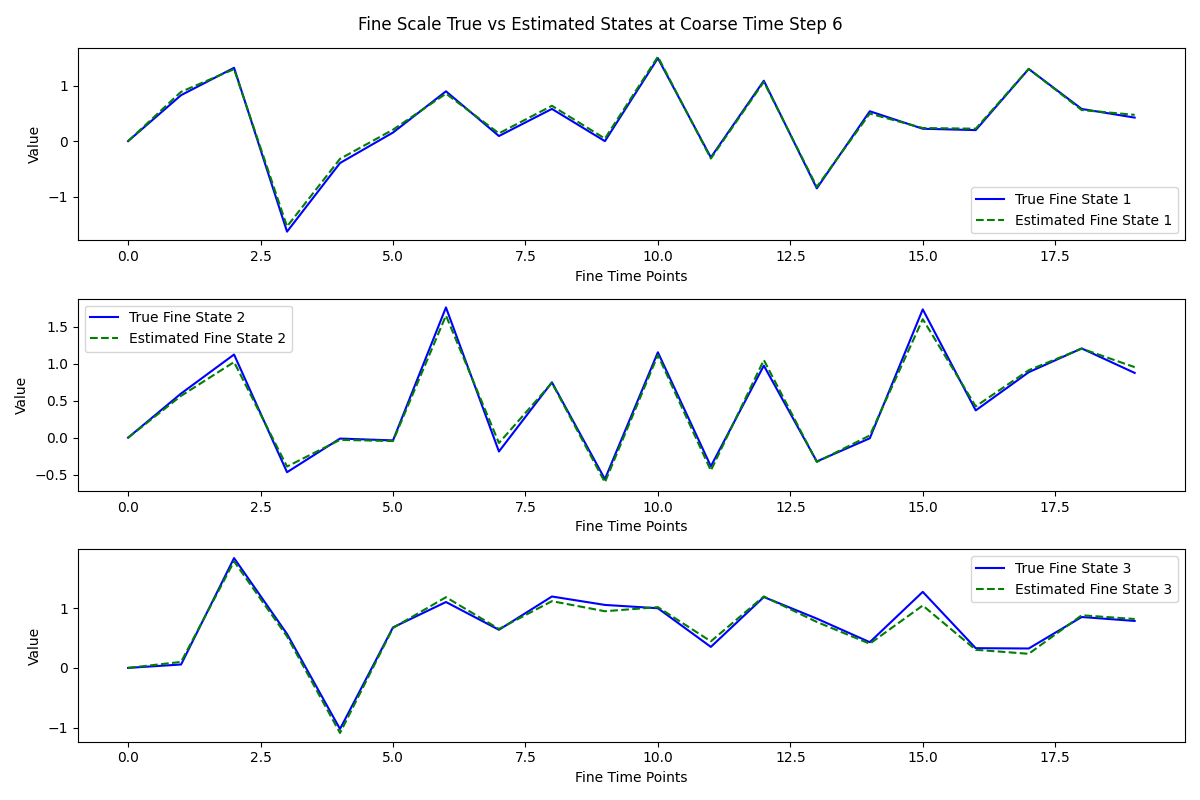}
        \caption{Individual $d=3$}
        \label{fig:d3fine}
    \end{subfigure}

    \caption{True vs. estimated fine time scale trajectories at coarse time step $t=6$ for individuals $d=0, d=1, d=2$, and $d=3$.}
    \label{fig:fine_trajectories}
\end{figure}
\begin{figure}[h!]
    \centering
    \begin{subfigure}[b]{0.45\textwidth}
        \centering
        \includegraphics[width=\textwidth]{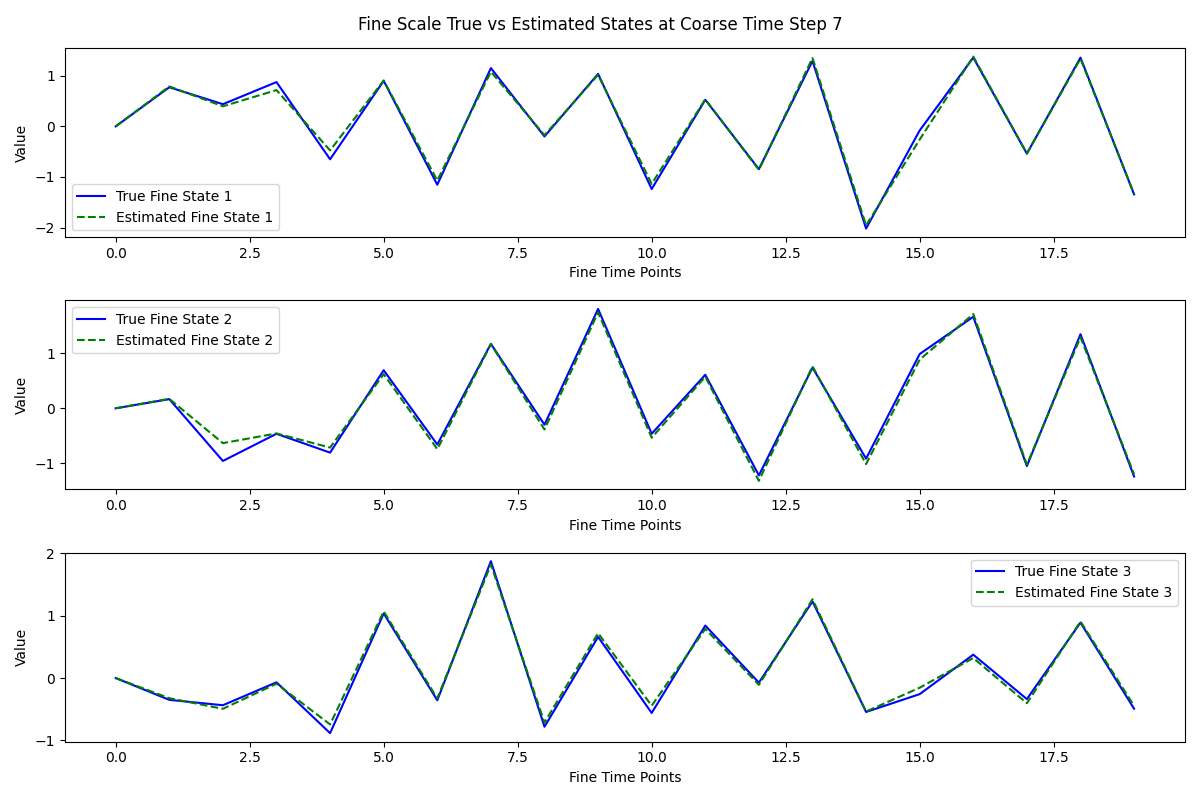}
        \caption{Individual $d=0$}
        \label{fig:d0fine}
    \end{subfigure}
    \hfill
    \begin{subfigure}[b]{0.45\textwidth}
        \centering
        \includegraphics[width=\textwidth]{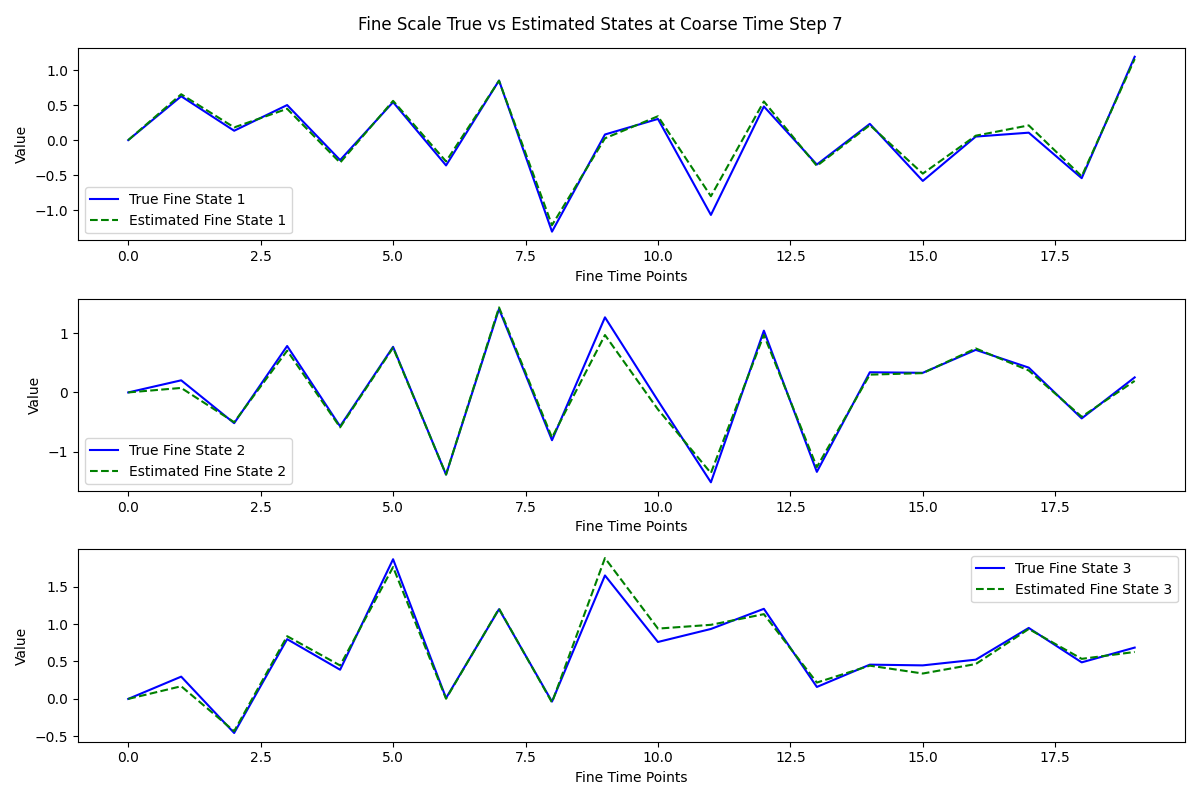}
        \caption{Individual $d=1$}
        \label{fig:d1fine}
    \end{subfigure}

    \vspace{0.3cm} 

    \begin{subfigure}[b]{0.45\textwidth}
        \centering
        \includegraphics[width=\textwidth]{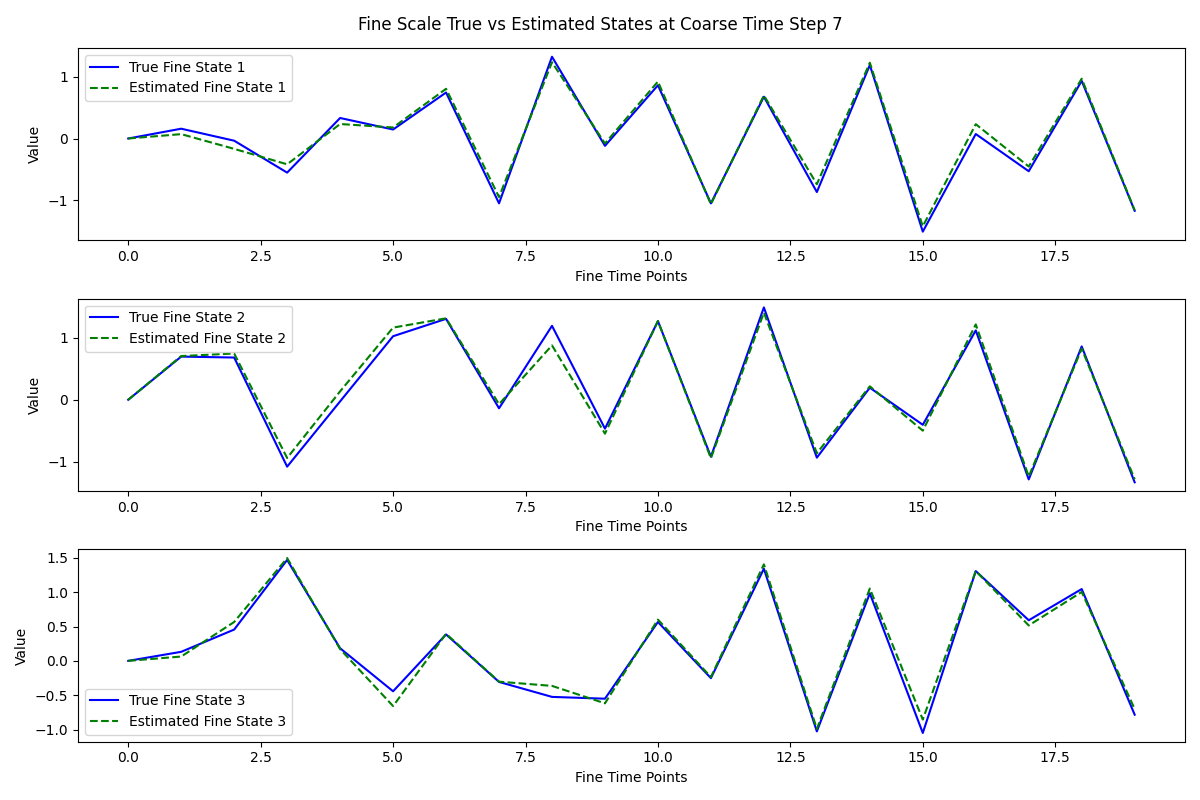}
        \caption{Individual $d=2$}
        \label{fig:d2fine}
    \end{subfigure}
    \hfill
    \begin{subfigure}[b]{0.45\textwidth}
        \centering
        \includegraphics[width=\textwidth]{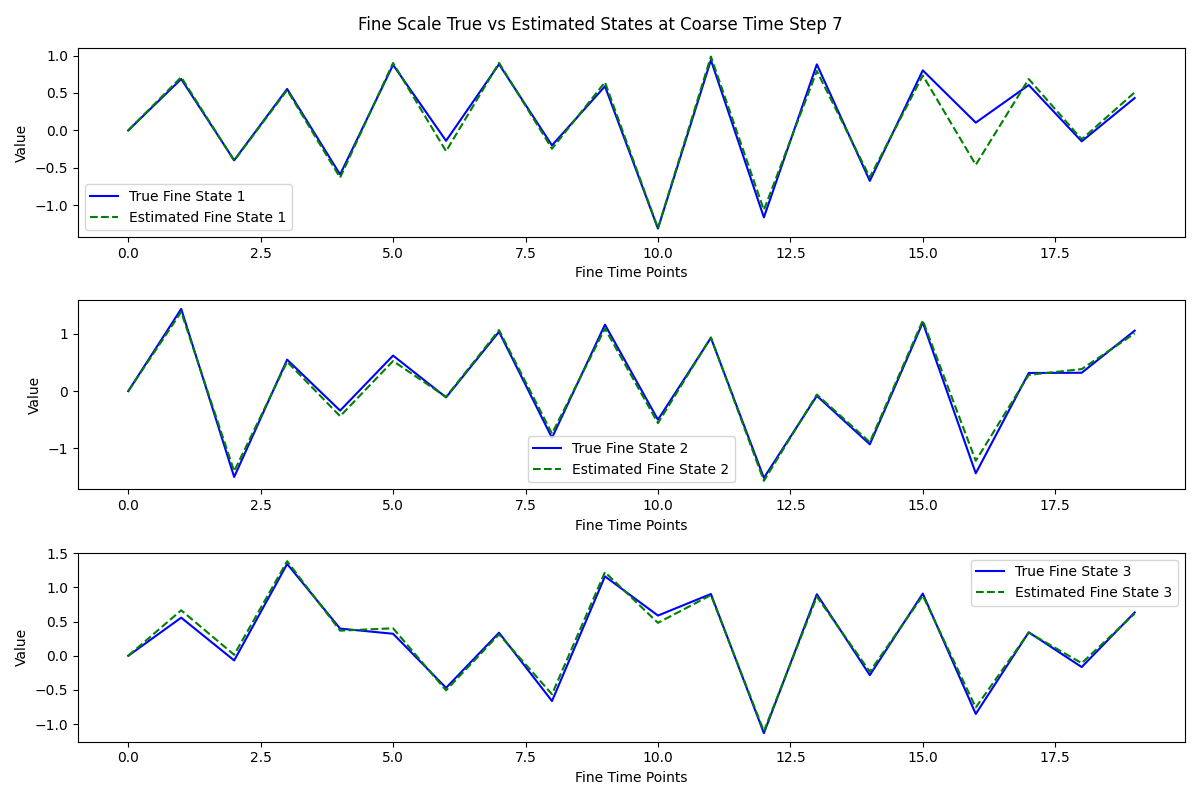}
        \caption{Individual $d=3$}
        \label{fig:d3fine}
    \end{subfigure}

    \caption{True vs. estimated fine time scale trajectories at coarse time step $t=7$ for individuals $d=0, d=1, d=2$, and $d=3$.}
    \label{fig:fine_trajectories}
\end{figure}
\begin{figure}[h!]
    \centering
    \begin{subfigure}[b]{0.45\textwidth}
        \centering
        \includegraphics[width=\textwidth]{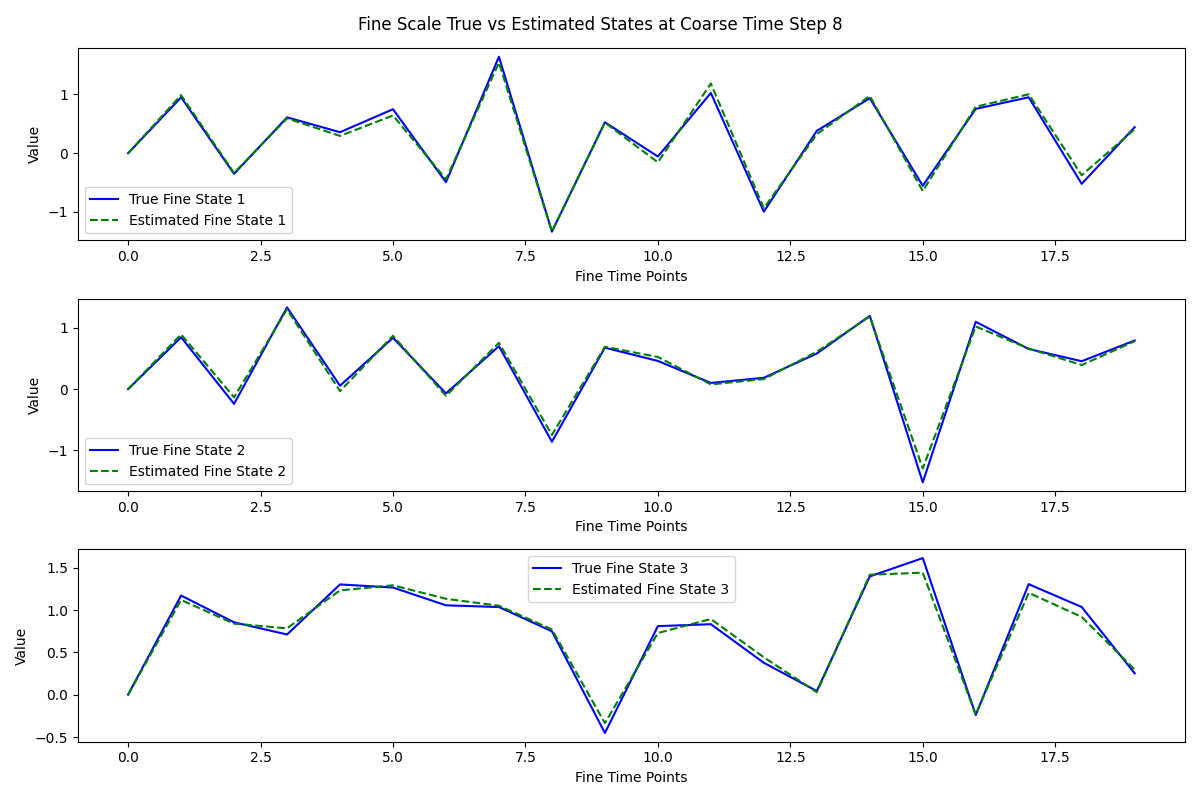}
        \caption{Individual $d=0$}
        \label{fig:d0fine}
    \end{subfigure}
    \hfill
    \begin{subfigure}[b]{0.45\textwidth}
        \centering
        \includegraphics[width=\textwidth]{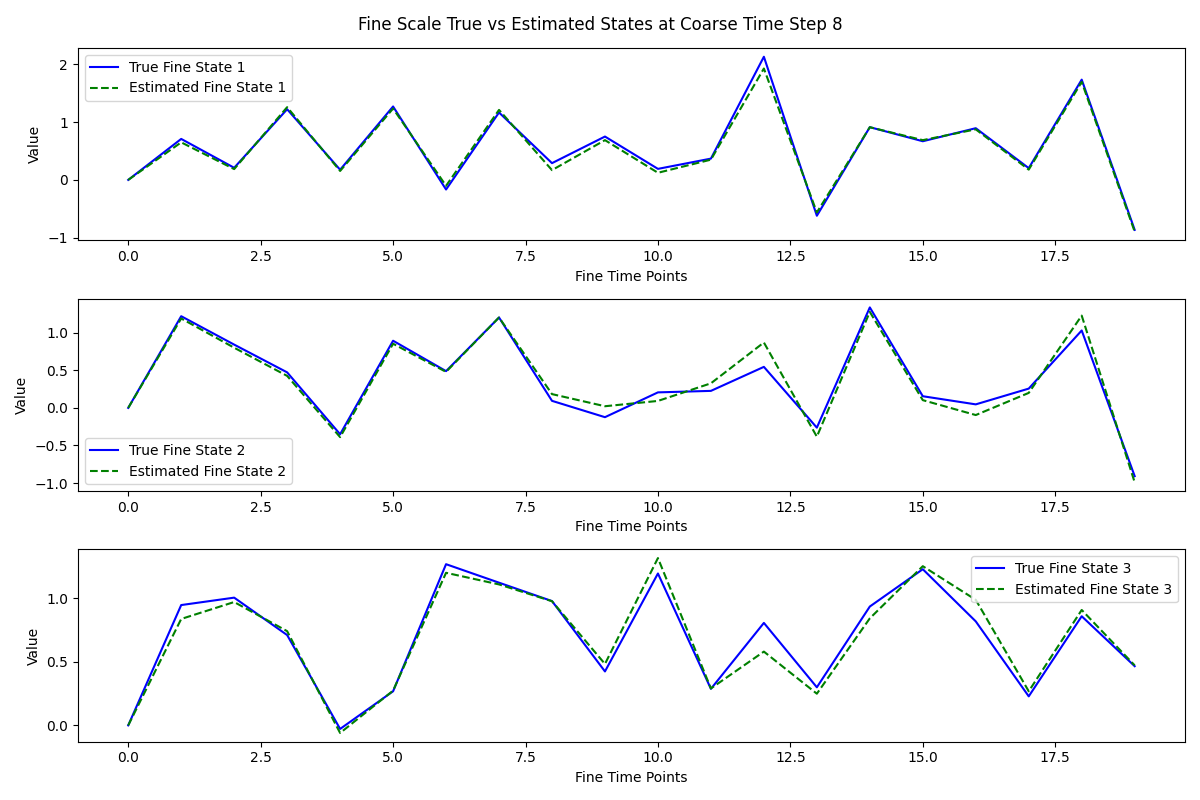}
        \caption{Individual $d=1$}
        \label{fig:d1fine}
    \end{subfigure}

    \vspace{0.3cm} 

    \begin{subfigure}[b]{0.45\textwidth}
        \centering
        \includegraphics[width=\textwidth]{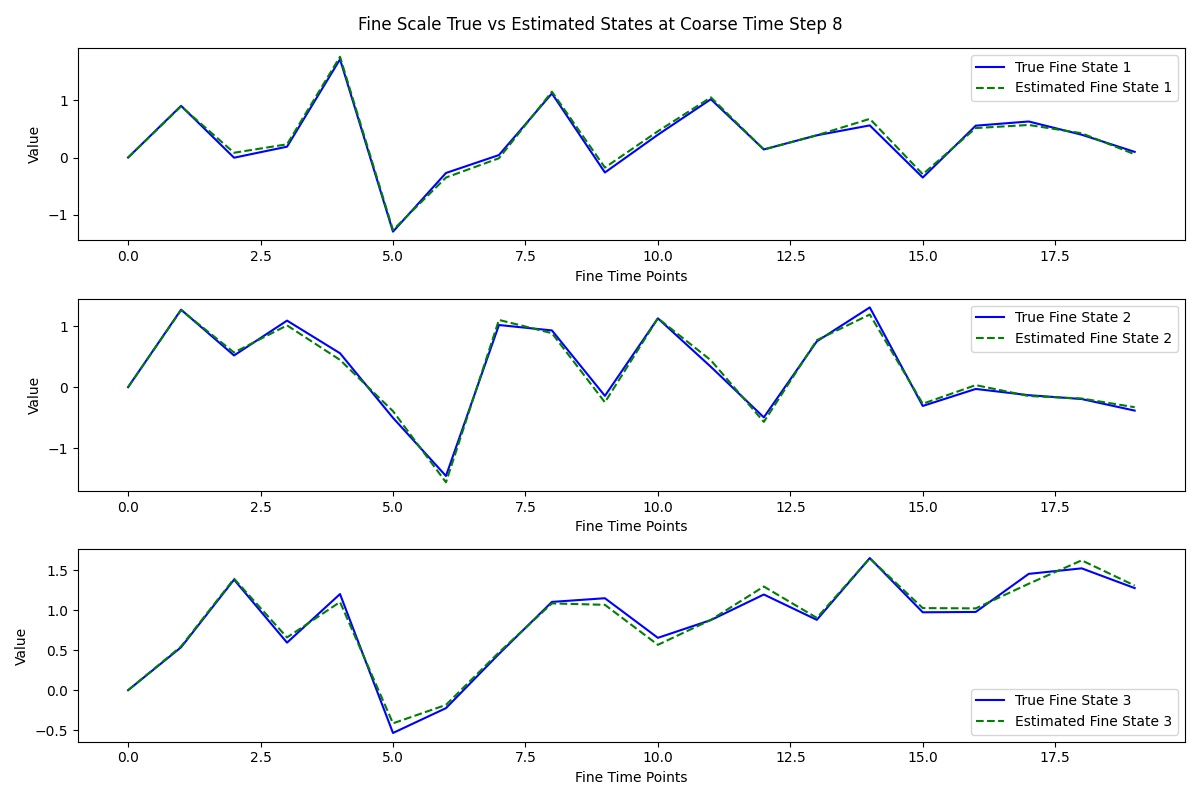}
        \caption{Individual $d=2$}
        \label{fig:d2fine}
    \end{subfigure}
    \hfill
    \begin{subfigure}[b]{0.45\textwidth}
        \centering
        \includegraphics[width=\textwidth]{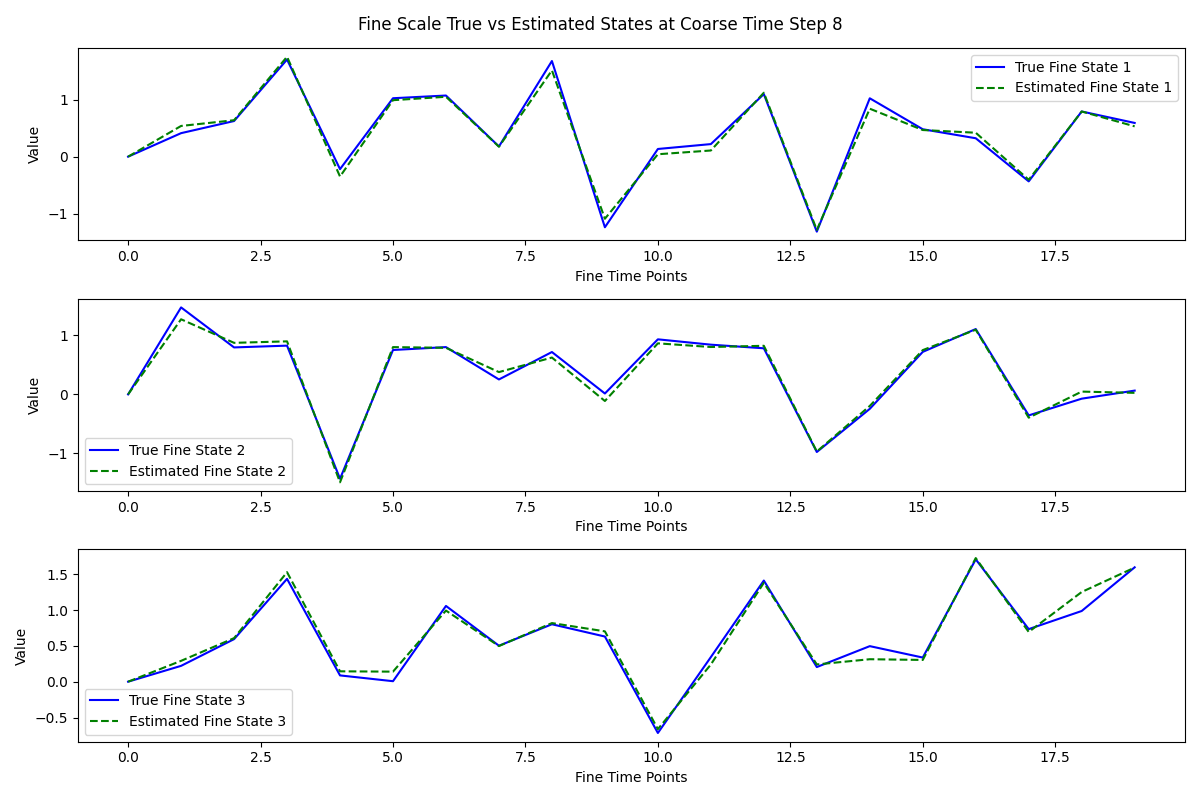}
        \caption{Individual $d=3$}
        \label{fig:d3fine}
    \end{subfigure}

    \caption{True vs. estimated fine time scale trajectories at coarse time step $t=8$ for individuals $d=0, d=1, d=2$, and $d=3$.}
    \label{fig:fine_trajectories}
\end{figure}
\begin{figure}[h!]
    \centering
    \begin{subfigure}[b]{0.45\textwidth}
        \centering
        \includegraphics[width=\textwidth]{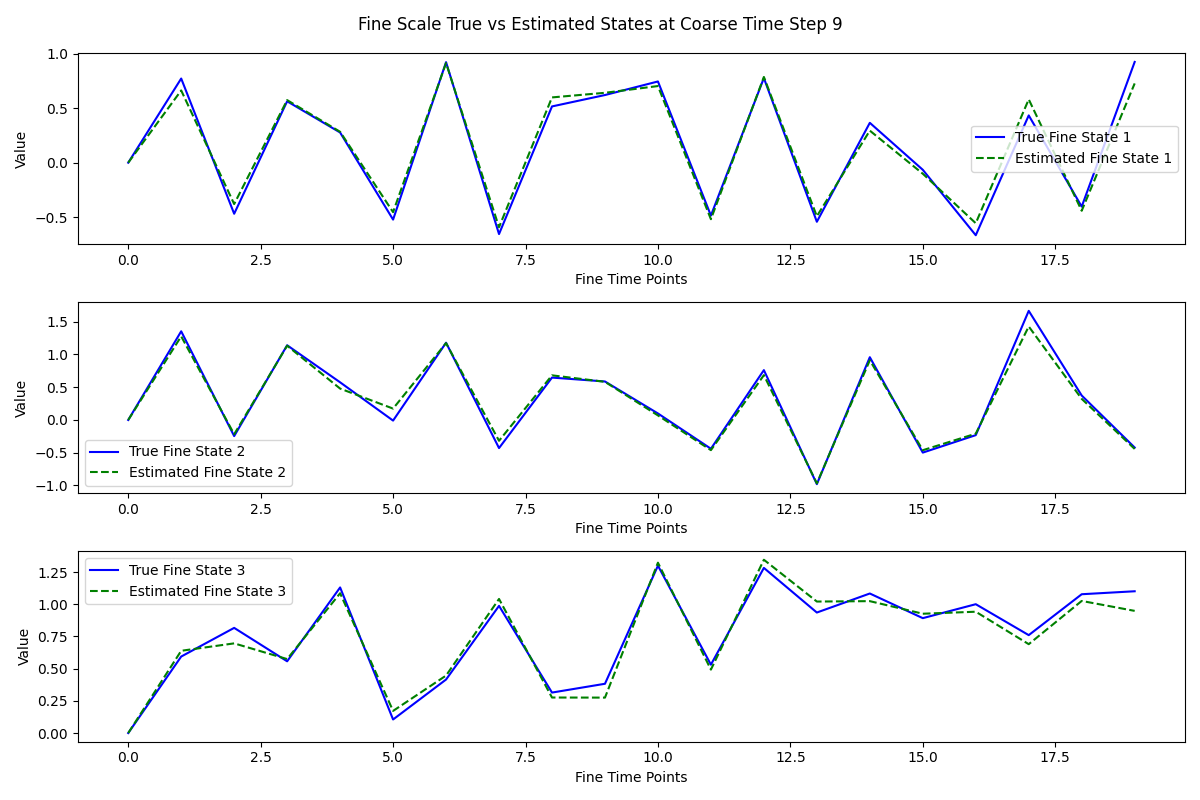}
        \caption{Individual $d=0$}
        \label{fig:d0fine}
    \end{subfigure}
    \hfill
    \begin{subfigure}[b]{0.45\textwidth}
        \centering
        \includegraphics[width=\textwidth]{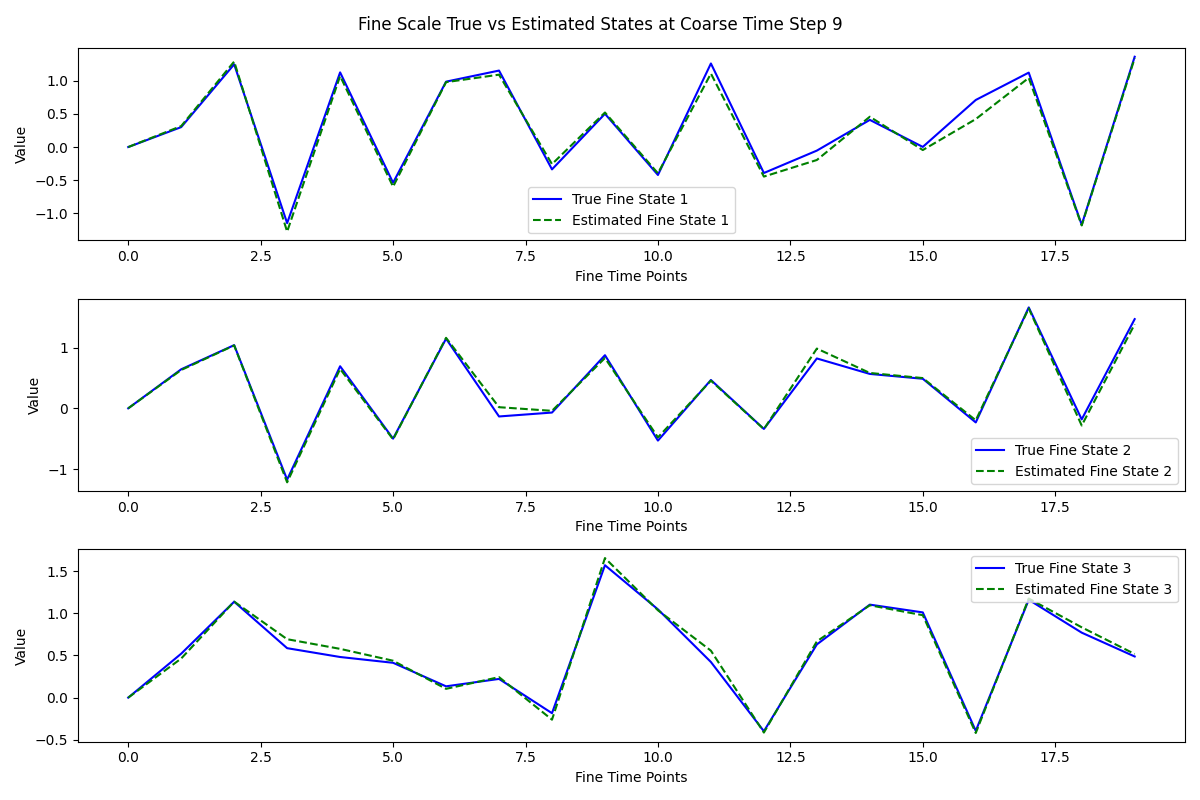}
        \caption{Individual $d=1$}
        \label{fig:d1fine}
    \end{subfigure}

    \vspace{0.3cm} 

    \begin{subfigure}[b]{0.45\textwidth}
        \centering
        \includegraphics[width=\textwidth]{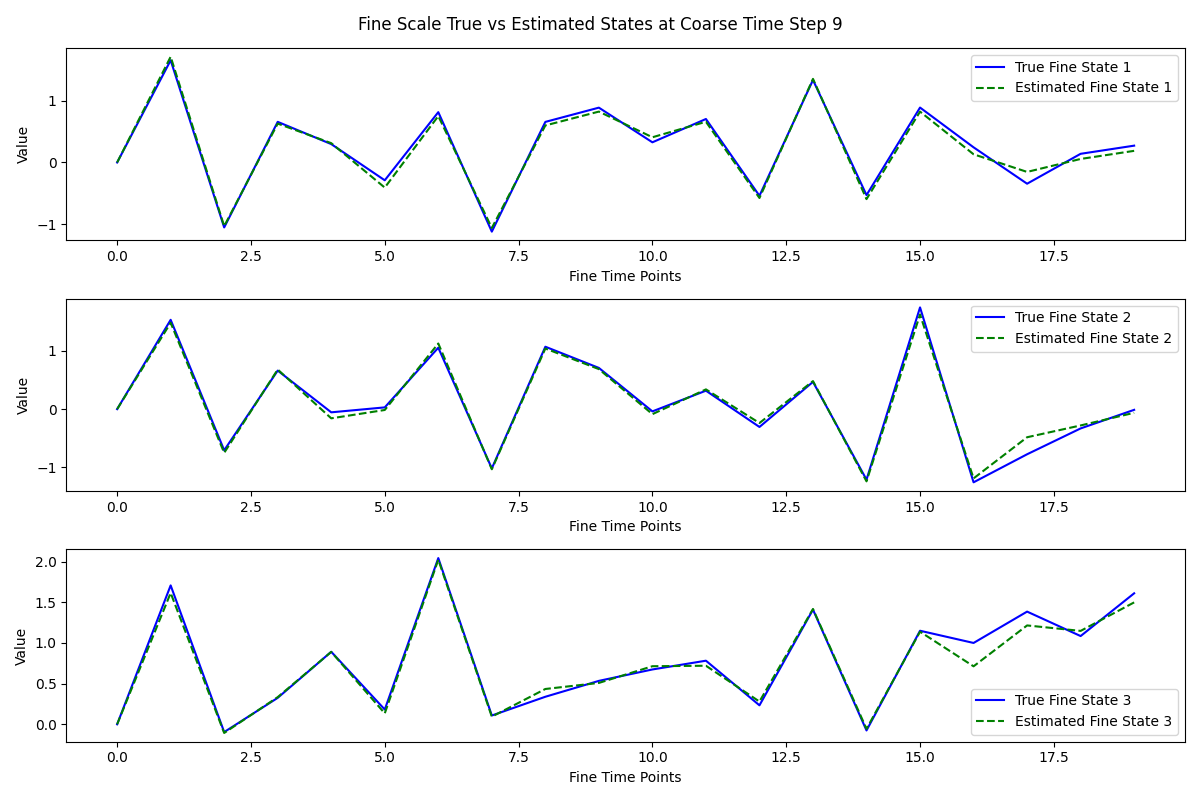}
        \caption{Individual $d=2$}
        \label{fig:d2fine}
    \end{subfigure}
    \hfill
    \begin{subfigure}[b]{0.45\textwidth}
        \centering
        \includegraphics[width=\textwidth]{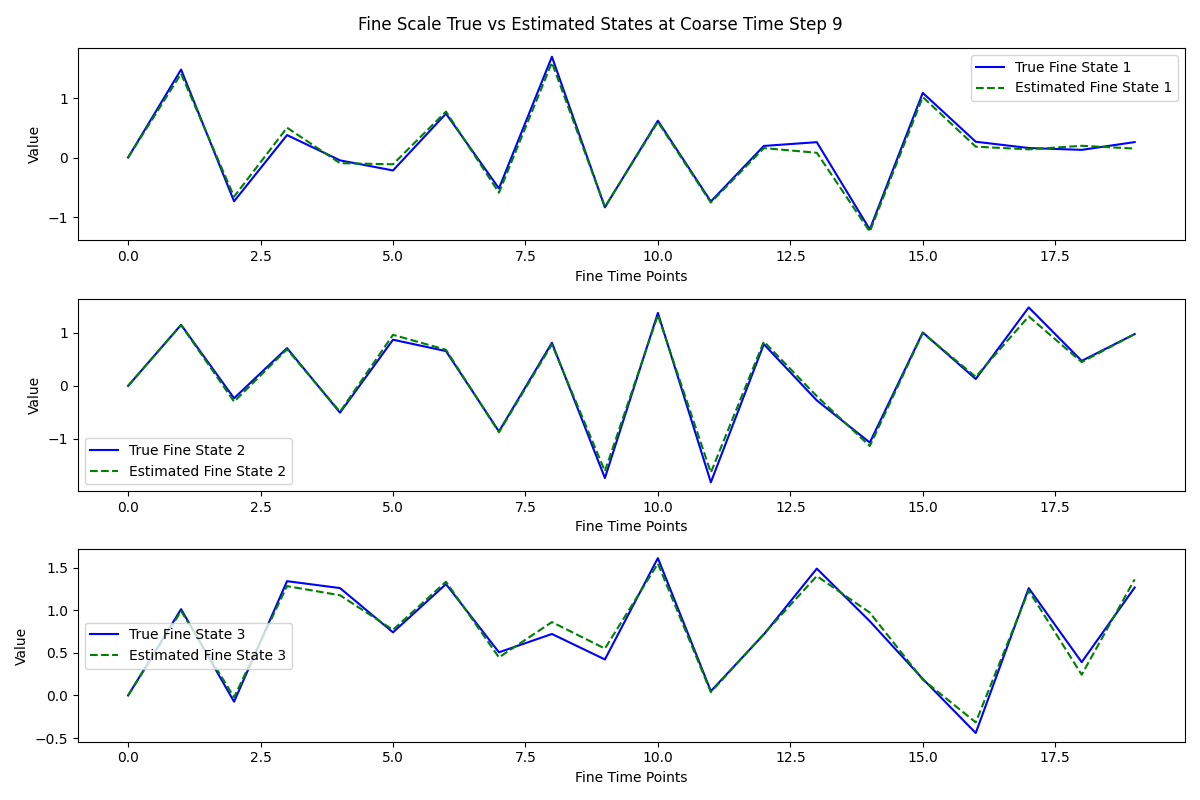}
        \caption{Individual $d=3$}
        \label{fig:d3fine}
    \end{subfigure}

    \caption{True vs. estimated fine time scale trajectories at coarse time step $t=9$ for individuals $d=0, d=1, d=2$, and $d=3$.}
    \label{fig:fine_trajectories}
\end{figure}
\begin{figure}[h!]
    \centering
    \begin{subfigure}[b]{0.45\textwidth}
        \centering
        \includegraphics[width=\textwidth]{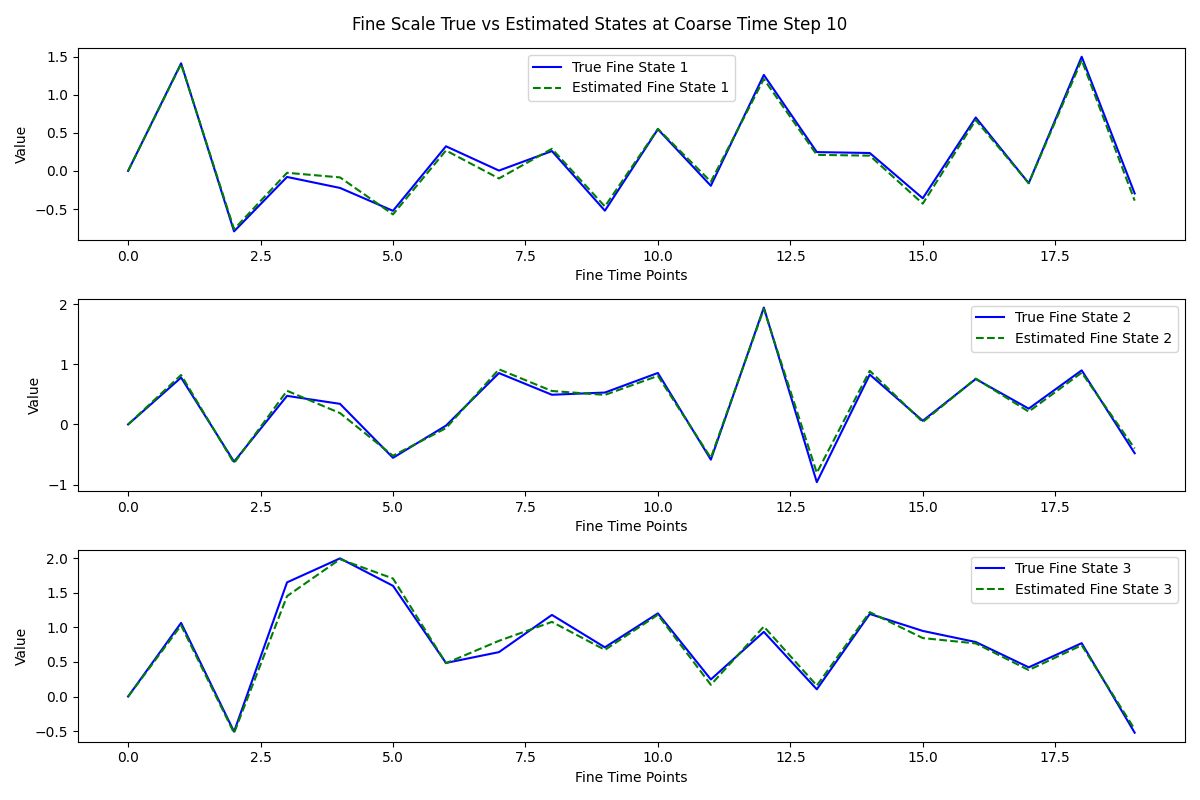}
        \caption{Individual $d=0$}
        \label{fig:d0fine}
    \end{subfigure}
    \hfill
    \begin{subfigure}[b]{0.45\textwidth}
        \centering
        \includegraphics[width=\textwidth]{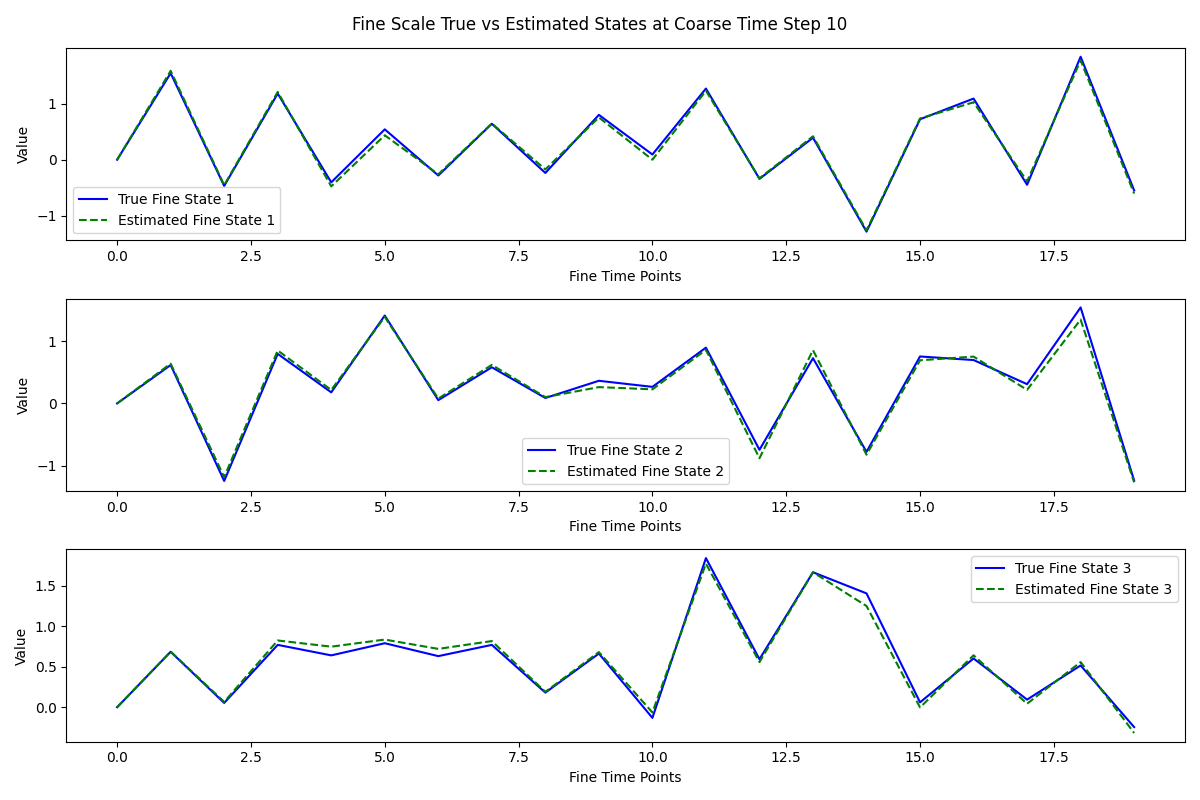}
        \caption{Individual $d=1$}
        \label{fig:d1fine}
    \end{subfigure}

    \vspace{0.3cm} 

    \begin{subfigure}[b]{0.45\textwidth}
        \centering
        \includegraphics[width=\textwidth]{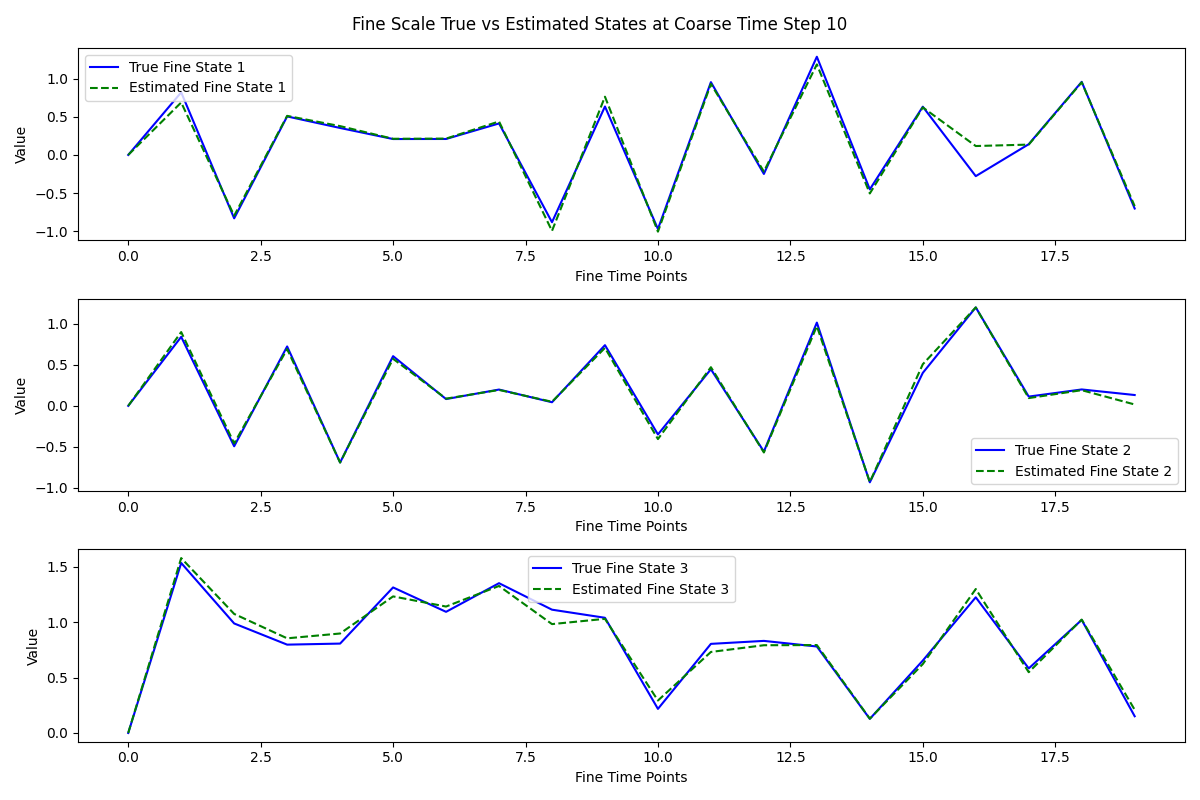}
        \caption{Individual $d=2$}
        \label{fig:d2fine}
    \end{subfigure}
    \hfill
    \begin{subfigure}[b]{0.45\textwidth}
        \centering
        \includegraphics[width=\textwidth]{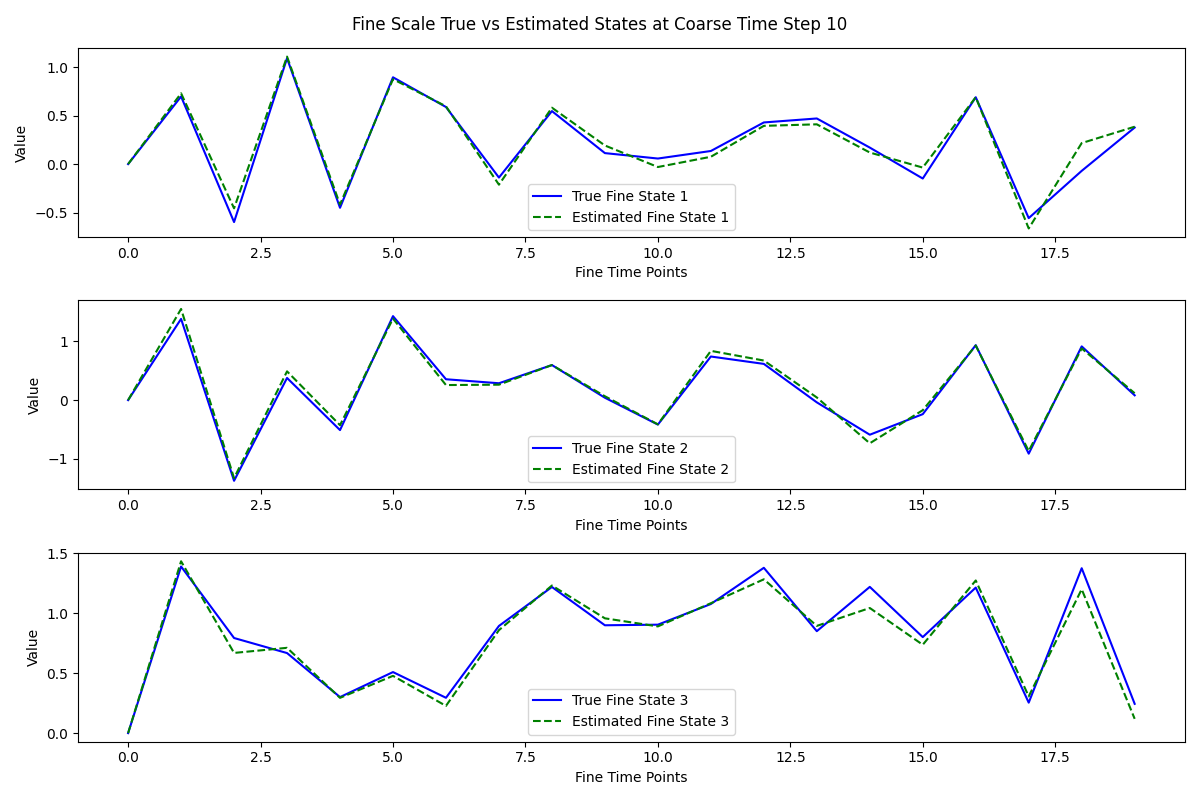}
        \caption{Individual $d=3$}
        \label{fig:d3fine}
    \end{subfigure}

    \caption{True vs. estimated fine time scale trajectories at coarse time step $t=10$ for individuals $d=0, d=1, d=2$, and $d=3$.}
    \label{fig:fine_trajectories}
\end{figure}
\begin{figure}[h!]
    \centering
    \begin{subfigure}[b]{0.45\textwidth}
        \centering
        \includegraphics[width=\textwidth]{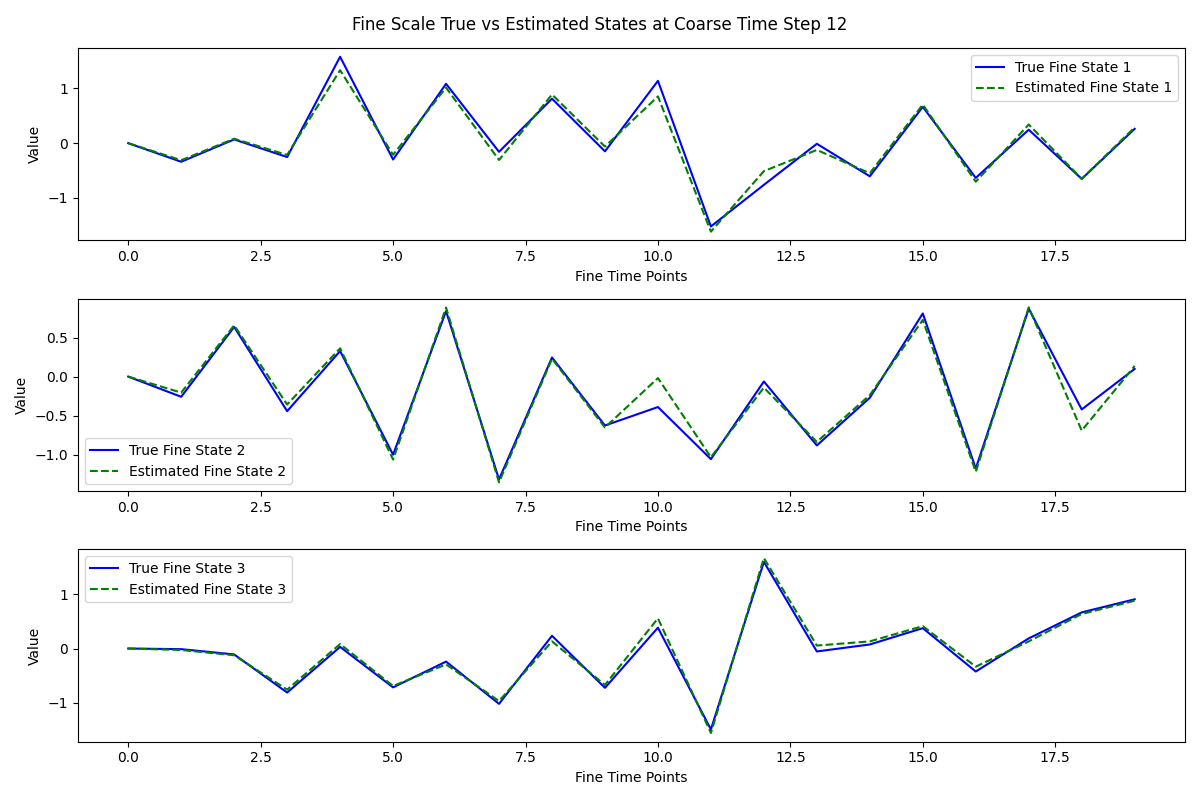}
        \caption{Individual $d=0$}
        \label{fig:d0fine}
    \end{subfigure}
    \hfill
    \begin{subfigure}[b]{0.45\textwidth}
        \centering
        \includegraphics[width=\textwidth]{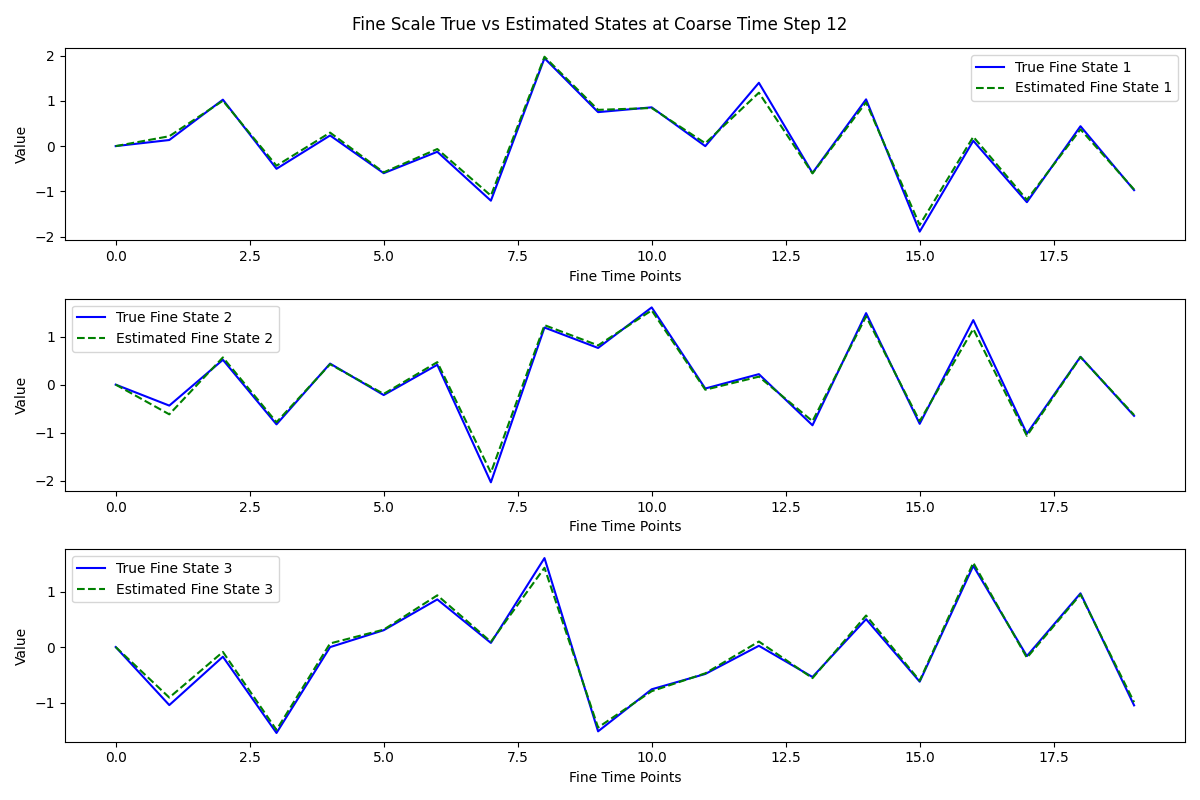}
        \caption{Individual $d=1$}
        \label{fig:d1fine}
    \end{subfigure}

    \vspace{0.3cm} 

    \begin{subfigure}[b]{0.45\textwidth}
        \centering
        \includegraphics[width=\textwidth]{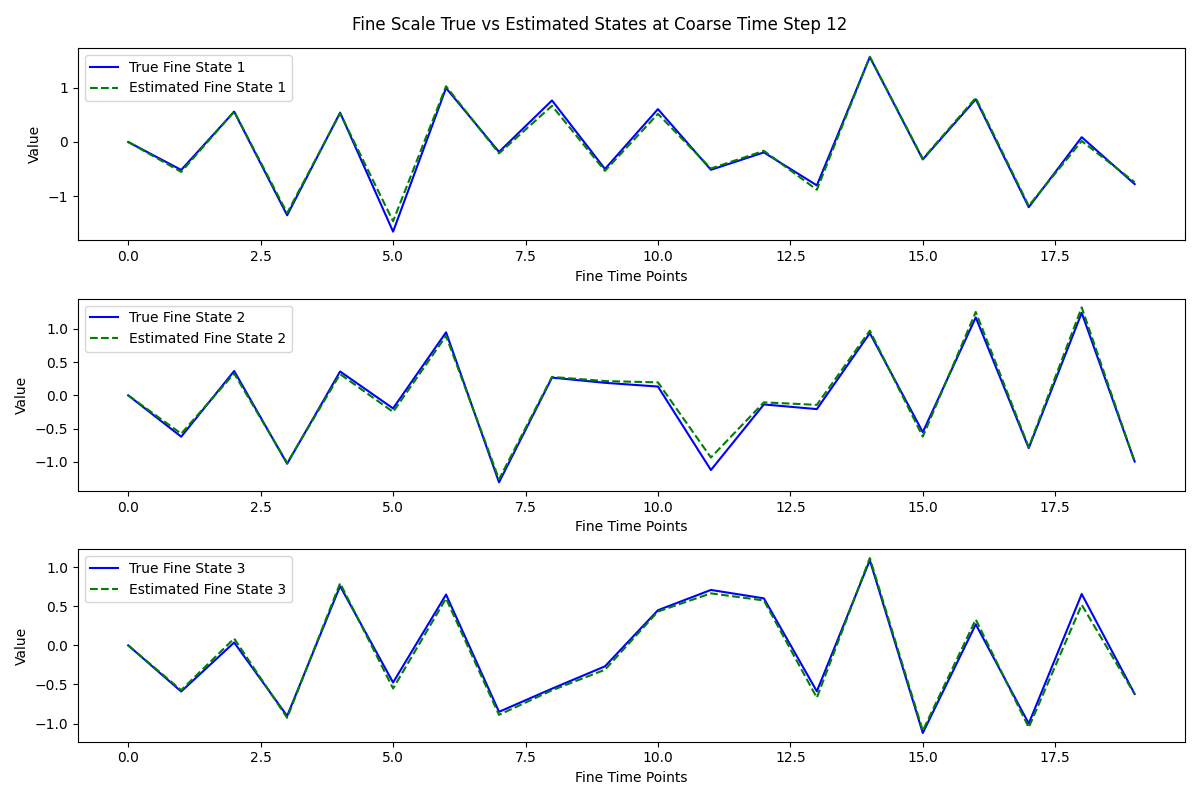}
        \caption{Individual $d=2$}
        \label{fig:d2fine}
    \end{subfigure}
    \hfill
    \begin{subfigure}[b]{0.45\textwidth}
        \centering
        \includegraphics[width=\textwidth]{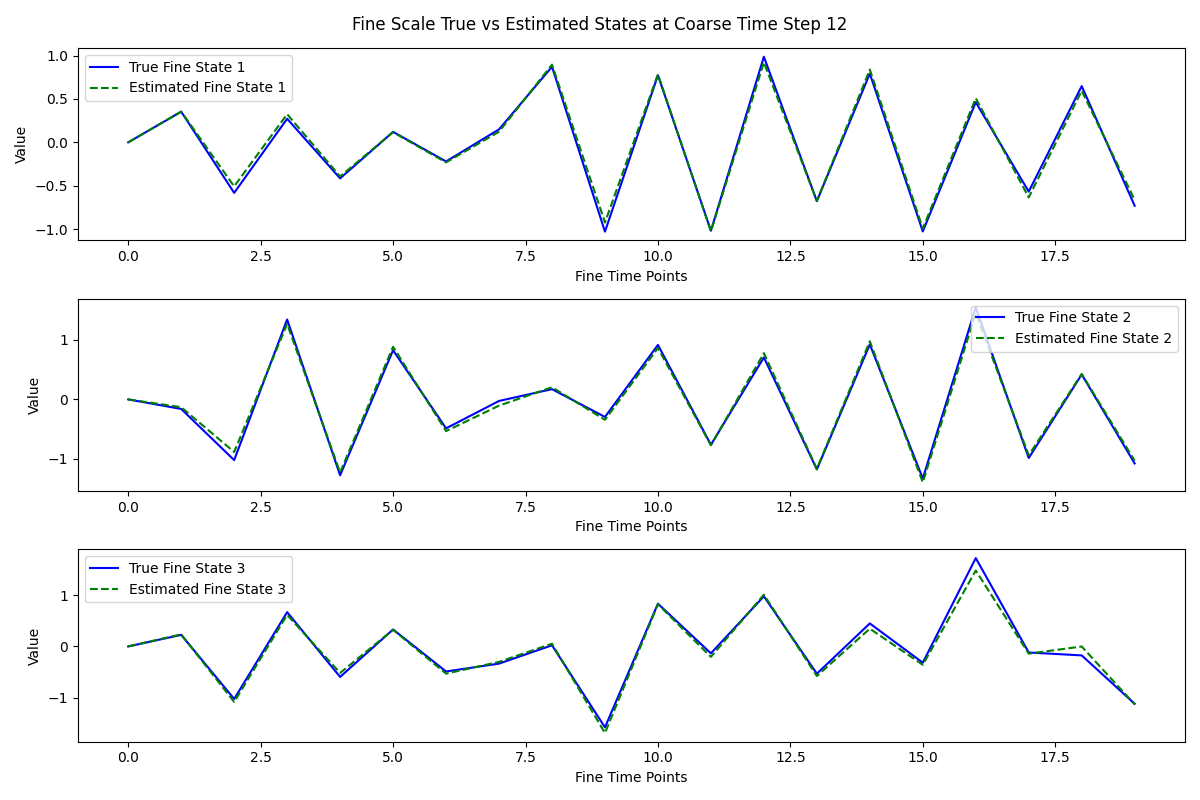}
        \caption{Individual $d=3$}
        \label{fig:d3fine}
    \end{subfigure}

    \caption{True vs. estimated fine time scale trajectories at coarse time step $t=12$ for individuals $d=0, d=1, d=2$, and $d=3$.}
    \label{fig:fine_trajectories}
\end{figure}
\begin{figure}[h!]
    \centering
    \begin{subfigure}[b]{0.45\textwidth}
        \centering
        \includegraphics[width=\textwidth]{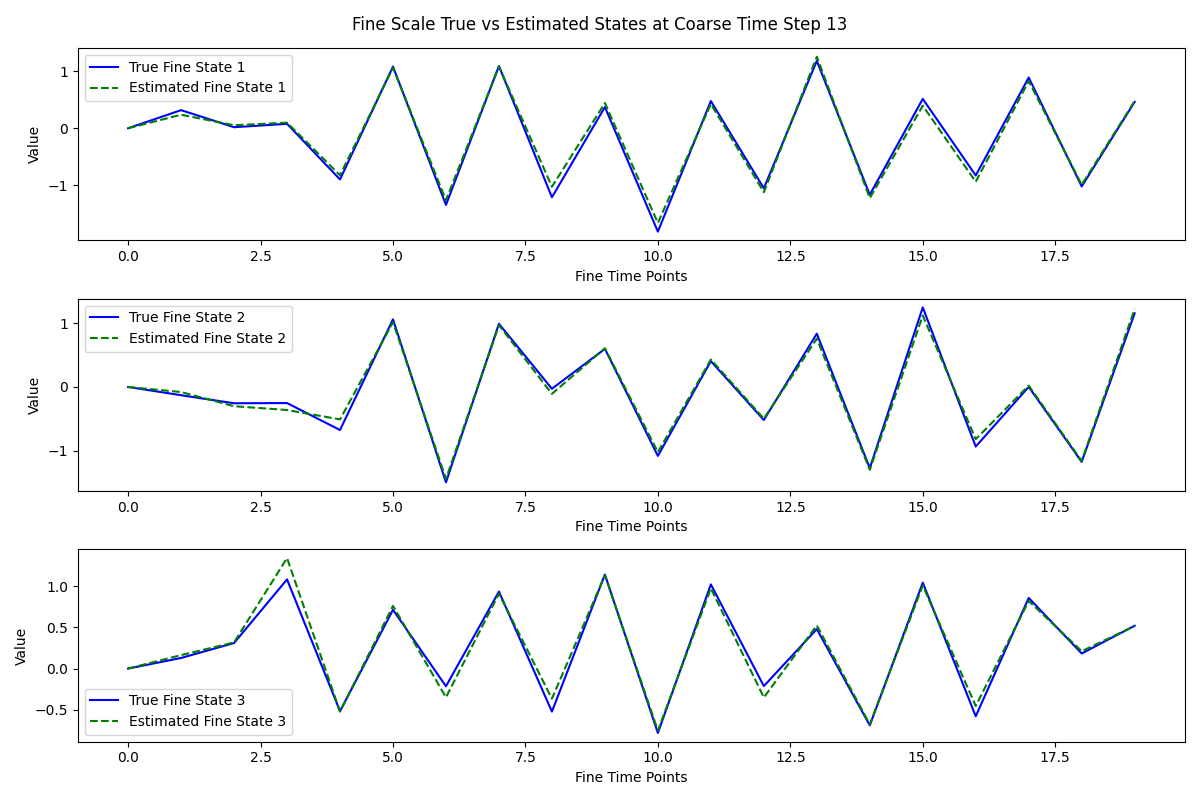}
        \caption{Individual $d=0$}
        \label{fig:d0fine}
    \end{subfigure}
    \hfill
    \begin{subfigure}[b]{0.45\textwidth}
        \centering
        \includegraphics[width=\textwidth]{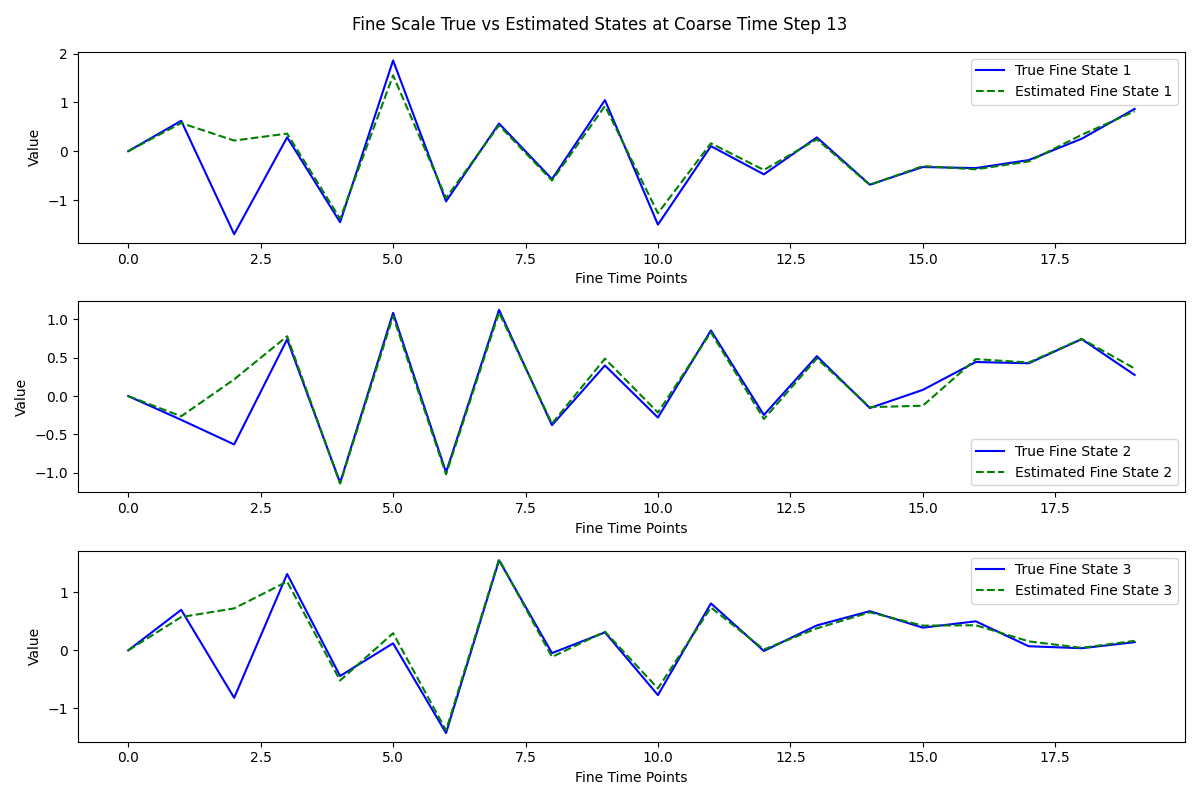}
        \caption{Individual $d=1$}
        \label{fig:d1fine}
    \end{subfigure}

    \vspace{0.3cm} 

    \begin{subfigure}[b]{0.45\textwidth}
        \centering
        \includegraphics[width=\textwidth]{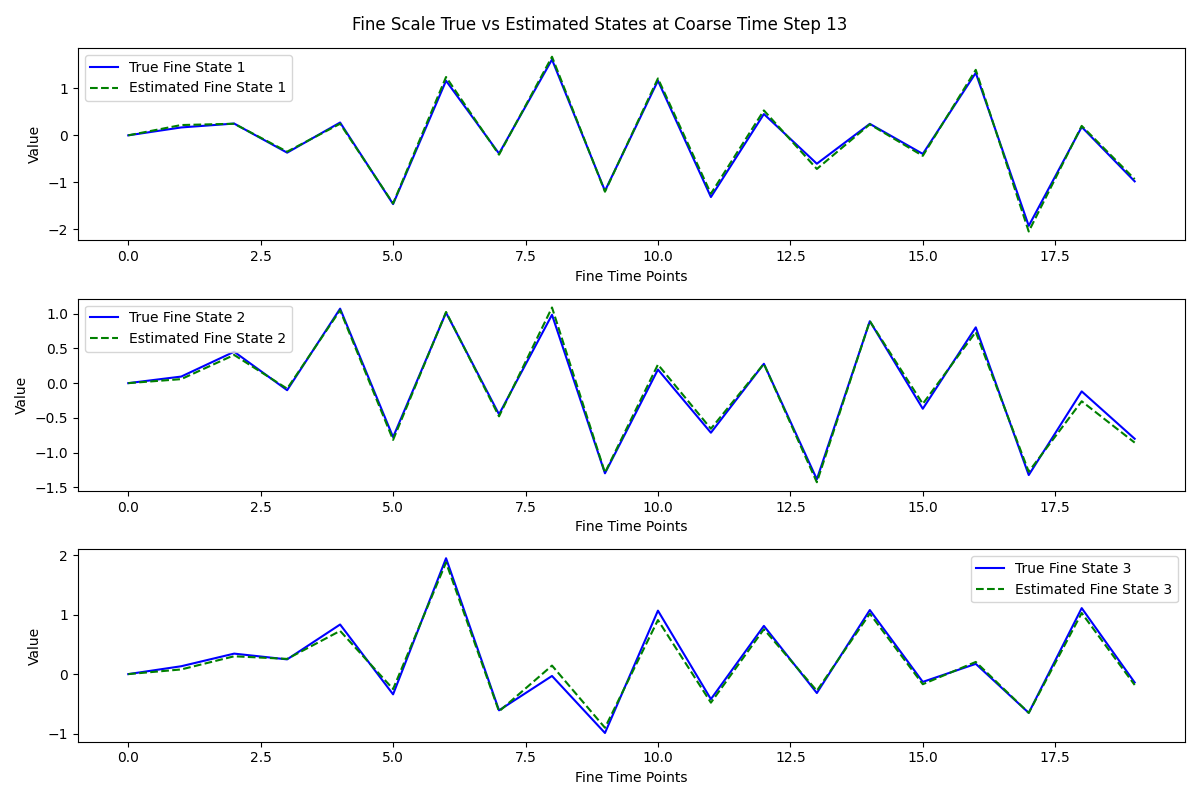}
        \caption{Individual $d=2$}
        \label{fig:d2fine}
    \end{subfigure}
    \hfill
    \begin{subfigure}[b]{0.45\textwidth}
        \centering
        \includegraphics[width=\textwidth]{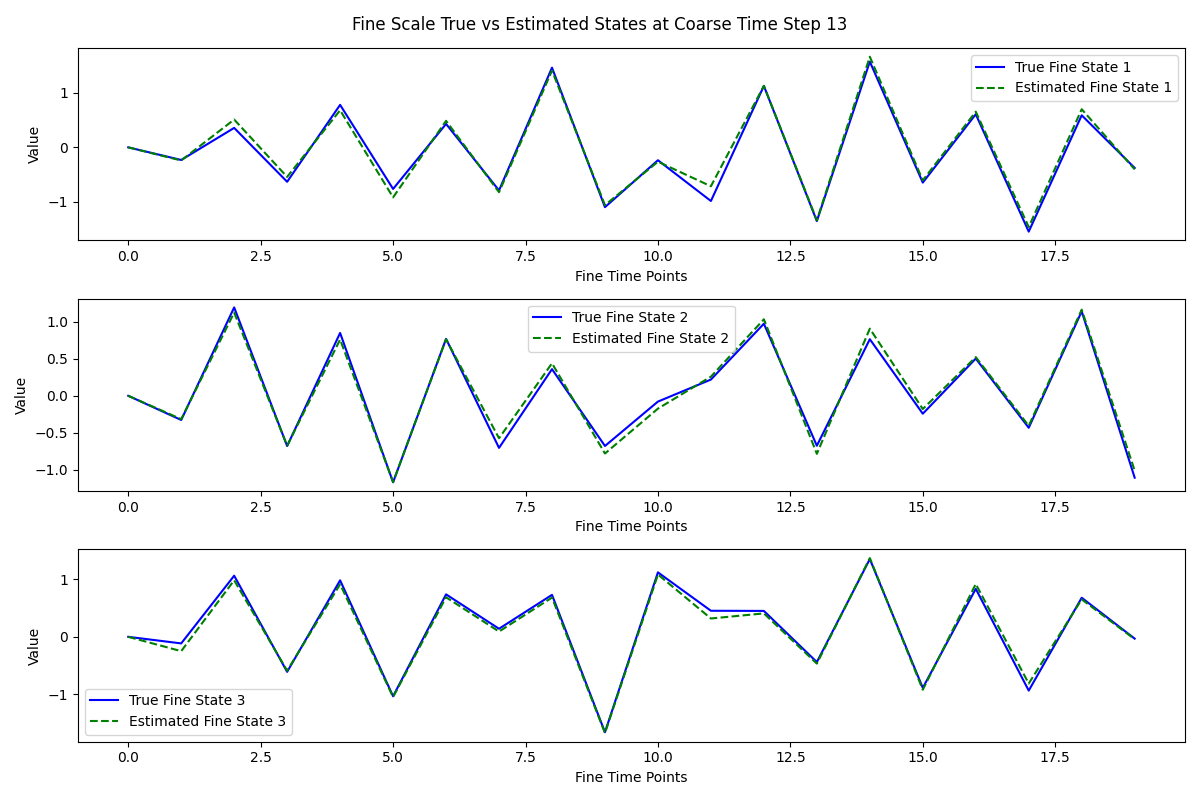}
        \caption{Individual $d=3$}
        \label{fig:d3fine}
    \end{subfigure}

    \caption{True vs. estimated fine time scale trajectories at coarse time step $t=13$ for individuals $d=0, d=1, d=2$, and $d=3$.}
    \label{fig:fine_trajectories}
\end{figure}
\begin{figure}[h!]
    \centering
    \begin{subfigure}[b]{0.45\textwidth}
        \centering
        \includegraphics[width=\textwidth]{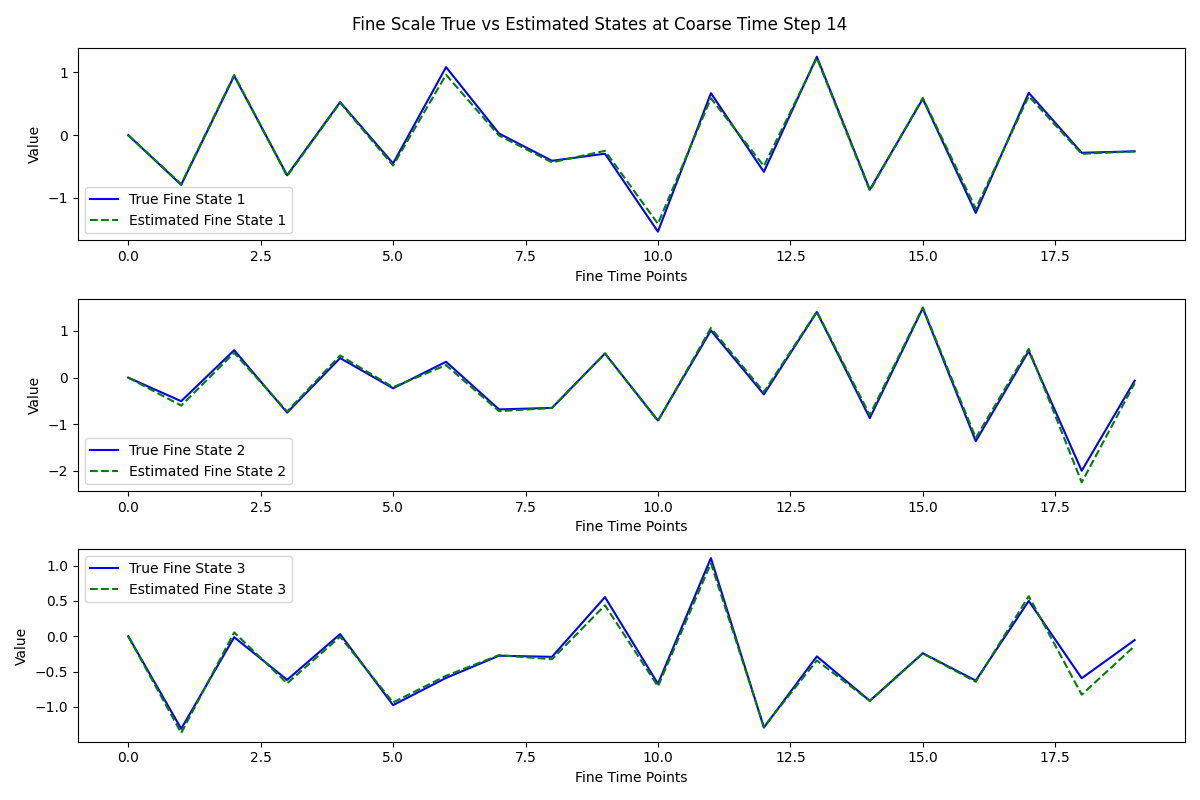}
        \caption{Individual $d=0$}
        \label{fig:d0fine}
    \end{subfigure}
    \hfill
    \begin{subfigure}[b]{0.45\textwidth}
        \centering
        \includegraphics[width=\textwidth]{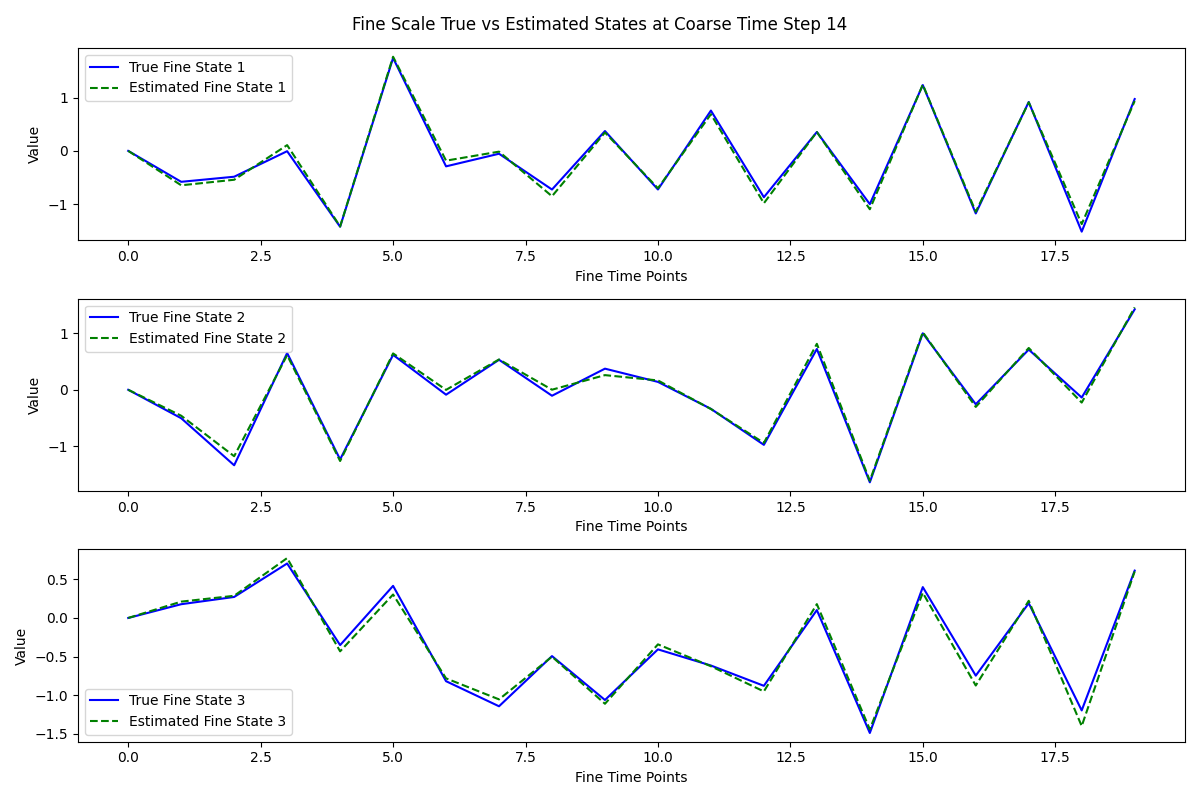}
        \caption{Individual $d=1$}
        \label{fig:d1fine}
    \end{subfigure}

    \vspace{0.3cm} 

    \begin{subfigure}[b]{0.45\textwidth}
        \centering
        \includegraphics[width=\textwidth]{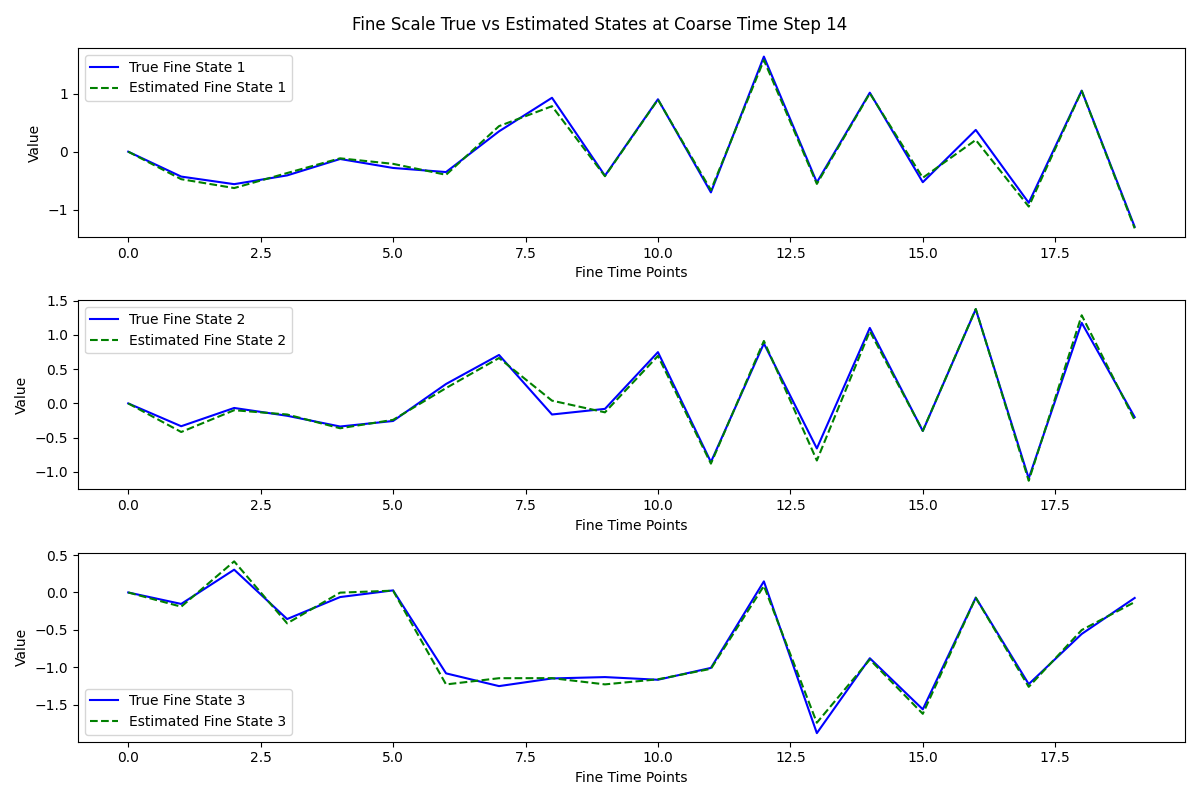}
        \caption{Individual $d=2$}
        \label{fig:d2fine}
    \end{subfigure}
    \hfill
    \begin{subfigure}[b]{0.45\textwidth}
        \centering
        \includegraphics[width=\textwidth]{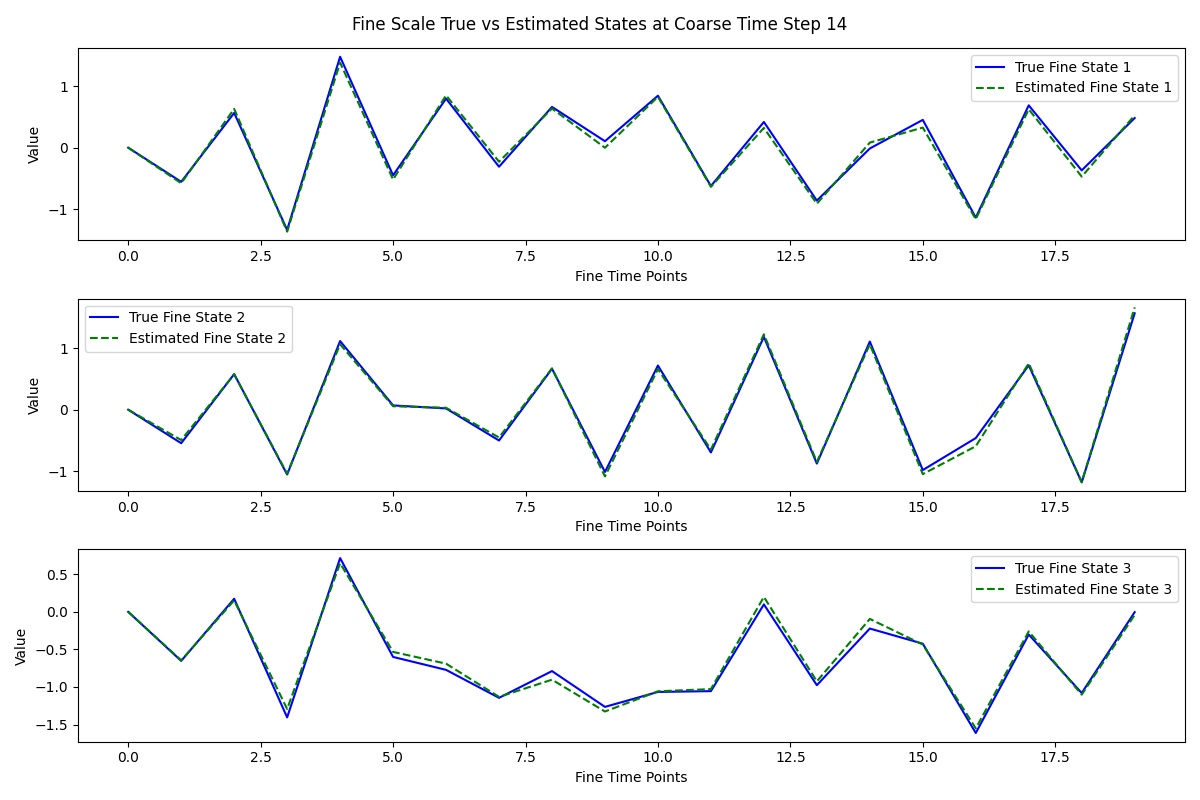}
        \caption{Individual $d=3$}
        \label{fig:d3fine}
    \end{subfigure}

    \caption{True vs. estimated fine time scale trajectories at coarse time step $t=14$ for individuals $d=0, d=1, d=2$, and $d=3$.}
    \label{fig:fine_trajectories}
\end{figure}
\begin{figure}[h!]
    \centering
    \begin{subfigure}[b]{0.45\textwidth}
        \centering
        \includegraphics[width=\textwidth]{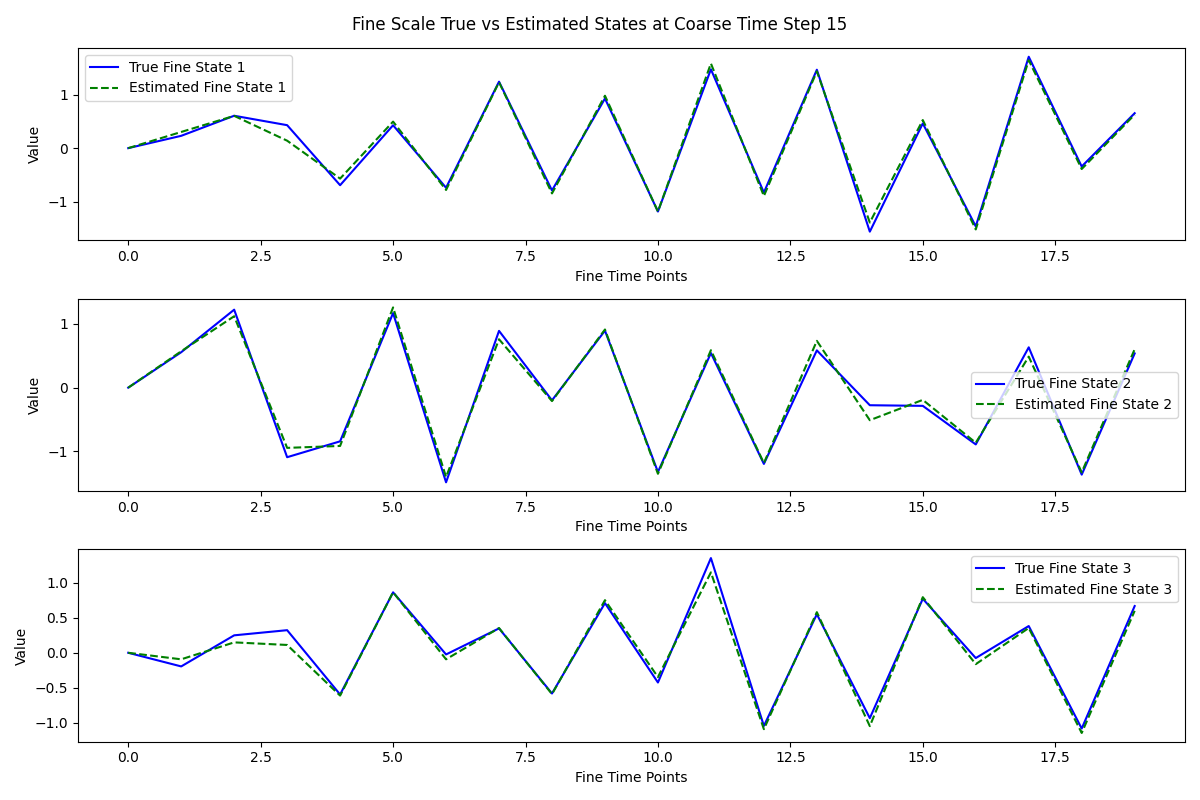}
        \caption{Individual $d=0$}
        \label{fig:d0fine}
    \end{subfigure}
    \hfill
    \begin{subfigure}[b]{0.45\textwidth}
        \centering
        \includegraphics[width=\textwidth]{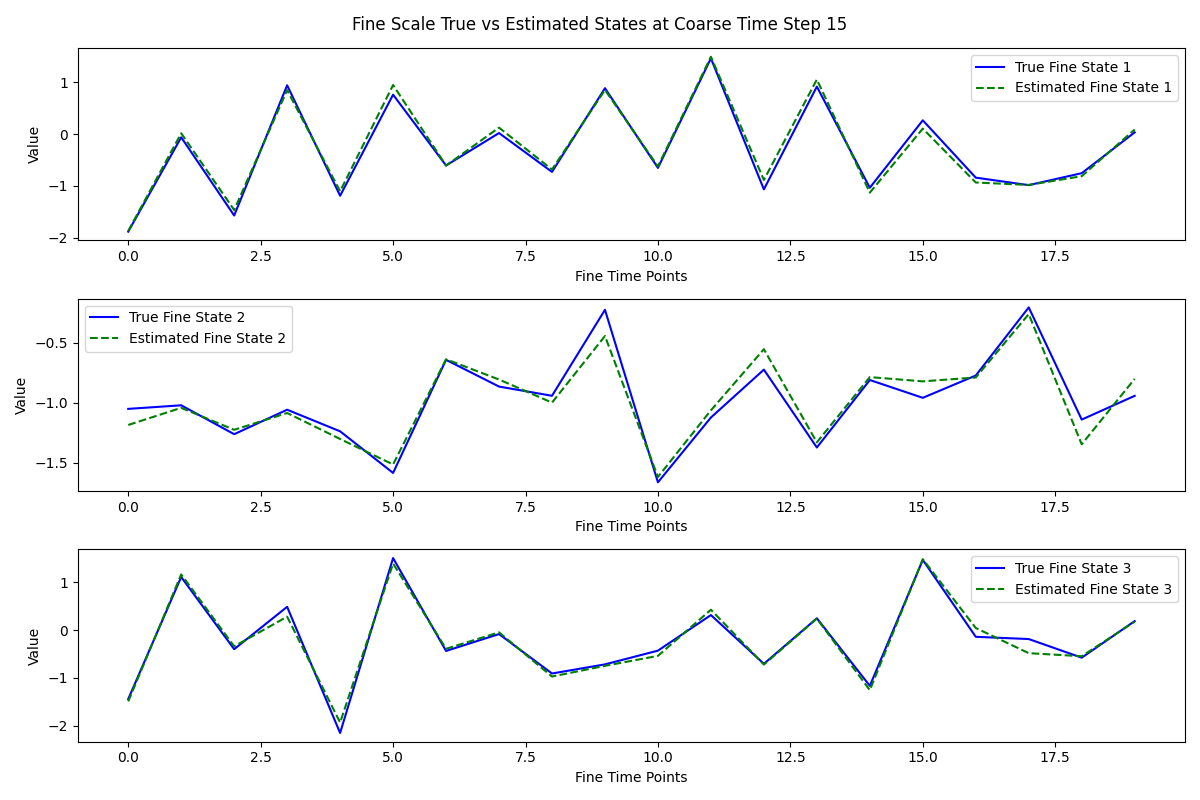}
        \caption{Individual $d=1$}
        \label{fig:d1fine}
    \end{subfigure}

    \vspace{0.3cm} 

    \begin{subfigure}[b]{0.45\textwidth}
        \centering
        \includegraphics[width=\textwidth]{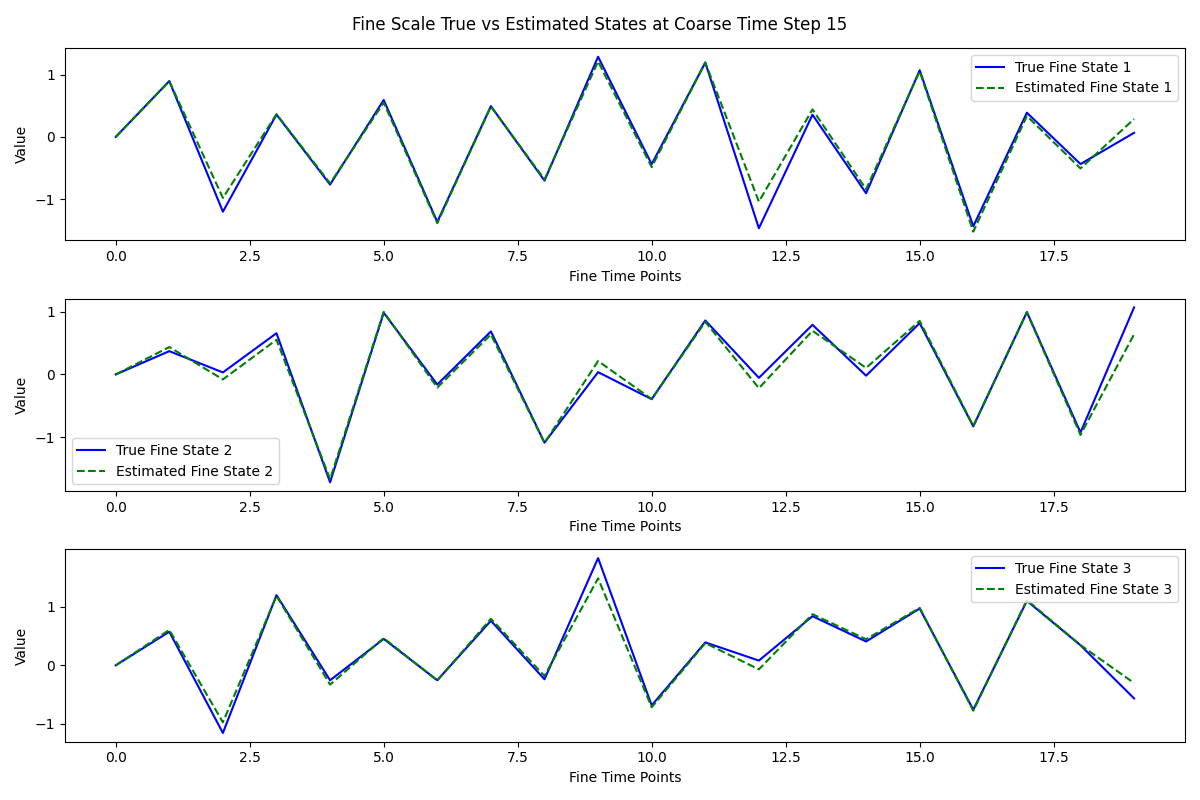}
        \caption{Individual $d=2$}
        \label{fig:d2fine}
    \end{subfigure}
    \hfill
    \begin{subfigure}[b]{0.45\textwidth}
        \centering
        \includegraphics[width=\textwidth]{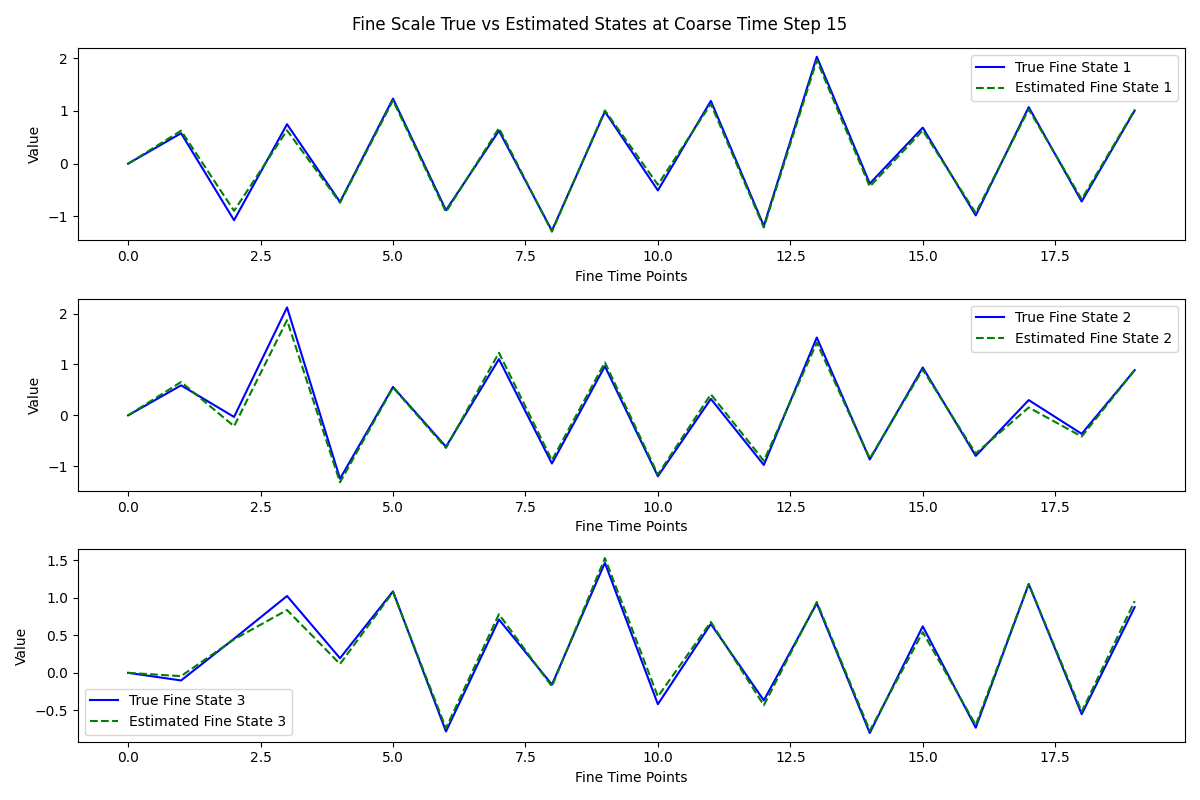}
        \caption{Individual $d=3$}
        \label{fig:d3fine}
    \end{subfigure}

    \caption{True vs. estimated fine time scale trajectories at coarse time step $t=15$ for individuals $d=0, d=1, d=2$, and $d=3$.}
    \label{fig:fine_trajectories}
\end{figure}
\begin{figure}[h!]
    \centering
    \begin{subfigure}[b]{0.45\textwidth}
        \centering
        \includegraphics[width=\textwidth]{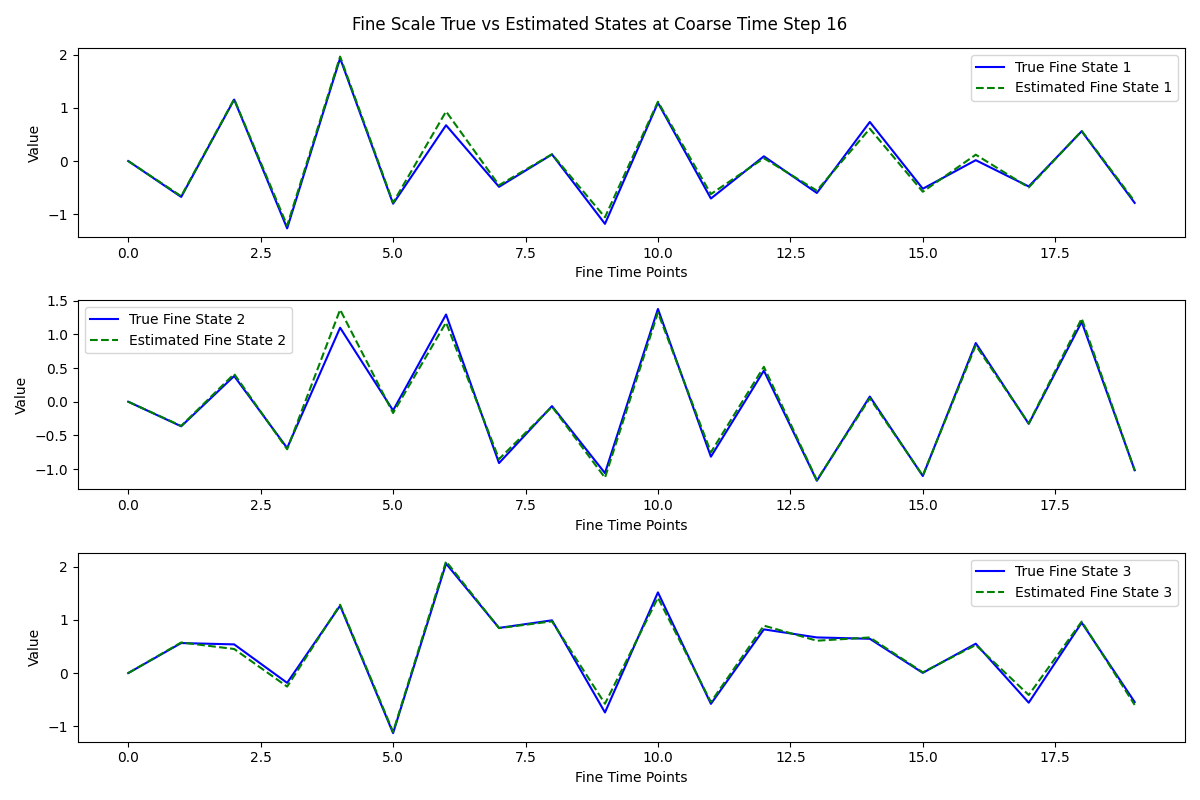}
        \caption{Individual $d=0$}
        \label{fig:d0fine}
    \end{subfigure}
    \hfill
    \begin{subfigure}[b]{0.45\textwidth}
        \centering
        \includegraphics[width=\textwidth]{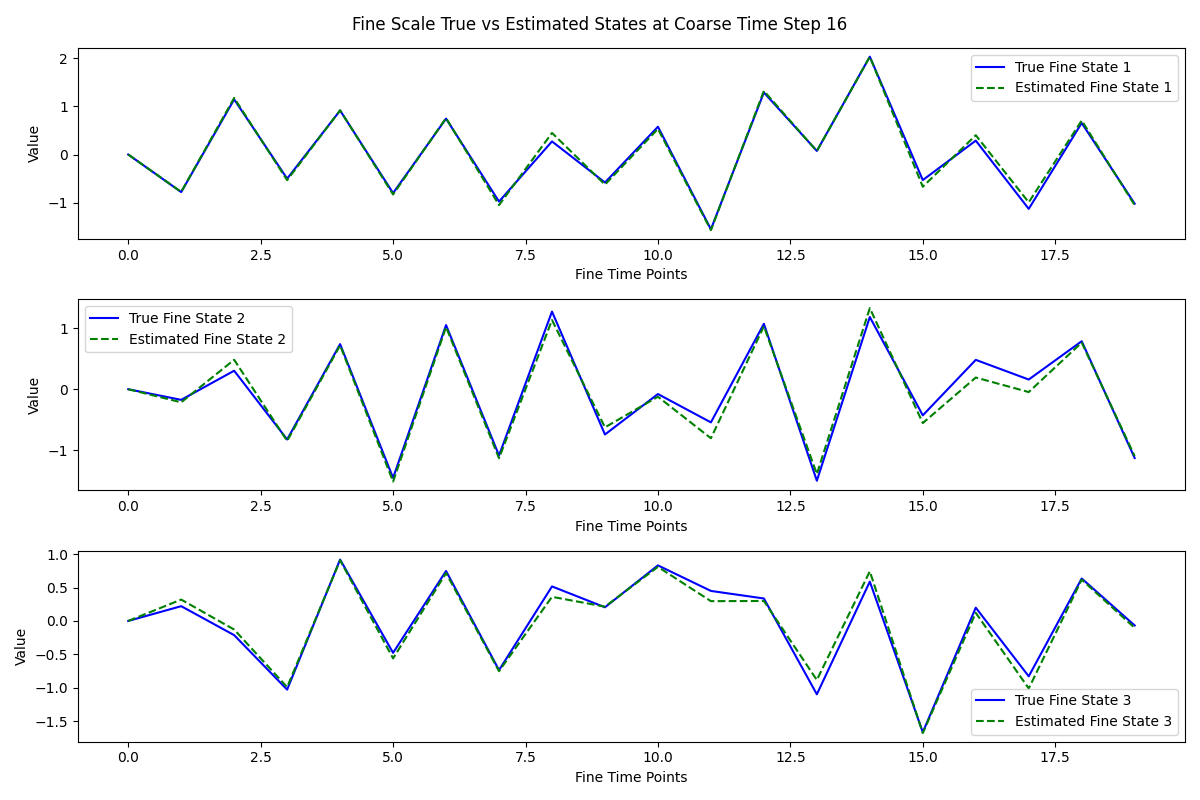}
        \caption{Individual $d=1$}
        \label{fig:d1fine}
    \end{subfigure}

    \vspace{0.3cm} 

    \begin{subfigure}[b]{0.45\textwidth}
        \centering
        \includegraphics[width=\textwidth]{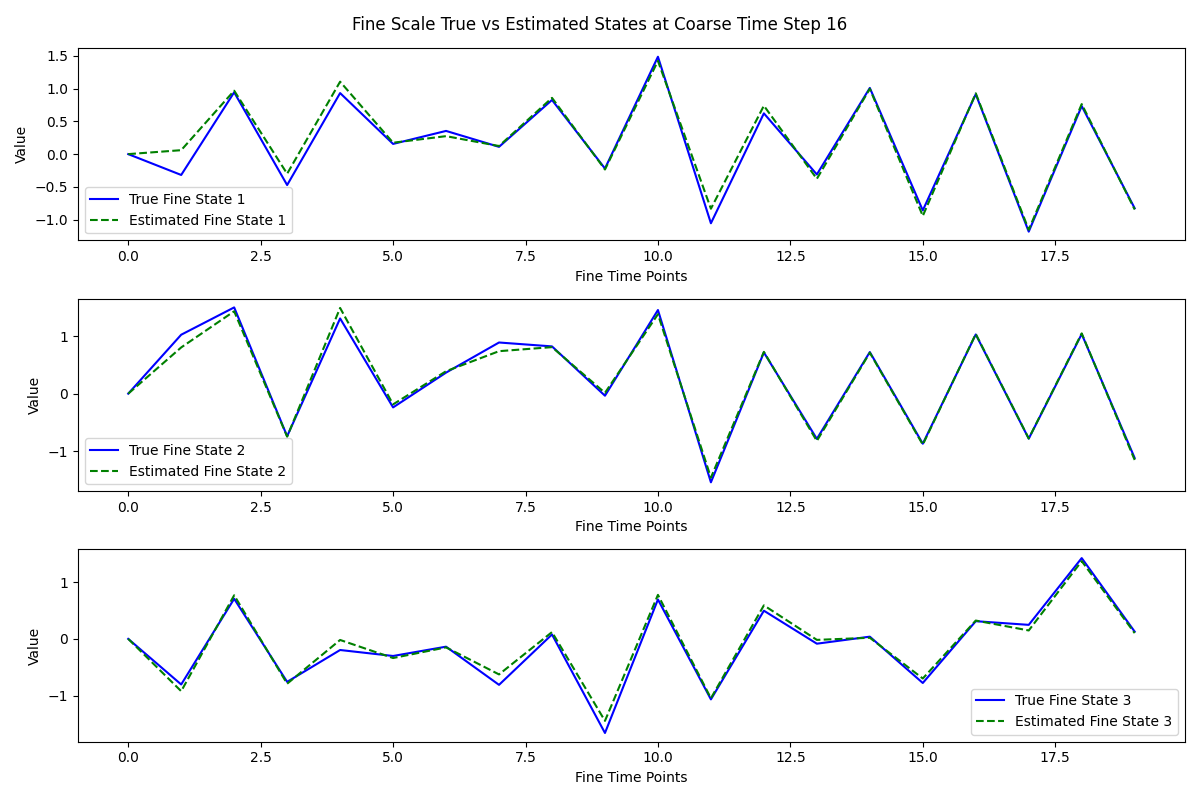}
        \caption{Individual $d=2$}
        \label{fig:d2fine}
    \end{subfigure}
    \hfill
    \begin{subfigure}[b]{0.45\textwidth}
        \centering
        \includegraphics[width=\textwidth]{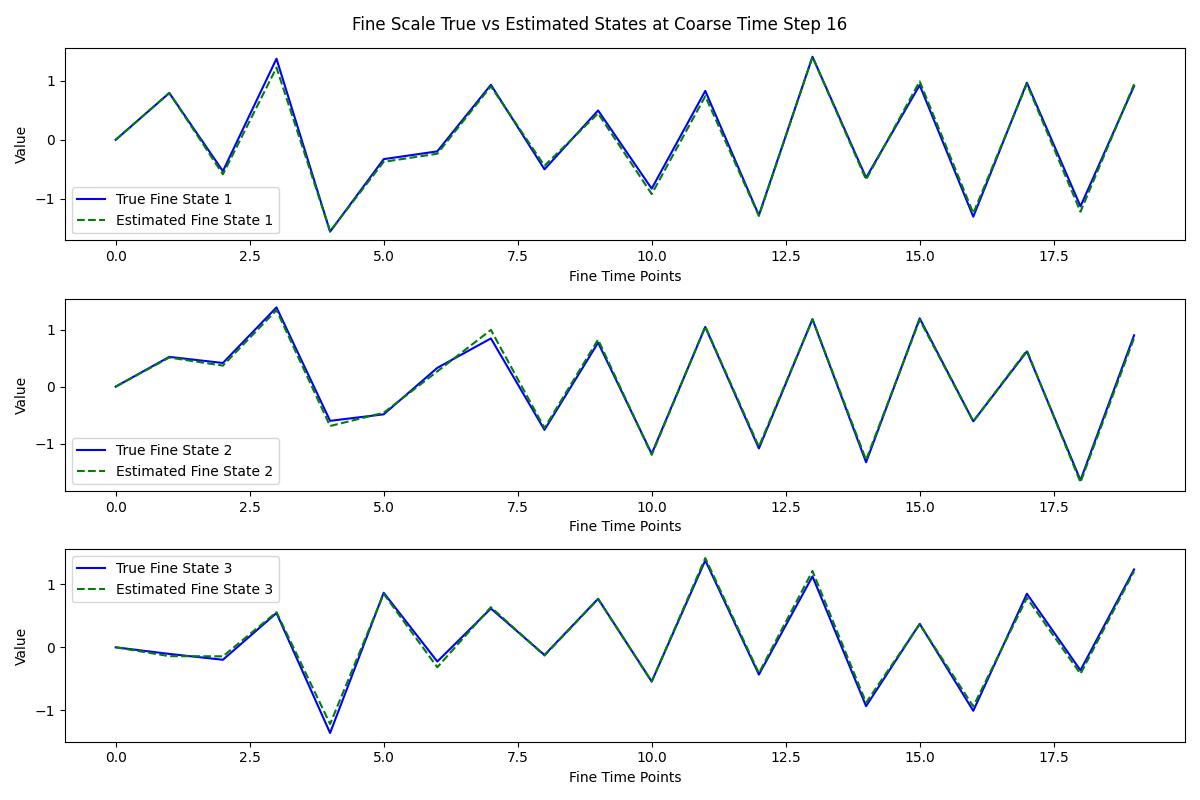}
        \caption{Individual $d=3$}
        \label{fig:d3fine}
    \end{subfigure}

    \caption{True vs. estimated fine time scale trajectories at coarse time step $t=16$ for individuals $d=0, d=1, d=2$, and $d=3$.}
    \label{fig:fine_trajectories}
\end{figure}
\begin{figure}[h!]
    \centering
    \begin{subfigure}[b]{0.45\textwidth}
        \centering
        \includegraphics[width=\textwidth]{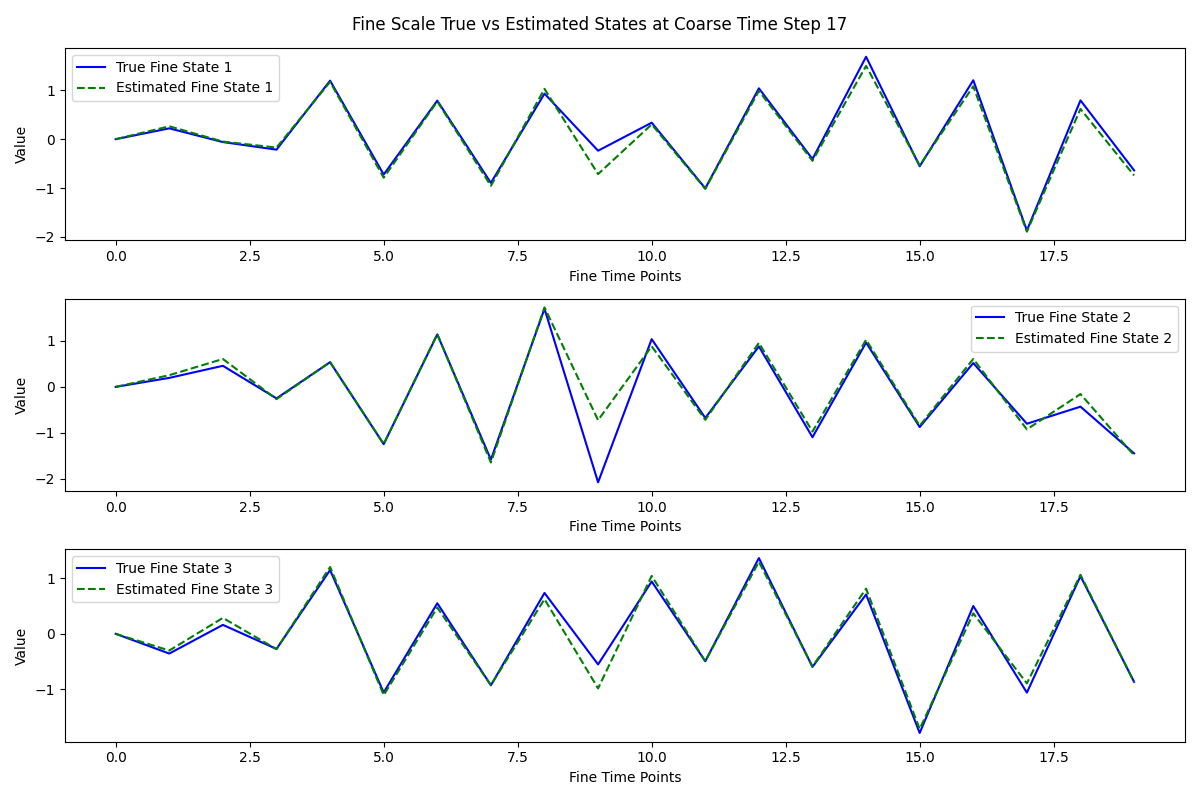}
        \caption{Individual $d=0$}
        \label{fig:d0fine}
    \end{subfigure}
    \hfill
    \begin{subfigure}[b]{0.45\textwidth}
        \centering
        \includegraphics[width=\textwidth]{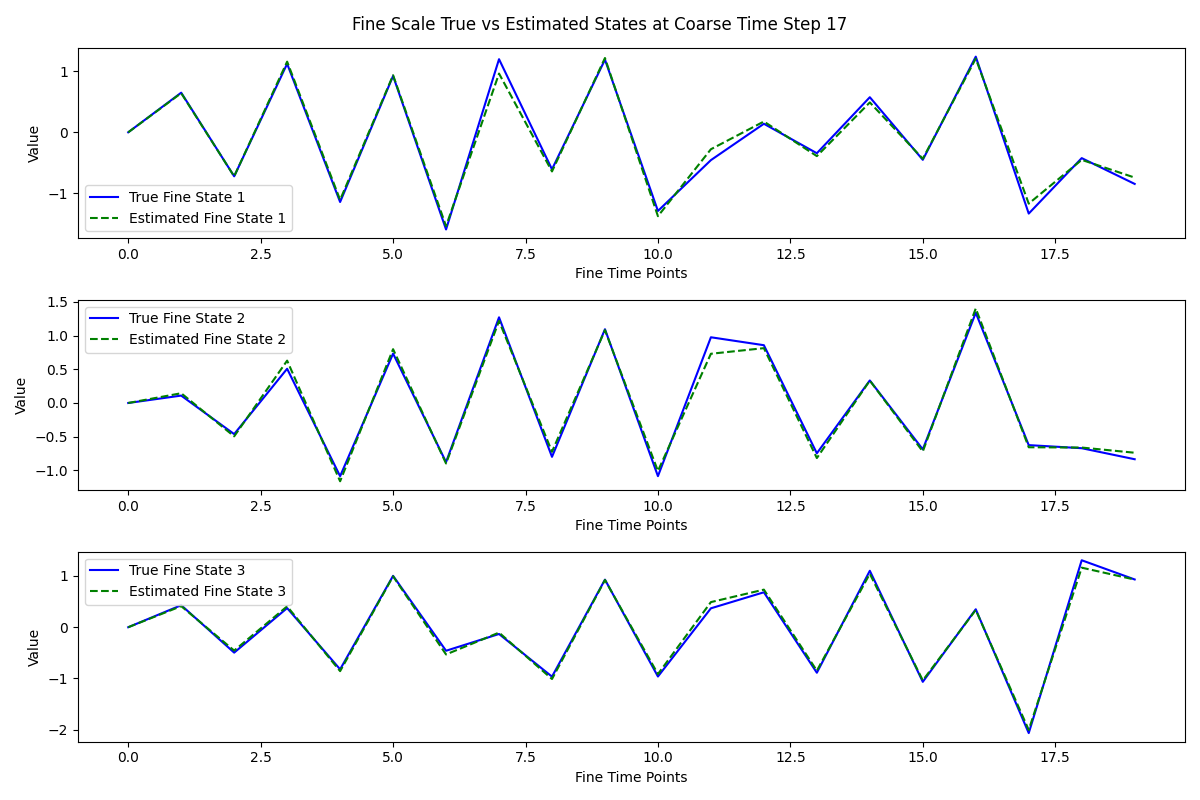}
        \caption{Individual $d=1$}
        \label{fig:d1fine}
    \end{subfigure}

    \vspace{0.3cm} 

    \begin{subfigure}[b]{0.45\textwidth}
        \centering
        \includegraphics[width=\textwidth]{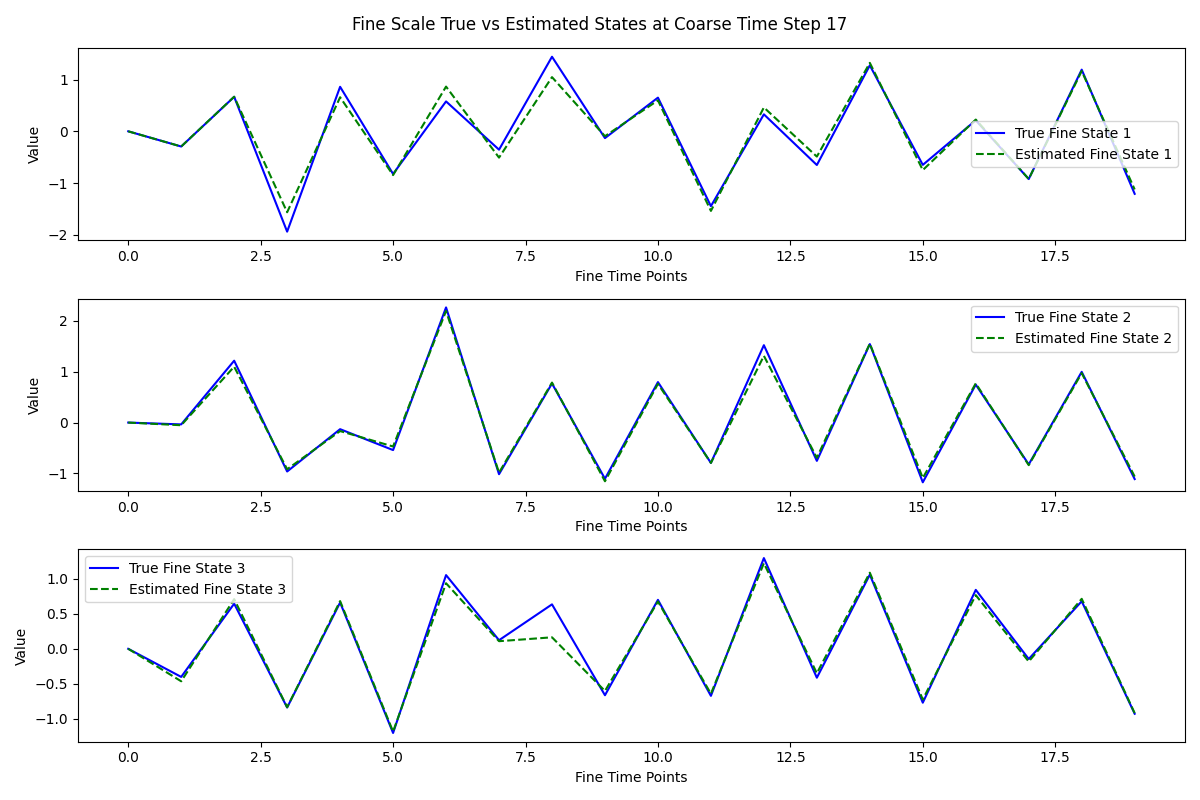}
        \caption{Individual $d=2$}
        \label{fig:d2fine}
    \end{subfigure}
    \hfill
    \begin{subfigure}[b]{0.45\textwidth}
        \centering
        \includegraphics[width=\textwidth]{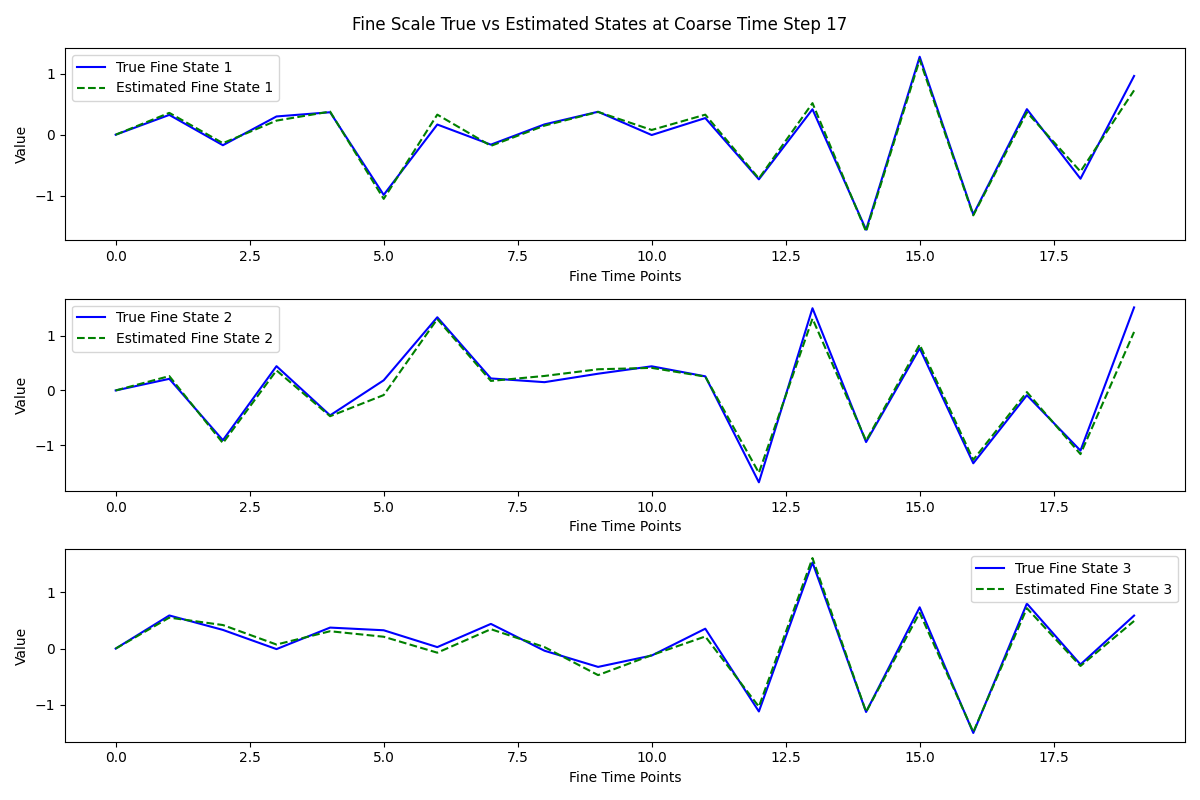}
        \caption{Individual $d=3$}
        \label{fig:d3fine}
    \end{subfigure}

    \caption{True vs. estimated fine time scale trajectories at coarse time step $t=17$ for individuals $d=0, d=1, d=2$, and $d=3$.}
    \label{fig:fine_trajectories}
\end{figure}
\begin{figure}[h!]
    \centering
    \begin{subfigure}[b]{0.45\textwidth}
        \centering
        \includegraphics[width=\textwidth]{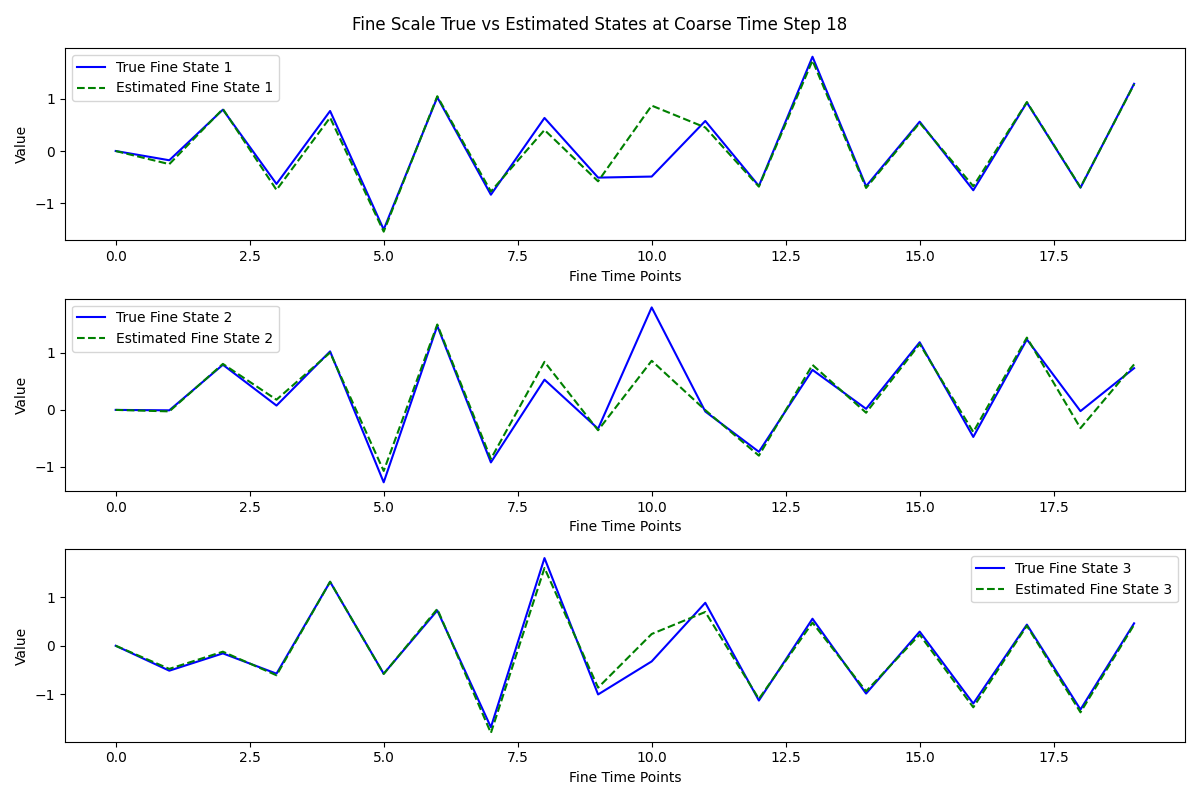}
        \caption{Individual $d=0$}
        \label{fig:d0fine}
    \end{subfigure}
    \hfill
    \begin{subfigure}[b]{0.45\textwidth}
        \centering
        \includegraphics[width=\textwidth]{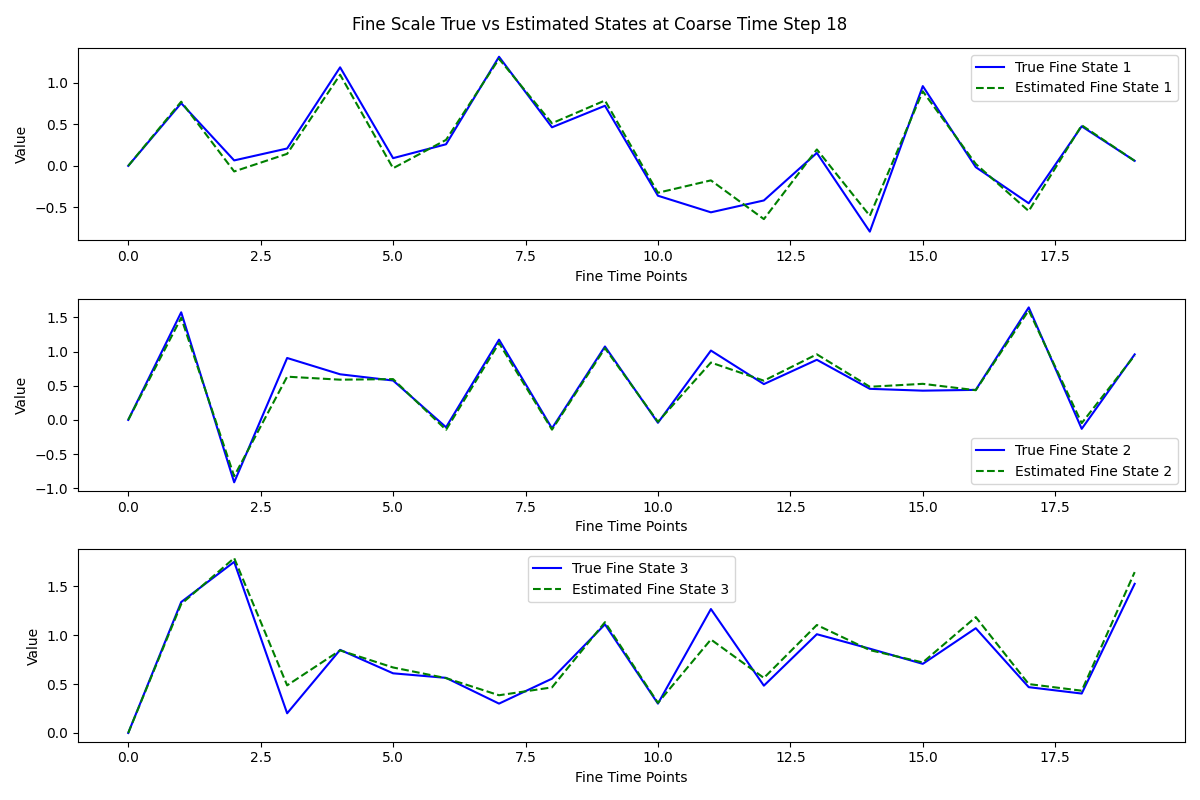}
        \caption{Individual $d=1$}
        \label{fig:d1fine}
    \end{subfigure}

    \vspace{0.3cm} 

    \begin{subfigure}[b]{0.45\textwidth}
        \centering
        \includegraphics[width=\textwidth]{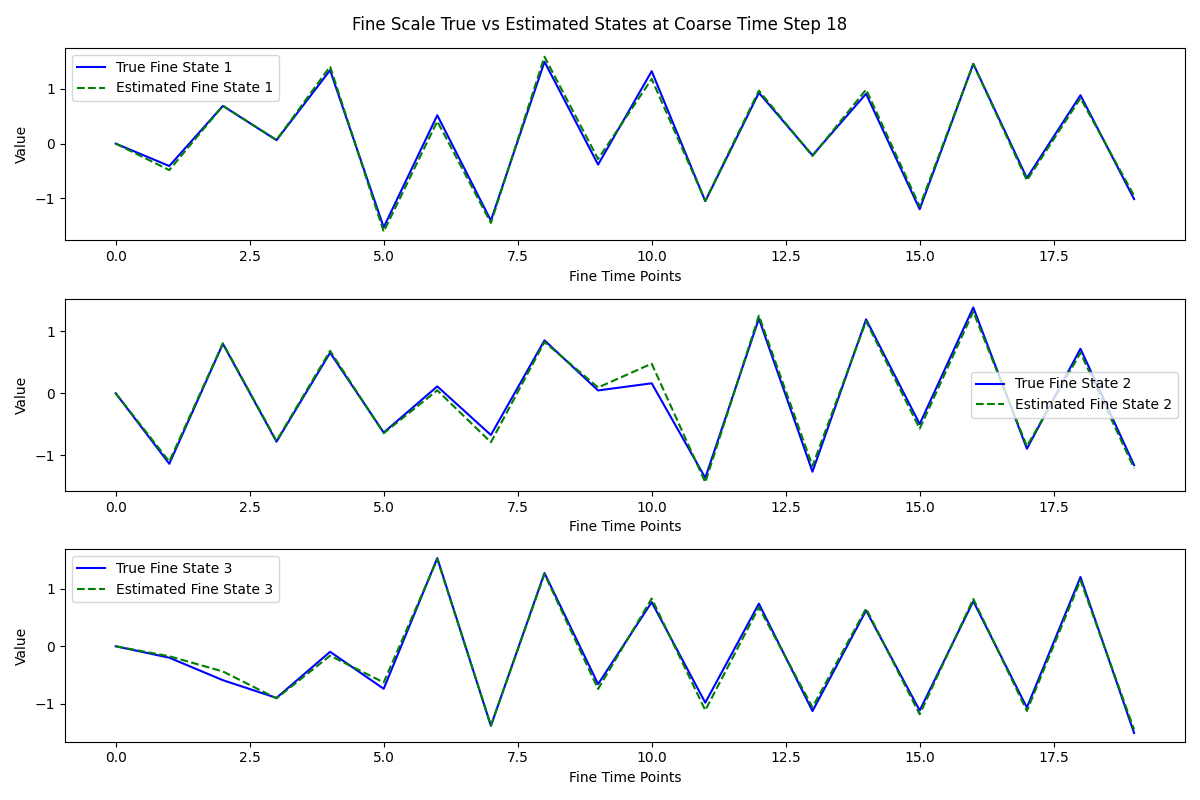}
        \caption{Individual $d=2$}
        \label{fig:d2fine}
    \end{subfigure}
    \hfill
    \begin{subfigure}[b]{0.45\textwidth}
        \centering
        \includegraphics[width=\textwidth]{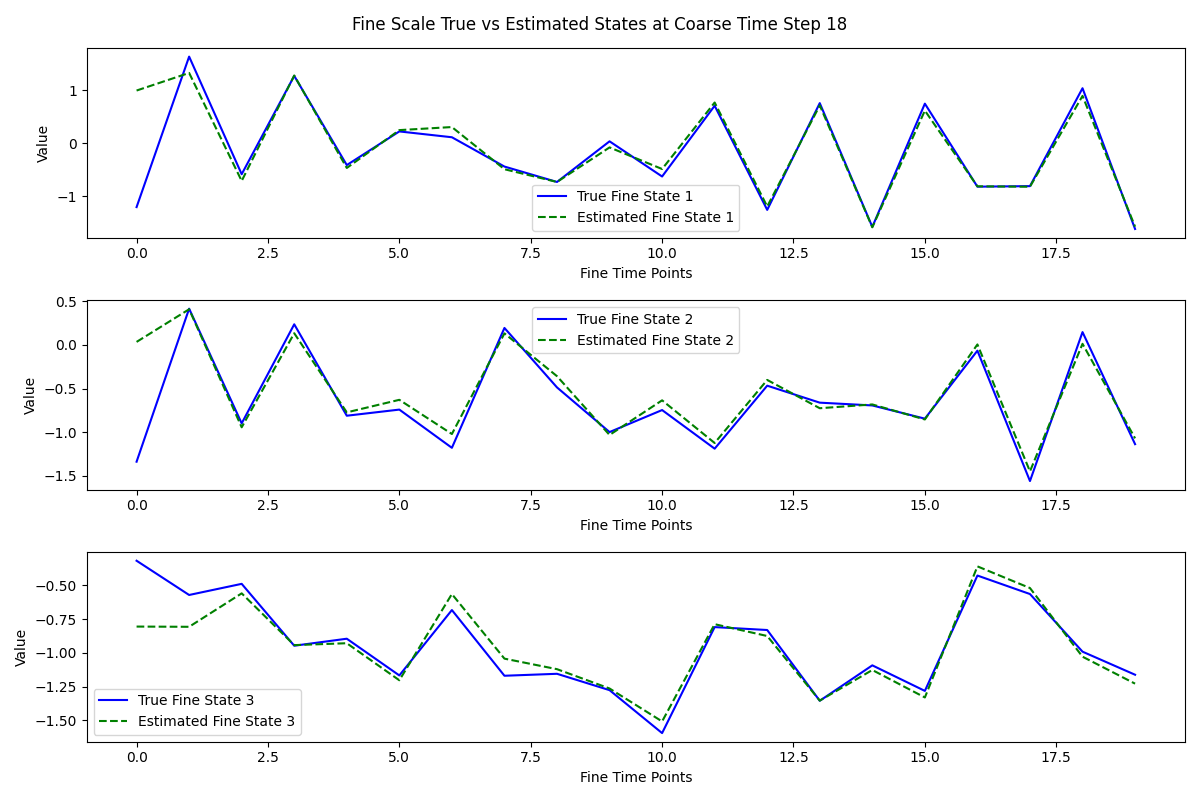}
        \caption{Individual $d=3$}
        \label{fig:d3fine}
    \end{subfigure}

    \caption{True vs. estimated fine time scale trajectories at coarse time step $t=18$ for individuals $d=0, d=1, d=2$, and $d=3$.}
    \label{fig:fine_trajectories}
\end{figure}
\begin{figure}[h!]
    \centering
    \begin{subfigure}[b]{0.45\textwidth}
        \centering
        \includegraphics[width=\textwidth]{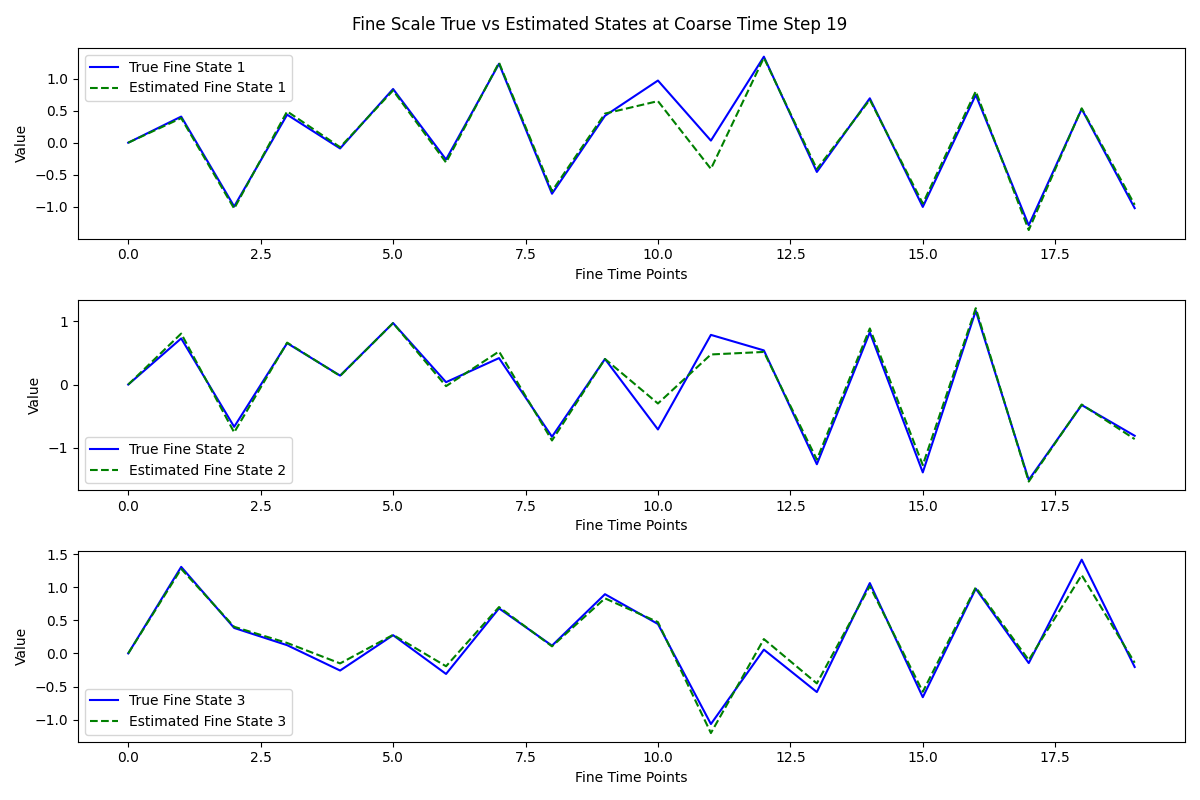}
        \caption{Individual $d=0$}
        \label{fig:d0fine}
    \end{subfigure}
    \hfill
    \begin{subfigure}[b]{0.45\textwidth}
        \centering
        \includegraphics[width=\textwidth]{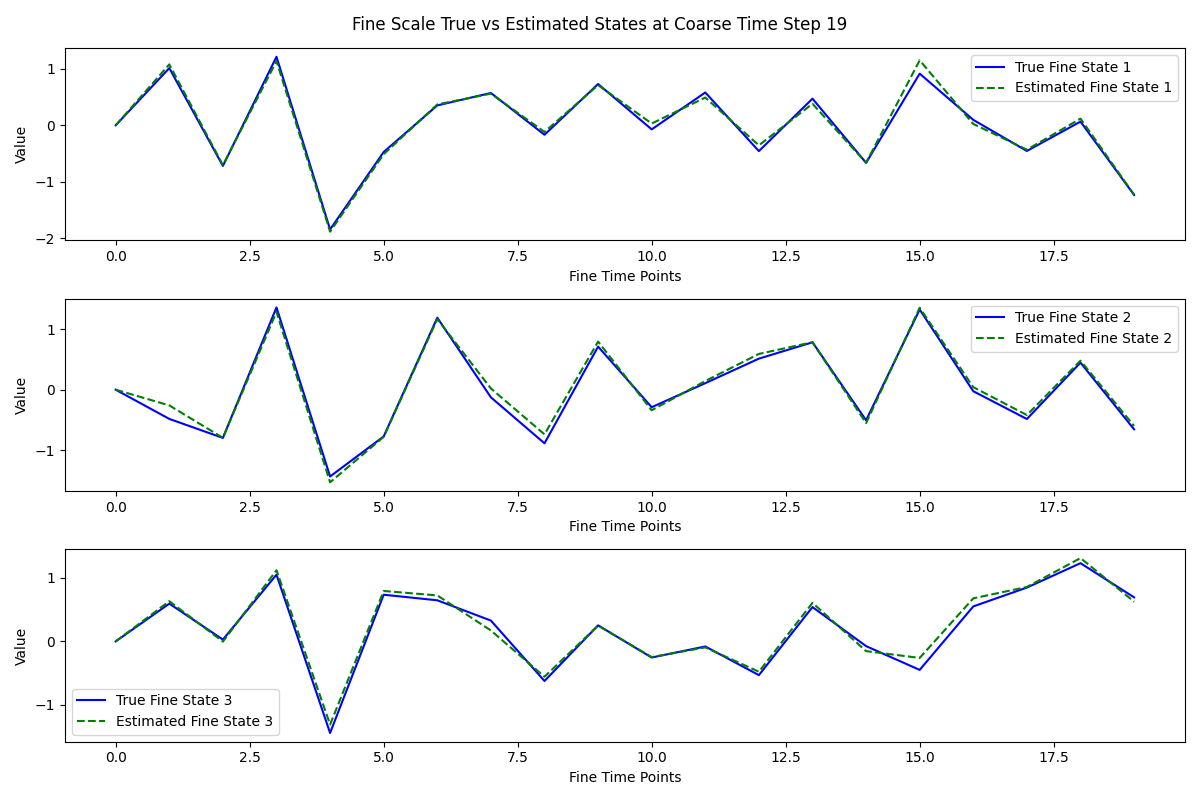}
        \caption{Individual $d=1$}
        \label{fig:d1fine}
    \end{subfigure}

    \vspace{0.3cm} 

    \begin{subfigure}[b]{0.45\textwidth}
        \centering
        \includegraphics[width=\textwidth]{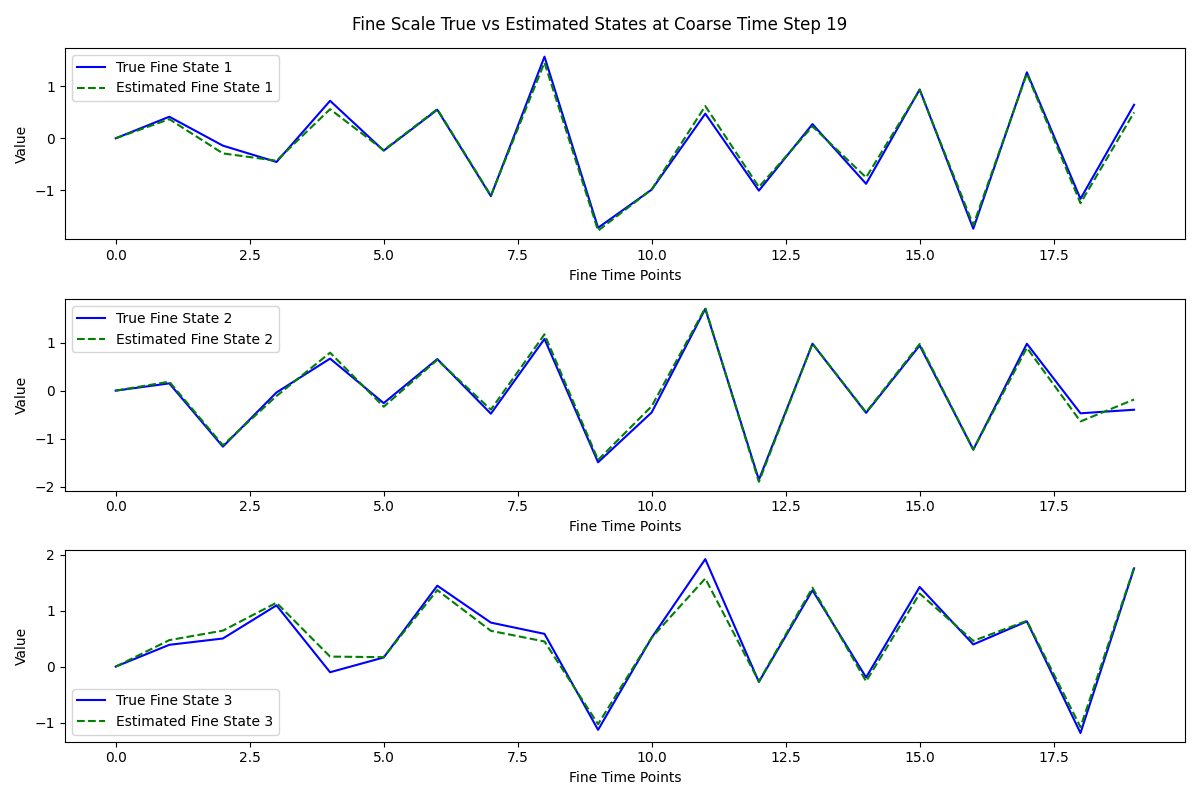}
        \caption{Individual $d=2$}
        \label{fig:d2fine}
    \end{subfigure}
    \hfill
    \begin{subfigure}[b]{0.45\textwidth}
        \centering
        \includegraphics[width=\textwidth]{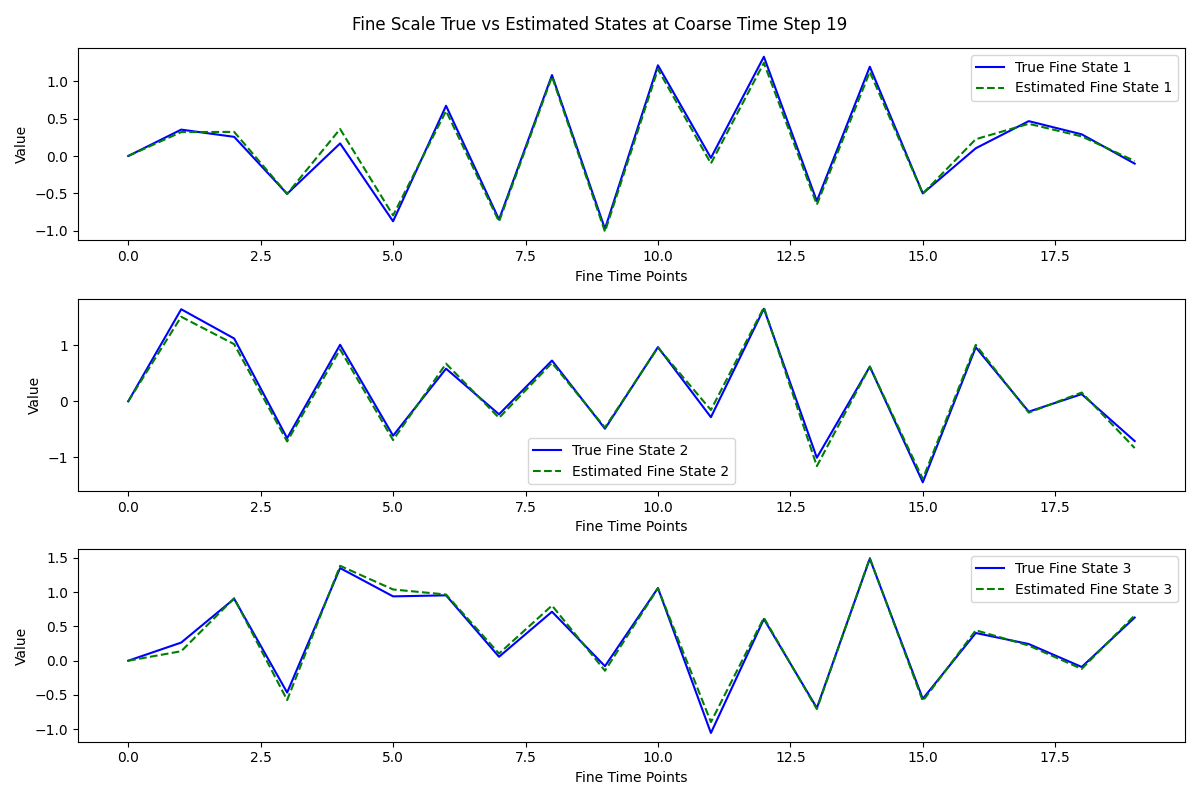}
        \caption{Individual $d=3$}
        \label{fig:d3fine}
    \end{subfigure}

    \caption{True vs. estimated fine time scale trajectories at coarse time step $t=19$ for individuals $d=0, d=1, d=2$, and $d=3$.}
    \label{fig:fine_trajectories}
\end{figure}

\end{document}